%% file: renormalize.tex
\documentclass[aps,twocolumn,prb]{revtex4} 

\usepackage{graphicx}
\usepackage{amssymb}

\include{definitions}

\begin{document}

\title{Renormalization of the One-Loop Theory of Fluctuations  \\
       in Polymer Blends and Diblock Copolymer Melts}
\author{ Piotr Grzywacz, Jian Qin, and David C. Morse }
\affiliation{ 
   Department of Physics and
   Department of Chemical Engineering \& Materials Science,
   University of Minnesota, 421 Washington Ave. S.E.,
   Minneapolis, MN 55455 }
\date{\today}

\begin{abstract}
Attempts to use coarse-grained molecular theories to calculate corrections 
to the random-phase approximation (RPA) for correlations in polymer mixtures
have been plagued by an unwanted sensitivity to the value of an arbitrary 
cutoff length, {\it i.e.}, by an ultraviolet (UV) divergence.  We analyze 
the UV divergence of the inverse structure factor $S^{-1}(k)$ predicted by 
a `one-loop' approximation similar to that used in several previous studies. 
We consider both miscible homopolymer blends and disordered diblock copolymer 
melts. We show, in both cases, that all UV divergent contributions can be 
absorbed into a renormalization of the values of the phenomenological 
parameters of a generalized self-consistent field theory (SCFT). This 
observation allows the construction of a UV convergent theory of corrections 
to SCFT phenomenology. The UV-divergent one-loop contribution to $S^{-1}(k)$ 
are shown to be the sum of: 
(i) a $k$-independent contribution that arises from a renormalization of 
the effective $\chi$ parameter, (ii) a $k$-dependent contribution that 
arises from a renormalization of monomer statistical segment lengths, 
(iii) a contribution proportional to $k^{2}$ that arises from a
square-gradient contribution to the one-loop fluctuation free energy, and 
(iv) a $k$-dependent contribution that is inversely proportional to the 
degree of polymerization, which arises from local perturbations in fluid
structure near chain ends and near junctions between blocks in block copolymers. 
\end{abstract}
\maketitle             


\section{Introduction}
\label{sec:Intro}

The statistical mechanics of polymer mixtures and block copolymer melts 
exhibit some universal features that are well described by self-consistent 
field theory (SCFT). The phase behavior of homopolymer mixtures is 
reasonably well described by Flory-Huggins theory. Various inhomogeneous
structures formed by flexible polymers, such as interfaces and ordered 
phases of block copolymers, are accurately described by a SCFT of 
inhomogeneous liquids, which reduces to Flory-Huggins theory in the
case of a homogeneous mixture. A self-consistent field approximation also 
underlies the so-called random phase approximation (RPA) for the structure 
function $S(k)$ in homogeneous mixtures: The RPA is obtained by using 
SCFT to calculate the susceptibility of a liquid to a hypothetical 
infinitesimal perturbation, and using the correlation-response theorem 
to relate this linear susceptibility to the corresponding correlation 
function.

SCFT is a highly successful theory, but not a perfect one. Among its 
limitations is the inability of the RPA to accurately describe 
fluctuations very near a critical point in a polymer blend or near an 
order-disorder transition (ODT) in a symmetric diblock copolymer melt. 
The range of temperatures over which deviations from the RPA are
significant is believed to decrease with increasing degree of 
polymerization $N$: The fractional width of this so-called Ginzburg 
region is predicted to decrease as $N^{-1}$ with molecular weight $N$ 
in a homopolymer blend \cite{deGennes1977,Joanny1978,Binder1984}, and 
as $N^{-1/3}$ in a symmetric diblock copolymer melt. 
\cite{Fredrickson1987} For molecular weights typical of experiments, 
however, fluctuation effects that are ignored by SCFT have significant 
observable consequences.

The form of SCFT that has proved useful for the analysis of experimental
data is a phenomenological theory. It contains several parameters whose 
values are sensitive to details of monomer scale structure, which must be 
determined by comparison to experiment. In the simplest form of SCFT for
systems with two types of monomers, these parameters are a Flory Huggins 
interaction parameter $\chi(T)$, and the statistical segment lengths of 
both species. SCFT does not predict how these parameters depend upon the 
details of molecular structure.  Its usefulness arises instead from its 
ability to predict phase behavior, equilibrium structures, and diffuse 
scattering for systems containing polymers that are comprised of the
same types of monomers, but that have different molecular weights 
and/or architectures.

Any attempt to systematically calculate corrections to this SCFT, 
however, must start with some sort of micro-mechanical model.  (We 
need a Hamiltonian to do statistical mechanics.) Here, as in several 
previous studies
\cite{delaCruz1988,Fredrickson1994,Fredrickson1995,Wang2002,Holyst1993,Holyst1994b}
we start from a coarse-grained model of Gaussian chains with 
pairwise-additive interactions between monomers. Each coarse-grained 
monomer in such a model represents a subchain of many chemical repeat 
units, within a chain that contains many such monomers. Such models 
are thus implicitly coarse-grained to some cutoff length intermediate 
between the chemical monomer size and the polymer coil size.  

The long wavelength composition fluctuations that become important 
near the critical point of a blend, or the ODT of a symmetric diblock 
copolymer, exhibit a universal phenomenology of their own. For a 
blend, sufficiently close to the critical point, this is the critical 
behavior of the Ising universality class.  One might hope that our 
theoretical description of these long wavelength fluctuations would 
be insensitive to the value chosen for a cutoff length, or to other 
arbitrary details of how our coarse-grained model behaves at very 
short length scales.  Unfortunately, this is not so, at least not 
in the simplest sense: Numerical values predicted for a variety of 
quantities turn out to be very sensitive to the value chosen for the 
cutoff length. The purpose of this paper is to show how this may be 
remedied by an appropriate renormalization scheme.

\subsection{Field Theory and Mean-Field Theory}
The calculations presented here make use of the auxiliary field 
representation of the partition function that was introduced 
into polymer physics by Edwards\cite{Edwards1965,Edwards1966}.  
This approach has been used in in several previous studies of 
fluctuation effects in polymer blends 
\cite{delaCruz1988,Fredrickson1994,Fredrickson1995,Wang2002}. 
Let
\begin{equation}
     \Zc \equiv \int D[\Rv] e^{-U[\Rv]}
\end{equation}
denote the partition function for a model with a potential energy $U$, 
where $\int D[\Rv]$ denotes an integral over all particle positions. 
The auxiliary field approach makes use of an exact transformation of 
the partition function for any model in which $U$ is a sum of 
intramolecular potential and a pairwise additive potential for 
interactions between monomers. This transformation can be applied to 
either the canonical or grand-canonical partition function. 
The transformation yields a representation of $\Zc$ (in either ensemble) 
as a functional integral of the form
\begin{equation}
    \Zc = \int D[J] \; e^{L[J]}
    \label{ZofJint}
\end{equation}
where $J$ is an auxiliary field (or a pair of such fields, one for 
each monomer type) that has units of monomer chemical potential.  This 
approach is discussed in more detail in Sec. \ref{sec:Functional}.

A saddle-point approximation to the auxiliary field functional
integral is known to yield a very simple form of mean-field theory. 
The free energy functional obtained in this saddle-point approximation 
is the sum of the free energy of an ideal gas of polymers plus a
mean field approximation for the interaction energy. The average
interaction energy obtained in this approximation is the same as
that obtained by a ``random mixing" approximation in which we 
ignore all correlations among the monomer positions. 
The underlying assumption of microscopically random mixing is 
analogous to that used in the Poisson-Boltzmann theory of electrolytes, 
or the original Flory-Huggins lattice model, which both thus
``mean-field" theories in the same sense. Like other microscopic
mean-field theories of this type, the resulting theory makes very 
simple, but generally inaccurate, predictions about the relationship 
between microscopic interaction parameters and macroscopic parameters, 
such as the effective $\chi$ parameter observed in scattering 
experiments. We will make a distinction in what follows between 
this microscopic mean-field theory and the phenomenological SCFT 
that is used to fit experimental data, which contains several 
adjustable parameters. 

\subsection{UV Divergence and Renormalization}
One appealing feature of the auxiliary field approach is that it allows 
the effects of fluctuations of the auxiliary field about this mean-field 
approximation to be treated by standard methods of diagrammatic 
perturbation theory, analogous to those used in the study of critical 
phenomena.\cite{Morse2006} Several studies have attempted to calculate 
corrections to the simple mean field theory for binary polymers blends 
\cite{delaCruz1988,Fredrickson1994,Fredrickson1995,Wang2002} by 
introducing a Gaussian approximation for distribution of fluctuations 
of the auxiliary $J$ about its saddle point. In any perturbative field 
theory, a Gaussian approximation for fluctuations about the saddle 
point can be expressed diagrammatically in terms of Feynman diagrams 
that involve only a single ``loop", or a single wavevector integration. 
For this reason, this approximation is often referred to (and will 
be referred to here) as a ``one-loop" approximation.

In all of these calculations, it was found that the predictions of
the one-loop approximation for corrections to the mean-field free 
energy density and for the inverse structure factor 
\cite{delaCruz1988,Fredrickson1994,Fredrickson1995,Wang2002} 
are dominated by the contributions of short-wavelength fluctuations, 
with wavelengths of order the coarse-graining or (equivalently) a
cutoff length. If the relevant Fourier integrals are cut off at a 
cutoff wavenumber $\Lambda$, predictions for the apparent $\chi$ 
parameter are found to contain terms proportional to $\Lambda$ and
$\Lambda^{3}$, which diverge in the limit $\Lambda \rightarrow 
\infty$.  In the jargon of field theory, the one-loop approximation 
was thus found to be ultraviolet (UV) divergent.  

This UV divergence is not peculiar to studies that rely on the 
auxiliary field approach, but arises in all coarse-grained models
of fluctuation effects in polymer liquids.  The most influential 
theory of fluctuation effects in diblock copolymer blends is that 
of Brazovskii\cite{Brazovskii}, Fredrickson and Helfand 
\cite{Fredrickson1987} (BFH).  The BFH theory and its descendants 
\cite{Dobrynin1991a,Fredrickson1991,delaCruz1991,Muthukumar1997,Stepanow2003}
are based on an effective Hamiltonian formalism in which the 
partition function is approximated as a functional integral with 
respect to a fluctuating monomer concentration field, rather than 
with respect to a fluctuating chemical potential. A saddle-point 
approximation for the functional integral used in the effective
Hamiltonian approach yields the same mean-field theory as that 
obtained by a saddle-point approximation in the auxiliary 
field approach. The BFH theory is based on a self-consistent 
one-loop approximations for fluctuations about this saddle-point. 
As emphasized by Kudlay and Stepanow\cite{Stepanow2003}, this 
approach leads to UV divergences analogous to those encountered
in the auxiliary field approach.

The physical reason for this sensitivity to the value chosen 
for the cutoff length $\Lambda^{-1}$ is not hard to understand. 
The total free energy of a polymer liquid is only subtly 
different from that of a corresponding liquid of oligomers, 
because both are dominated by the effects of monomer scale 
liquid structure.  It should thus be no surprise that the free 
energy density of a coarse-grained model is sensitive to our 
choice of cutoff length: The total free energy is sensitive 
to every detail of local fluid structure, at the shortest 
wavelength relevant to whatever model we consider.
 
Phenomenological SCFT assumes that this sensitivity to local 
structure can be encapsulated within a few phenomenological 
parameters.  Phenomenological SCFT is widely believed to be 
asymptotically exact in the high molecular-weight limit, 
except within a Ginzburg region very near the spinodal. 
If we asssume this to be true, however, we should nonetheless 
expect to obtain different estimates for the values of the 
SCFT phenomenological parameters from different mathematical 
approximations for the properties of a given microscopic
model. Specifically, we should expect to obtain different 
approximations for the $\chi$ parameter and statistical segment 
lengths from a one-loop approximation than those obtained from 
the simple mean-field approximation. 

One-loop corrections to the predictions of the simplest mean-field 
theory should thus be understood to contain two conceptually 
different kinds of corrections:
\begin{enumerate}

\item Corrections to mean-field estimates of the {\it parameters} 
of SCFT, {\it e.g.}, of the statistical segment lengths and 
effective interaction parameters. 

\item Corrections to the {\it phenomenology} of SCFT, which cannot 
be absorbed into corrections to these parameters. We expect these 
to become small in the limit $N \rightarrow \infty$, except within 
a narrow temperature window near the spinodal.

\end{enumerate}
We expect corrections to the SCFT parameters to be sensitive to 
our treatment of short wavelength correlations, and thus, in a 
coarse-grained theory, to our choice of cutoff wavenumber. 
Conversely, we expect corrections to the phenomenology of SCFT, 
such as corrections to Gaussian chain statistics or to RPA predictions 
of the functional form of $S(k)$, to arise primarily from longer
wavelength fluctuations, and to be independent of $\Lambda$. One 
goal of this paper is to provide mathematical evidence for the
consistency of this physical picture.
 
Our ultimate goal is to construct a renormalized perturbation 
theory that allows us to unambiguously calculate corrections to 
phenomenological SCFT. We assume (subject to confirmation) that a
SCFT with renormalized parameters is asymptotically exact in the 
limit $N \rightarrow \infty$.  We thus hope to construct a theory 
in which all corrections to this form of SCFT can be shown to 
vanish in the limit $N \rightarrow \infty$. The assumption that 
SCFT becomes exact as $N \rightarrow \infty$ implies, however, 
that all large, UV divergent contributions to the calculated 
correlation functions (which generally do not vanish in the limit 
$N \rightarrow \infty$) must preserve the dependence on wavevector, 
chain length, and architecture predicted by SCFT. This is possible 
only if all of the UV divergent terms in the one-loop approximation
can somehow be absorbed into corrections to the values of the 
phenomenological parameters used in SCFT.

This criteria for ``renormalizability" imposes some nontrivial 
constraints on the allowed functional forms of UV divergent 
corrections. It implies, for instance, that any UV divergent 
contribution to $S^{-1}(k)$ that we wish to interpret as a 
renormalization of $\chi$ be completely independent of $k$ and 
$N$, and that it have the same value in a polymer blend and a 
diblock copolymer melt of the same composition, like the $\chi$
parameter in SCFT.  Similarly, it requires that any contribution 
to $S^{-1}(k)$ that we associate with a renormalization of a
statistical segment length exhibit the nontrivial but predictable 
$k$-dependence of the change in the RPA prediction for $S^{-1}(k)$ 
that would be caused by a slight swelling or a contraction of 
Gaussian chains due to a change in statistical segment length. 
We show here that these criteria are actually satisfied by the 
one-loop approximation.

\subsection{UV Divergences in Prior Work}
The UV divergence of the one loop theory has inspired a variety of
responses in prior work. 

In studies of fluctuation effects in polymer blends, several authors 
have introduced a cutoff $\Lambda$ that is assumed to be of order the 
inverse monomer size, and simply reported the dominant UV divergent 
contributions as functions of $\Lambda$.
\cite{delaCruz1988,Fredrickson1994,Fredrickson1995,Wang1995}. These 
results have sometimes been interpreted as meaningful predictions for 
the dependence of the $\chi$ parameter 
\cite{delaCruz1988,Fredrickson1994,Fredrickson1995} or the statistical 
segment length \cite{Wang1995} upon microscopic structure. We believe 
that this interpretation is misguided: When a prediction of a 
coarse-grained model for some quantity depends upon a microscopic cutoff 
length, it indicates only that the value of that quantity is sensitive 
to details of local fluid structure that such a model should not be 
expected to accurately describe.

Holyst and Vilgis \cite{Holyst1993,Holyst1994b} have instead argued
for the introduction a cutoff length of order the polymer coil size 
$R \propto \sqrt{N}b$ as a way of controlling the UV divergence. 
Their reasoning is worth recounting: Holyst and Vilgis \cite{Holyst1993}
posed the question of whether the cutoff length that was needed as
a result of the UV divergence of their theory should be taken to be 
a length of order the monomer size $b$ or the coil size $R$. They 
noted that a cutoff length of order $b$ would lead to corrections to 
the mean field (i.e., saddle-point) theory that do not become small 
in the limit $N \rightarrow \infty$. They thus rejected this option
in favor of the introduction of an {\it ad hoc} cutoff length of 
order $R$.  These authors were thus led into a quandary, in part, 
because they did not make the conceptual distinction made here between 
the microscopic mean field theory, which is certainly not exact in 
the limit $N \rightarrow \infty$, and phenomenological SCFT theory, 
which (we argue) is. 

In Brazovskii's analysis of weakly first order crystallization, 
he considers a UV divergent integral expression expression for 
$S^{-1}(k)$, but never mentions its divergence. This UV divergence 
is also not mentioned by Fredrickson and Helfand \cite{Fredrickson1987}
in their application of Brazovskii's analysis to diblock copolymer 
melts. Instead, these authors all report the UV-convergent part of 
the integral, which develops an infrared (IR) divergence at the 
SCFT spinodal, and discard the UV divergent contribution without 
comment.  This approach is consistent with that normally taken in 
field theoretic studies of, e.g., the Ising critical point \cite{Amit1984}, 
in which it is well known that an analogous UV divergence of the 
one-loop theory can be absorbed into a renormalization of the critical 
temperature.  This interpretation of the divergence in the Brazovskii 
model was made more explicit by Dobrynin and Erukhimovich 
\cite{Dobrynin1991a}, who noted (in the appendix) it could also be 
absorbed into a shift in the critical temperature. 

This approach is the only appropriate one in the study of very 
generic models, such as the original Brazovskii model, in which the 
critical temperature is treated from the outset as an unknown 
parameter. It becomes problematic only when such a field theoretic 
treatment of fluctuations is grafted onto a SCFT theory that we 
expect to become exact in the limit $N \rightarrow \infty$.
SCFT predicts nontrivial relationships between critical temperatures 
and order-disorder temperatures of "homologous" systems containing
polymers that are constructed from the same monomers but that have 
different lengths or architectures (e.g., blends and diblock copolymer 
melts).  If we were to treat the transition temperatures for different 
such systems as completely independent parameters, we would forego 
the ability to say anything about corrections to SCFT predictions 
for these relationships. UV divergent contributions to the one-loop 
theory can be made truly benign only if they can be related to the
parameters of SCFT.

We are aware of two previous attempts to renormalize the theory 
of fluctuations in blends or copolymer melts, similar in spirit to 
that given here:

Kudlay and Stepanow \cite{Stepanow2003} attempted to renormalize a 
refined version of the one-loop approximation for $S^{-1}(k)$ in a 
diblock copolymer melt introduced by Fredrickson and Helfand.  
These authors proposed (in effect) that the UV divergence of
$S^{-1}(k)$ could be tamed if it could be absorbed into a 
renormalization of the Flory-Huggins $\chi$ parameter. They concluded, 
however, that this interpetation was not tenable for the theory that 
they considered, because the UV divergence of $S^{-1}(k)$ in this 
theory was shown to exhibit a nontrivial dependence on $k$, and 
because different results for the UV divergent contribution were 
obtained for a diblock copolymer melt and for a binary blend of the 
same composition. 

The first successful attempt to renormalize a prediction of a one-loop 
theory was given by Wang.\cite{Wang2002} Wang used the Edwards' auxiliary 
field approach to derive an expression for a one-loop correction to the 
$k=0$ limit of $S^{-1}(k)$ in a binary homopolymer blend. He showed that 
the result was the sum of a UV divergent part that he interpreted as a 
renormalization of $\chi$, and a UV convergent contribution that vanishes 
in the limit $N \rightarrow \infty$, as suggested by the physical picture 
discussed above. The main limitation of Wang's calculation is that the 
method that he and others \cite{delaCruz1988,Fredrickson1994,Fredrickson1995}
have used to calculate $S^{-1}(k=0)$ in a blend is not easily generalizable 
to allow the calculation of $S^{-1}(k)$ at $k \neq 0$, or to study 
fluctuations in block copolymer melts. Wang was thus not able to examine 
either the $k$-dependent or the architecture dependence of his results. 
As such, Wang's calculation left open the question of whether his proposed 
renormalization scheme would have passed the more stringent 
consistency requirements imposed by Kudlay and Stepanow. Wang's analysis 
is discussed in detail in Sec. \ref{sec:Overview}. Our analysis builds 
directly upon Wang's, and removes many of its limitations. 

\subsection{Outline}
The paper is organized as follows: Sec. \ref{sec:ModelDefinitions} 
introduces the coarse-grained model of interest, as well as several
mathematical and conceptual definitions that are needed to discuss
our results. Sec. \ref{sec:Overview} contains an overview of our 
renormalization procedure and results, beginning with a review 
of Wang's results.  We hope that readers who are interested in 
understanding only the main physical ideas and results will be 
able to follow to this point. 

Sec. \ref{sec:Functional}-\ref{sec:Defects} present the technical
details of our analysis. Sec. \ref{sec:Functional} reviews the 
auxiliary field method, and the Gaussian/one-loop approximation. 
Sec. \ref{sec:BinaryChi} reviews the one-loop calculation of the 
free energy density of a homogeneous binary blend, and a corresponding 
analysis of $S^{-1}(0)$, in which we retain some subdominant terms 
that previous authors ignored.  In \ref{sec:GaussianSq}, we give a
self-contained derivation of the one-loop correction to $S^{-1}(k)$ 
at arbitrary $k$ by functional differentiation of the Gaussian 
approximation for the free energy functional. In \ref{sec:Diagrams}, 
we review the general diagrammatic rules obtained in Ref. [13]. 
These provide an alternate path to the same expressions for $S(k)$,
and also allow us to separate intramolecular from intermolecular 
correlations.  In Sec. \ref{sec:Concentration}, we convert our 
expression for the one-loop correction to $S^{-1}(k)$ at fixed 
chemical potential (grand-canonical ensemble) to a corresponding 
correction for a closed system (canonical ensemble). In Sec. 
\ref{sec:Blend}, we analyze the UV divergence of the one-loop 
approximation for $S^{-1}(k)$ for a binary blend. In Sec. \ref{sec:Diblock}, 
we present the corresponding calculation for a diblock copolymer melt. 
In Sec. \ref{sec:Defects}, we show that ${\cal O}(1/N)$ corrections 
to the UV divergent part of $S^{-1}(k)$ can be consistently interpreted 
as the result of end effects, and of a square-gradient contribution to 
the interaction free energy.

Sec. \ref{sec:PowerCounting} presents a power counting analysis
of the order of UV divergences of an arbitrary diagram, at any
order in a loop expansion. We also show there that a renormalized
loop expansion will yield an asymptotic expansion of corrections
to SCFT in powers of $1/\sqrt{N}$. Sec. \ref{sec:Solutions} 
discusses the relationship between this work and the extensive 
literature on excluded volume problem in polymer solutions. 
Concluding remarks are given in Sec. \ref{sec:Conclusions}.

\section{Model and Definitions}
\label{sec:ModelDefinitions}

In what follows, we explicitly consider binary homopolymer blends
and diblock copolymer melts. We adopt a notation that allows for a 
mixture of any number of molecular species constructed from a palette 
of any number of monomer types.  Let $\cpol_{a}$ be the number 
concentration of molecules of species $a$ in a mixture.  Let 
$\cmon_{i}(\rv)$ be the fluctuating number concentration of monomers 
of type $i$ at point $\rv$. If $\Rv_{a m i}(s)$ denotes the position 
of monomer $s$ of monomer type $i$ on molecule number $m$ of species 
$a$, then
\begin{equation}
   \cmon_{i}(\rv) \equiv \sum_{sma}
   \delta( {\bf r} - {\bf R}_{a m i}(s) ) 
   \quad.
   \label{cmondef} 
\end{equation}
Here, the sum over $s$ is taken over monomers of type $i$ on 
molecules of species $a$, the sum over $m$ is taken over molecules 
of species $a$, and the sum over $a$ is taken over all species that
contain $i$ monomers.  We will consider nearly incompressible liquid 
with an average volume $v$ per monomer, in which $v$ is the same 
for all monomer types, and independent of composition in a mixture.


\subsection{A Coarse-Grained Model}
\label{sub:Model}
Consider a coarse-grained model for polymer liquids in which the 
total potential energy is a sum
\begin{equation}
   U = U_{\rm chain} + U_{\rm int} + U_{\rm ext}[h]
   \eqsp, \label{Utot}
\end{equation}
of an intramolecular potential energy $U_{\rm chain}$, a pair interaction
potential
\begin{equation}
   U_{\rm int} \equiv \frac{1}{2}
   \sum_{ij} \int\!d\rv\int\!d\rv' \;
   \Ur_{ij}(\rv-\rv') \cmon_{i}(\rv) \cmon_{j}(\rv')
   \label{Uint}
\end{equation}
and an external potential
\begin{equation}
  U_{\rm ext}[h] = -\sum_{i} 
          \int \! d\rv \; \hr_{i}(\rv) \cmon_{i}(\rv)
  \eqsp, \label{Uext}
\end{equation}
Here, $\Ur_{ij}(\rv-\rv')$ is a pair potential for interactions 
between monomers of types $i$ and $j$, and $h_i(\rv)$ is an external 
potential field conjugate to $\cmon_i(\rv)$.  The external potential 
fields are introduced as a mathematical convenience, and are set to 
zero in all final expressions.

In what follows, we sometimes use a compact notation in which a 
binary operator `*' is used to indicate integration over a shared 
coordinate (or wavevector) and summation over a shared monomer 
type index. For example,
\begin{eqnarray}
   U_{\rm int} & = & \frac{1}{2} \cmon*\Uh*\cmon \\
   U_{\rm ext} & = & -\hh*\cmon 
   \label{U_star}
\end{eqnarray}
in this notation.

We assume in what follows that $U_{\rm chain}$ is adequately approximated 
at the length scales of interest by the stretching energy of a 
continuous Gaussian chain. Let $b_{i}$ denote the statistical segment 
for monomers of type $i$ on an isolated chain (with $U_{\rm int}=0$), and
\begin{equation}
   l_{i} \equiv v/b_{i}^{2}
\end{equation}
be the corresponding packing length for monomers of type $i$. 

We consider a class of models in which the Fourier transform 
$\Uh_{ij}(\kv)$ of the pair potential $\Ur_{ij}(\rv-\rv')$ is 
of the form 
\begin{equation}
   \Uh_{ij}(\kv) = 
   \bar{U}_{ij} \hat{F}(\kv/\Lambda)
   \eqsp, \label{Uofgammak}
\end{equation}
where $\bar{U}_{ij}$ is a matrix of interaction strengths with
dimensions of volume, $\Lambda$ is an inverse range of interaction, 
and $\hat{F}(\kv/\Lambda)$ is a function that approaches $1$ as 
$\kv/\Lambda \rightarrow 0$. This corresponds to a pair potential
in coordinate space
\begin{equation}
   \Ur_{ij}(\rv-\rv') = 
   \bar{U}_{ij}\Lambda^{3} F(\Lambda |\rv-\rv'|)
   \eqsp, \label{Uofgammar}
\end{equation}
where the function $F$ is the inverse Fourier transform of 
$\hat{F}$, and satisfies normalization condition $\int d\xv \; 
F(\xv) = 1$.  The interaction matrix $\bar{U}_{ij}$ is taken 
to be of the form
\begin{equation}
   \bar{U}_{ij} = v \left [
   \begin{array}{cc}
   \barBZero  & \barBZero  + \barchiZero \\
   \barBZero  + \barchiZero & \barBZero 
   \end{array} \right ] \;
   \eqsp. \label{barUApprox}
\end{equation}
The resulting potential energy reduces in the limit of slow 
spatial variations, in which the characteristic scale for 
gradients is much larger than $\Lambda^{-1}$, to a continuum 
approximation 
\begin{equation}
   U_{\rm int}  \simeq  v \int \! d\rv \; 
   \left \{ \frac{1}{2}\barBZero  (\cmon_{1} +\cmon_{2})^{2} 
   + \barchiZero \cmon_{1}\cmon_{2} \right \} 
   \eqsp, \label{UintApprox}
\end{equation}
in which $\barBZero/v$ is the mean-field compression modulus.

The analysis presented here can be carried out in either canonical 
or grand-canonical ensemble, with only minor differences. We will
work in grand-canonical ensemble.  Let $\Zp[\hr]$ denote the 
grand canonical partition function, for some choice of chemical 
potentials. Let $\Gm[\la \cmon\ra]$ be the corresponding free 
energy expressed as a functional of the average monomer 
concentration fields. This is defined by the Legendre transform
\begin{equation}
    \Gm[\la \cmon \ra] = -\ln \Zp[\hr] + \hr*\la\cmon\ra
    \eqsp, \label{Gmdef}
\end{equation}
Here, $\hr$ is the external field required to establish 
a monomer concentration $\la\cmon\ra$, which is related to 
$\Gm$ by a functional derivative
\begin{equation}
    h_{i}(\rv) = \frac{\delta \Gm[\la \cmon \ra]}
                      {\delta \la \cmon_{i}(\rv) \ra}
    \eqsp. \label{hdGmdcmon}
\end{equation}
In Eq. (\ref{Gmdef}), and hereafter, we use energy units
in which $k_{B}T=1$.

\subsection{Correlation Functions}
\label{sub:CorrelationFunctions}
We focus on the calculation of the correlation function
\begin{eqnarray}
   \Scc_{ij}(\rv,\rv') & = & 
   \langle \delta \cmon_{i}(\rv)\delta \cmon_{j}(\rv') \rangle 
   \label{Scc2def}
\end{eqnarray}
and its Fourier transform
\begin{equation}
   \Scc_{ij}(\kv) \equiv 
   \int d\rv' \; \Scc_{ij}(\rv,0) e^{i\qv\cdot\rv}
   \eqsp, \label{Scc2kdef}
\end{equation}
where 
$\delta \cmon_{i}(\rv) \equiv \cmon_{i}(\rv) - \la \cmon_{i}(\rv) \ra$.
This quantity obeys the identities
\begin{eqnarray}
   \Scc_{ij}(\rv,\rv') & = & 
   \frac{\delta^{2} \ln \Zp[\hr]}
        {\delta \hr_{i}(\rv) \delta \hr_{j}(\rv')}
   \label{Sccd2lnZ} \\
   \Scc_{ij}^{-1}(\rv,\rv') & = & 
   \frac{\delta^{2} \Gm[\la\cmon\ra]}
        {\delta \la\cmon_{i}(\rv)\ra \delta \la \cmon_{j}(\rv')\ra}
    \label{Sccinvd2Gm} \quad.
\end{eqnarray}
The inverse $\Scc^{-1}$ is defined in coordinate space by 
requiring that $\Scc*\Scc^{-1} = \delta$, where $\delta$ denotes 
$\delta(\rv,\rv')\delta_{ik}$, or in the Fourier space by 
requiring (for a homogenous liquid) that
$\sum_{j}\Scc_{ij}^{-1}(\kv)\Scc_{jk}(\kv) = \delta_{jk}$. 

We define an intramolecular correlation
\begin{equation}
  \Ssh_{a, ij}(\rv,\rv') \equiv \sum_{m} \la
  \cmon_{a m i}(\rv) \; \cmon_{a m j}(\rv') \ra
  \eqsp, \label{Ssha2def}
\end{equation}
that describes correlations between pairs of monomers on the
same molecule $m$ of a specified species $a$, in which 
\begin{equation}
   \cmon_{a m i}(\rv)  \equiv  \sum_{s}
   \delta( {\bf r} - {\bf R}_{a m i}(s) ) 
   \label{cmonamidef}
\end{equation}
is the concentration of monomers of type $i$ on a specific
molecule $m$ of species $a$. The sum over $m$ in Eq. (\ref{Ssha2def}) 
is over all molecules of type $a$. The sum over $s$ in Eq. 
(\ref{cmonamidef}) is over all monomers of type $i$ on 
molecule $m$.

In a molecular mixture, let $\Ssh_{ij}(\rv,\rv')$, with no
species index $a$, denote the total contribution 
\begin{equation}
   \Ssh_{ij}(\rv,\rv') \equiv \sum_{a}
   \Ssh_{a,ij}(\rv,\rv') 
   \eqsp, \label{Ssh2sumdef}
\end{equation}
of intramolecular correlations to $\Scc_{ij}(\rv,\rv')$. Here,
the sum is taken over all species $a$ that contain both $i$ and 
$j$ monomers. In a binary mixture of homopolymers of species 
$a=1$ and $2$, the only nonzero element of $\Ssh_{1,ij}$ for 
homopolymer $1$ is the element $i=j=1$, while the only nonzero 
elements of $\Ssh_{ij}$ are the diagonal elements, with $i=j$. 
In a single component copolymer melt, $\Ssh_{a,ij}$ and $\Ssh_{ij}$
are identical.

\subsection{Self-Consistent Field Theory}
\label{sub:SCFT}
Polymer SCFT is a density functional theory of inhomogeneous polymer 
liquids. It is based upon an approximation of the free energy functional 
$\Gm[\langle c \rangle]$ as a sum
\begin{equation}
    \Gm[\la c \ra] = 
    \Gm_{\rm chain}[\la c \ra] + 
    \Gm_{\rm int}[\la c \ra] 
    \eqsp, \label{GmtotSCFT}
\end{equation}
in which $\Gm_{\rm chain}$ is the free energy of a hypothetical 
reference system of non-interacting polymers with a specified 
average concentration profile, and $\Gm_{\rm int}[\langle c\rangle]$ 
is an additional ``interaction" free energy functional. 

The functional $\Gm_{\rm chain}[\la c \ra]$ is the free energy of 
a hypothetical system of non-interacting chains, with $U_{\rm int}=0$, 
in which a monomer concentration $\la \cmon \ra$ is maintained 
by a potential $U_{\rm ext}[\hSCF]$, with an applied field $\hSCF$.  
The field $\hSCF$ thus satisfies an identity
\begin{equation}
   \hSCF_{i}(\rv) = 
   \frac{\delta \Gm_{\rm chain}[\langle c \rangle]}
        {\delta \la \cmon_{i}(\rv) \ra}
   \label{wdGmchaindcmon}
\end{equation}
as a special case of Eq. (\ref{hdGmdcmon}). 
Applying Eq. (\ref{hdGmdcmon}) to Eq. (\ref{GmtotSCFT}) for 
$F[\la c \ra]$ yields a self-consistent field equation 
\begin{equation}
   \hSCF_{i}(\rv) = h_{i}(\rv) -
   \frac{\delta \Gm_{\rm int}[\la \cmon \ra]}
        {\delta \la \cmon_{i}(\rv) \ra}
   \eqsp, \label{wSCF}
\end{equation}
Here, $h_{i}(\rv)$ is the external field that must be applied to 
the interacting fluid to establish a monomer concentration field
$\la \cmon \ra$, $\hSCF_{i}(\rv)$ is the field required to establish 
the same concentration field in the non-interacting reference system, 
and $\delta \Gm_{\rm int}/\delta \la\cmon\ra$ is an ``internal" field 
contribution.

By itself, Eq. (\ref{GmtotSCFT}) is merely a definition of $\Gm_{\rm int}$, 
rather than a predictive theory.  The standard coarse-grained model 
for dense multi-component liquids of flexible polymers \cite{Matsen2002a} 
may be obtained by supplementing this with the following physical 
assumptions: 
\begin{itemize}
\item[a)] {\it Gaussian Chains}: Polymer conformations are 
adequately described at the mesoscopic scales of interest 
as Gaussian random walks.  
\item[b)] {\it Locality}: The interaction free energy $F_{\rm int}$ 
may be adequately approximated by a local functional, of the form
\begin{equation}
    \Gm_{\rm int}[\la c \ra] = \int d\rv \; \gm_{\rm int}(c_1(\rv),c_2(\rv))
    \eqsp, \label{locality}
\end{equation}
where $\gm_{\rm int}(c_{1}(\rv),c_{2}(\rv))$ is a free energy density 
at point $\rv$ that depends only upon the monomer concentrations at 
point $\rv$.  More precisely, it is assumed that the range of any 
nonlocality in $F_{\rm int}$ is of order the monomer size, and that this 
is small compared to the length scales of interest in applications 
of the coarse-grained theory. 

\item[c)]{\it Incompressibility}: It is often assumed that liquid 
is essentially incompressible at the length scales of interest. 
If coarse-grained $1$ and $2$ monomers have been defined so as to 
occupy the same volume $v$, this may be summarized by a constraint
\begin{equation}
   v^{-1} = \la c_{1}(\rv) \ra + \la c_{2}(\rv) \ra
    \eqsp. \label{incompressible}
\end{equation}
\end{itemize}
The simplest form of phenomenological SCFT for systems with two 
types of monomers assumes, in addition, that $f_{\rm int}$ in an
incompressible liquid may be adequately described by the
Flory-Huggins expression,
$   \gm_{\rm int} = v \barchi \langle c_A(\rv) \rangle 
             \langle c_B(\rv)\rangle $
with an interaction parameter $\barchi$. This simple assumed 
composition dependence is not a necessary or essential part of 
the theory.

To precisely define the decomposition of $\Gm$ into $\Gm_{\rm chain}$ 
and $\Gm_{\rm int}$ in Eq. (\ref{GmtotSCFT}), we must make a choice
of what single-chain reference Hamiltonian to use to define the ideal 
gas free energy $\Gm_{\rm chain}$.  The physical reasoning underlying 
SCFT suggests that this reference Hamiltonian should yield chain 
conformations that, in a homogenous state, are similar to those in 
the liquid of interest. Even chains that are approximately Gaussian 
in a dense liquid generally have statistical segment lengths that are 
slightly different from those of a corresponding system of 
non-interacting chains.\cite{Wang1995} In what follows, we will thus 
define $\Gm_{\rm chain}$ to be the free energy of a hypothetical 
system of non-interacting chains in which the single-chain reference 
Hamiltonian is chosen so as to yield exactly the same single-chain 
probability distribution as that found for chains in the homogeneous 
state of the liquid of interest. The intramolecular reference 
Hamiltonian used to calculate $\Gm_{\rm chain}$ should thus be 
understood to be a single-chain potential of mean force, rather 
than the bare intramolecular potential $U_{\rm chain}$. Our results 
indicate that the UV divergent part of the one-loop contribution to 
the remaining free energy $\Gm_{\rm int}$ is actually a local 
functional only if $\Gm_{\rm chain}$ and $\Gm_{\rm int}$ are 
defined in this way.

\subsection{Ornstein-Zernicke Relation}
\label{sub:OrnsteinZernicke}
It is useful to introduce a generalized Ornstein-Zernicke (OZ) relation
\cite{Erukhimovich1979,Benoit1984,Schweizer1988,Schweizer1989}
\begin{equation}
  \Scc^{-1}_{ij}(\rv,\rv') = 
  \Ssh^{-1}_{ij}(\rv,\rv') - 
  \Ch_{ij}(\rv,\rv') \label{OZdef}
\end{equation}
or
\begin{equation}
  \Scc^{-1}_{ij}(\kv) = \Ssh^{-1}_{ij}(\kv) - \Ch_{ij}(\kv)
\end{equation}
in a homogeneous liquid. Eq. (\ref{OZdef}) defines the direct 
correlation function $\Ch_{ij}$. 
  
The single chain correlation function $\Ssh_{ij}(\kv)$ is related 
to the reference free energy $\Gm_{\rm chain}[\la\cmon\ra]$, as 
defined above, by an identity 
\begin{equation}
   \Ssh_{ij}^{-1}(\rv,\rv') 
    =  \frac{\delta^{2} \Gm_{\rm chain}[\la\cmon\ra]}
       {\delta \la\cmon_{i}(\rv)\ra \delta \la \cmon_{j}(\rv')\ra}
    \quad, \label{Sshinvd2Gmchain}
\end{equation}
analogous to Eq (\ref{Sccinvd2Gm}). Differentiation of the free 
energy of a reference system of non-interacting molecules yields the
intramolecular correlation function because the only correlations in 
such an ideal gas are intramolecular. Differentiation yields the true 
intramolecular correlation function $\Ssh_{ij}^{-1}(\rv,\rv')$ in 
the dense liquid of interest, rather than that of a gas of molecules 
with the bare intramolecular potential $U_{\rm chain}$, as a result 
of the definition of $\Gm_{\rm chain}$ discussed above.

The direct correlation function is intimately related to the 
interaction free energy $\Gm_{\rm int}$ defined by Eq. (\ref{GmtotSCFT}). 
Combining Eqs.  (\ref{Sccinvd2Gm}) and (\ref{Sshinvd2Gmchain}) 
with Eq. (\ref{GmtotSCFT}) for $\Gm$ yields an identity
\begin{equation}
   \Ch_{ij}(\rv,\rv')
    = - \frac{\delta^{2} \Gm_{\rm int}[\la\cmon\ra]}
        {\delta \la\cmon_{i}(\rv)\ra \delta \la \cmon_{i}(\rv')\ra}
    \quad.
\end{equation}
Assuming that $F_{\rm int}$ is a local functional thus implies that 
the Fourier transform $\Ch_{ij}(\kv)$ should be independent of 
$\kv$.

In the case of a binary blend of two homopolymers, the 
$\kv \rightarrow 0$ limit of the transform $\Ch_{ij}(\kv)$ is related 
to the the composition dependence of the free energy density of a 
homogeneous mixture.  Consider a homogeneous blend of two homopolymers, 
in which $N_{i}$ is the degree of polymerization of $i$ homopolymers, 
$c_{i} = N_{i}\cpol_{i}$ is the macroscopic (i.e., spatial average)
concentration of $i$ monomers, and $f_{\rm int}(c_{1},c_{2})$ is the
interaction free energy density of the mixture. Then
\begin{equation}
   \lim_{\kv \rightarrow 0}
   \Ch_{ij}(\kv) 
    = - \frac{\partial^{2} f_{\rm int}}
       {\partial \cmon_{i} \partial \cmon_{j}}
    \eqsp. \label{Chk=0}
\end{equation}

\subsection{Incompressible Limit}
\label{sub:Incompressible}
A simplified expression for $S_{ij}(\kv)$ may be obtained in the 
limit of an incompressible liquid.  In a nearly incompressible liquid 
containing two types of monomer of equal volume, the $2 \times 2$ 
matrix $\Scc_{ij}(\kv)$ has two widely disparate eigenvalues:  In 
this limit, one eigenvector of $\Scc_{ij}(\kv)$ must approach a 
pure ``composition" fluctuation mode, 
$[\delta \la c_{1}(\kv) \ra, \delta \la c_{2}(\kv) \ra]\propto [1,-1]$, 
which satisfies the constraint 
$\delta \la c_{1}(\kv) \ra +  \delta \la c_{2}(\kv) \ra = 0$, and 
thus avoids the large free energy penalty for changes in total monomer
density. The other eigenvector must have a vanishing inner product 
with the first, and so must approach a pure ``compression" mode, 
$[\delta \la c_{1}(\kv) \ra, \delta \la c_{2}(\kv) \ra] \propto [1,1]$.

The incompressible limit of $\Scc_{ij}(\kv)$ may be obtained by assuming 
that the eigenvectors approach the limits described above, and taking the 
eigenvalue of the compression mode to vanish. Alternatively, it may be 
obtained by assuming $\Ch_{ij}(\kv)$ to be of the form 
$\Ch_{ij}(\kv) = -\bar{B} + \delta \Ch_{ij}(\kv)$, and taking $\bar{B}$ 
to infinity while keeping $\delta\Ch_{ij}(\kv)$ finite. Either method
yields a matrix correlation function of the form
\begin{equation}
   \Sccb(\kv) = S(\kv)
   \left [ \begin{array}{cc}
   +1  &  -1  \\
   -1 &   +1 \end{array} \right ] \;
    \eqsp. \label{SccInc}
\end{equation}
with a scalar correlation function 
\begin{equation}
   S(\kv) = 
   \frac{|\Ssh(\kv)|}{\Ssh_{+}(\kv) - 2 v \barchia(\kv)|\Ssh(\kv)|}
\end{equation}
where
\begin{eqnarray}
   \Ssh_{+}(\kv) & \equiv  &
   \Ssh_{11}(\kv) + \Ssh_{22}(\kv)+\Ssh_{12}(\kv)+\Ssh_{21}(\kv)
   \nonumber \\
   |\Ssh(\kv)| & \equiv &
   \Ssh_{11}(\kv)\Ssh_{22}(\kv)- \Ssh_{12}(\kv)\Ssh_{21}(\kv)
   \label{SshPlusDet}
\end{eqnarray}
are the sum of elements of $\Ssh_{ij}(\kv)$ and its determinant,
respectively, and where
\begin{equation}
    \barchia(\kv) \equiv 
    \frac{1}{2 v}
    [ \Ch_{11}(\kv) + \Ch_{22}(\kv)  - 2\Ch_{12}(\kv) ]
    \label{chiadef}
\end{equation}
is a wavenumber-dependent ``apparent" $\barchi$ parameter. 
This definition was introduced by Schweizer and Curro 
\cite{Schweizer1988,Schweizer1989}, using similar reasoning.

In an incompressible binary homopolymer blend, the long-wavelength 
limit 
\begin{equation}
    \barchia(0) \equiv \lim_{\kv \rightarrow 0}
    \barchia(\kv) 
\end{equation}
may be expressed as a derivative
\begin{equation}
    \barchia(0) = - \frac{v}{2}
    \frac{\partial^{2} f_{\rm int}(\phi_{1})}
         {\partial \phi_{1}^{2}}
     \quad, \label{chia_k=0}
\end{equation}
where $f_{\rm int}$ is expressed as a function of the total volume 
fraction $\phi_{1} \equiv v \cmon_{1}$ of one of the two
homopolymers.

\section{Overview}
\label{sec:Overview}

In this section, we provide an overview of our renormalization procedure, 
and summarize our main results.

\subsection{Mean Field Theory}
A saddle-point approximation to the Edwards auxiliary field theory 
yields a very simple mean field theory.  This is a form of SCFT in 
which $\Gm_{\rm chain}[\la \cmon \ra]$ is approximated by the free 
energy functional for a gas of non-interacting chains, with a 
single-chain Hamiltonian $U_{\rm chain}$, and in which
\begin{equation}
   \Gm_{\rm int} = \frac{1}{2}\int \! d\rv \int \! d\rv' \;
   \la \cmon_{i}(\rv)\ra \Ur_{ij}(\rv-\rv') \la \cmon_{j}(\rv') \ra 
   \quad.
\end{equation}
This approximation yields a self-consistent (saddle-point) field
\begin{equation}
   \hSCF_{i}(\rv) = \hr_{i}(\rv) - 
   \int d\rv' \; \Ur_{ij}(\rv-\rv') \la \cmon_{j}(\rv')\ra 
   \eqsp. \label{hSCF}
\end{equation}
In this approximation, the internal contribution to $\hSCF_{i}(\rv)$ 
(i.e., the convolution integral in the above) is literally a ``mean" 
field, insofar as it is approximated by the ensemble average of the 
fluctuating potential field $-\int d\rv' \Ur_{ij}(\rv-\rv')\cmon_{j}(\rv')$ 
at point $\rv$. 

The characteristic features of this microscopic mean-field theory 
(however it is obtained) are that: 
i) $\Gm_{\rm int}$ is approximated by an expression for the average 
interaction energy $\la U_{\rm int} \ra$ that neglects all correlations 
in monomer density, and ii) intramolecular correlations are taken to be 
identical to those of a gas of non-interacting molecules. This theory
makes very simple predictions about the relationship between SCFT 
parameters and the microscopic parameters: It yields a direct correlation 
function $\Ch_{ij}(k) = -\Uh_{ij}(k)$, an apparent $\chi$-parameter 
$\barchia(k) \simeq \barchiZero$ for $k \ll \Lambda$, and statistical 
segment lengths equal to those of the non-interacting chains.


\subsection{One-loop Approximation for Free Energy} 
The treatment of fluctuation effects given in this paper is based on a 
one-loop (or Gaussian) approximation. Several authors 
\cite{delaCruz1988,Fredrickson1994,Fredrickson1995,Wang2002} 
have previously obtained a one-loop approximation for $\barchia(0)$ by 
calculating the free energy of a homogeneous blend as a function of
composition, and applying Eq. (\ref{chia_k=0}).
All of these studies started from an idealized model in which chains
are treated as continuous Gaussian threads, the pair interaction is 
taken to be nominally point-like, as in the Edwards model of excluded 
volume interaction in solution, and in which the liquid is taken to 
be nominally incompressible. When these assumptions are all taken 
literally, the resulting one-loop correction to $\barchia(0)$ is 
given by a Fourier integral that diverges at large wavenumber, i.e., 
that is UV divergent. The divergence can be removed either by 
introducing a discrete chain model or by introducing a nonzero range 
for two-body interactions.\cite{Morse2006} In all of these previous 
studies, the integral expression for $\barchia(0)$ was regularized by 
restricting the integral to wavenumbers less than a cutoff wavenumber
$\Lambda$.

This regularized one-loop approximation for binary homopolymer blends, 
which is presented in Sec.  \ref{sec:BinaryChi}, yields an apparent
$\chi$ parameter of the form
\begin{equation}
   \barchia(0) = \barchiZero 
   + A \Lambda^{3} + B \barchiZero \Lambda 
   + \sum_{i=1}^{2}\frac{H_{i}}{N_{i}}\Lambda 
   + \delta \barchi^{*}(0)
   \label{binaryChik=0Diverge}
\end{equation}
where 
\begin{eqnarray}
   A & = & 
   \frac{ (l_1- l_2)^2  v}{24\pi^2\bar{l}^2}
   \nonumber \\
   B & = & 
   -\frac{6 l_1^2  l_2^2 }{ \pi^2 \bar{l}^3}
   \nonumber \\
   H_{1} & = & - l_1 (l_1 - 1_2)
   \frac{3  l_1 l_2 }{ 2\pi^2\bar{l}^3 }   
   \nonumber \\
   H_{2} & = & - l_2 (l_2 - l_1)
   \frac{3  l_1 l_2 }{ 2\pi^2\bar{l}^3 }   
\end{eqnarray} 
and
\begin{equation}
  \bar{l}(\phi_{1}) \equiv \phi_{1}l_1 + \phi_{2}l_{2}
\end{equation}
Here, $\delta \barchi^{*}(0)$ is a contribution that remains 
finite in the limit $\Lambda \rightarrow \infty$ (i.e., that is 
UV convergent). 

The dominant UV divergent parts of this result have been reported
previously.  De la Cruz {\it et al.} \cite{delaCruz1988} considered 
the structurally symmetric case $b_{1}=b_{2}$, for which $A=0$, and 
reported the contribution of the form $B\barchi\Lambda$. Fredrickson 
and Liu \cite{Fredrickson1994,Fredrickson1995} instead considered
the athermal case $b_{1} \neq b_{2}$ and $\barchiZero=0$, and so 
found the $A\Lambda^{3}$ contribution.  Wang \cite{Wang2002} retained 
both of these contributions. The terms proportional to $\Lambda/N$ 
have not been retained or analyzed in previous work. 

Wang was the only one to attempt to renormalize this theory, in the 
sense proposed here, by absorbing the strongly cutoff dependent
contributions to $\barchi_{a}$ into a redefinition of the SCFT
$\barchi$ parameter. Wang proposed that the $\barchi$ parameter used 
in phenomenological SCFT theory be identified, within the one-loop 
approximation, with a sum 
\begin{equation}
   \barchieff \equiv \barchiZero 
   + A \Lambda^{3} + B \barchi \Lambda 
   \quad. \label{chieffdef}
\end{equation}
This definition of $\barchieff$ is independent of chain length and
architecture, and is thus consistent with the physical picture of 
$\chi$ in SCFT as a parameter that is sensitive to details of local 
fluid structure, but insensitive to changes in chain connectivity
at longer length scales. 

By absorbing the dominant cutoff-dependent terms into a renormalization 
of $\barchi$, while ignoring the contributions of order $\Lambda/N$, 
Wang was able to isolate the remaining UV convergent contribution 
$\delta \barchi^{*}(0)$. It is this quantity that contains predictions 
of the renormalized theory for corrections to SCFT phenomenology.
By introducing several further approximations, Wang obtained an 
approximate analytic expression for $\delta \barchi^{*}(0)$.  Both 
Wang's analytic approximation $\delta \chi^{*}(0)$ and the full
one-loop integral expression for this quantity may be expressed in
the non-dimensionalized form 
\begin{equation}
   N \delta \barchi^{*}(0) = 
   \frac{1}{\bar{N}^{1/2}} \;
   \delta \hat{\chi}^{*}(\chi N, \phi_{1}, N_{1}/N_{2}, b_{1}/b_{2}) 
   \quad, \label{barchi0star_nondim}
\end{equation}
in which $\delta \hat{\chi}^{*}$ is a dimensionless function of all of 
the dimensionless variables relevant to SCFT. Here, $N$ is a reference 
degree of polymerization (e.g., typically $N_{1}$ or $N_{2}$), 
\begin{equation}
   \bar{N} = Nb^{6}/v^{2}
\end{equation}
is an invariant degree of polymerization, and $b$ is a reference
statistical segment length (e.g., $b_{1}$ or $b_{2}$). The value of 
$\chi$ used in the integral that defines the RHS of Eq. 
(\ref{barchi0star_nondim}) may be taken to be either $\barchieff$, 
to obtain a renormalized perturbation theory, or may be replaced by 
$\barchia(0)$ to obtain the type of self-consistent one-loop approximation 
considered by Wang. In either variant of the theory, the resulting 
correction to SCFT vanishes in the limit $N \rightarrow \infty$ as 
a result of the prefactor of $\bar{N}^{-1/2}$.

After showing how to remove the UV divergence from this calculation,
Wang focused primarily on a discussion of $\delta\barchi^{*}(0)$ 
near the spinodal, which determines the critical behavior of a blend. 
We will defer our own discussion of this subject, and other physical 
predictions of the one-loop theory, to a subsequent publication.

\subsection{One-loop Approximation for $S(k)$}
In Secs. \ref{sec:Blend} and \ref{sec:Diblock}, we analyze the UV 
divergent contributions to $S^{-1}(k)$ in homopolymer blends and 
diblock copolymers, respectively. The physical assumptions underlying 
phenomenological SCFT allow for the possibility that both the 
$\barchi$ parameter and the statistical lengths may be different 
in a one-loop approximation from that in mean-field theory.  
Renormalization of a local $\chi$ parameter is expected to introduce 
a $k$-independent change in $S^{-1}(k)$.  Changes in statistical 
segment length would give rise to changes in $S^{-1}(k)$ with a 
non-trivial, but foreseeable, wavenumber dependence.  To distinguish 
these effects, we use the OZ expression for $S^{-1}_{ij}(k)$, and 
calculate separate one-loop contributions to the single chain correlation 
function $\Ssh_{ij}(k)$ and to the direct correlation function $\Ch_{ij}(k)$. 
As in the simpler calculation of $\barchia(k=0)$ discussed above, 
we divide the one-loop contribution to each of these quantities into 
a UV-divergent part that we calculate explicitly, and a remaining 
UV-convergent part that will be examined elsewhere.

\subsubsection{Intramolecular Correlations}
The standard SCFT assumes that polymers are approximately Gaussian 
in a dense mixture, but not that the statistical segment lengths are 
necessarily the same as those in a reference system of non-interacting 
chains.  Let $\Sshi_{ij}(k;b)$ denote the single chain correlation 
function for a Gaussian chain with a specified statistical segment 
length $b$, or (for a diblock copolymer) with a specified pair of 
statistical segment, $b \equiv (b_{1},b_{2})$. For a homopolymer, 
$\Sshi_{ij}(k)$ with $i=j$ is proportional to a Debye function. If 
the only effect of interactions upon the single chain correlation 
function $\Ssh_{ij}(k)$ were to change the values of the statistical 
segment lengths, we would expect to find 
\begin{equation}
   \Ssh_{ij}(k) \simeq \Sshi_{ij}(k; b_{0} + \delta b)
   \quad,
\end{equation}
where $b_{0}$ is the ``bare" statistical segment length (or lengths) 
for non-interacting chains, and $\delta b$ is a correction arising 
from interactions. This correction is calculated here to first order 
in a loop expansion.  
To first order in an expansion in powers of $\delta b$, or to first 
order in a loop expansion, we would expect a renormalization of $b$ 
to yield a result of the form
\begin{equation}
   \Ssh_{ij}(k) \simeq \Sshi_{ij}(k; b_{0})  +
   \sum_{k}\frac{\partial \Sshi_{ij}(k;b)}{\partial b_{k}}
   \delta b_{k} 
   \quad. \label{SshRenorm}
\end{equation}
In the case of a homopolymer blend, the only nonzero terms in the 
sum are those with $i=j=k$. The more general notation is required 
for a diblock copolymer melt, in which $\Sshi_{12}(k;b)$ is a nonzero 
function of both $b_{1}$ and $b_{2}$. 
In addition to this renormalization of the statistical length, we 
expect to find small corrections to Gaussian chain statistics.  We 
thus expect to find a total one-loop contribution to $\Ssh_{ij}(\kv)$ 
of the form 
\begin{equation}
   \delta \Ssh_{ij}(k) \simeq 
   \sum_{k}\frac{\partial \Sshi_{ij}(k;b)}{\partial b_{k}} 
   \delta b_{k} +
   \delta \Ssh_{ij}^{*}(k) 
\end{equation}
in which $\delta \Ssh_{ij}^{*}(k)$ is a UV convergent correction 
that vanishes in the limit $N \rightarrow \infty$.

Our results for the one-loop contribution to $\Ssh_{ij}(k)$ are 
completely consistent with the above discussion: We find that the 
UV-divergent parts of the one-loop contribution to $\Ssh_{ij}(k)$ 
in both homopolymer blends and diblock copolymer melts have precisely 
the wavenumber dependence suggested by Eq. (\ref{SshRenorm}). The 
calculated fractional change in statistical segment length $b_{i}$, 
to first order in a loop expansion, is given by an expression
\begin{equation}
     \frac{\delta b_i}{b_{i}} = 
     \frac{l_{i}^{2}}{\pi^{2}\bar{l}} \;\Lambda
     \label{deltabi}
\end{equation}
that increases linearly with $\Lambda$.  Identical expressions for 
$\delta b_{k}$ are obtained from calculations of $\Sshi_{ij}(k)$ in 
a homopolymer blend and in a disordered diblock copolymer melt with 
the same overall composition.  Eq. (\ref{deltabi}) for $\delta b_{k}$ 
was obtained previously by Wang \cite{Wang1995} by considering the 
effect of fluctuations upon the end-to-end vector of a single chain
in a binary blend. 

The UV convergent one-loop contribution $\delta\Ssh^{*}_{ij}$ can be 
isolated by subtracting the above UV-divergence from the total one-loop
contribution. Our result for this quantity is given by a convergent 
Fourier integral that can be expressed, in either blends or diblock 
copolymer melts, in non-dimensional form 
\begin{equation}
   \frac{v}{N} \; \delta \Ssh_{ij}^{*}(k) =
   \frac{1}{\bar{N}^{1/2}} \;
   \delta \hat{\Ssh}_{ij}^{*}(kR,\chi N, \ldots, )
   \label{dSsh_nondim}
\end{equation}
in which $\delta \hat{\Ssh}_{ij}^{*}$ is a dimensionless function of 
all of the dimensionless variables of SCFT (i.e., $\chi N$, $\phi$, 
$N_{1}/N_{2}$, and $b_{1}/b_{2}$ in a blend, or of $\chi N$, $f$, and 
$b_{1}/b_{2}$ in a diblock copolymer melt), and of a non-dimensionalized
wavevector $k R$. Here, $R = \sqrt{N}b$, $N$ is a reference degree of 
polymerization (i.e., $N_{1}$ or $N_{2}$ in a blend or $N$ in a diblock 
copolymer melt), and $b$ is a reference statistical segment length 
(i.e., $b_{1}$ or $b_{2}$).

\subsubsection{Inter-molecular Correlations} 
We now consider inter-molecular correlations, as characterized by the 
direct correlation function $\Ch_{ij}(k)$. We showed in subsection
\ref{sub:OrnsteinZernicke} that $\Ch_{ij}(k)$ can be expressed as a 
functional derivative of the SCFT interaction free energy
$F_{\rm int}$.  The standard SCFT assumes that $F_{\rm int}$ is a
local functional of monomer concentration fields, and is independent 
of chain length and architecture. If these assumptions were rigorously
correct, the quantities $\Ch_{ij}(k)$ and $\barchia(k)$ would be 
completely independent of $k$, chain length and architecture, but
could depend upon composition.  

In Sec. \ref{sec:Blend}, we show that the one-loop approximation for 
$\barchia(k)$ in a binary blend can be expressed as a sum
\begin{equation}
   \barchia(k) = \barchieff
   + \sum_{i=1}^{2}\frac{H_{i}(kR_{i})}{N_{i}}\Lambda 
   + \delta \barchi^{*}(k)
   \label{binaryChikDiverge}
\end{equation}
in which $\barchieff$ is the renormalized $\chi$ parameter given 
explicitly in Eq. (\ref{chieffdef}), which is independent of $k$.
Here, $\delta \barchi^{*}(k)$ is a UV convergent contribution that 
becomes significant only near a spinodal, while $H_{i}(kR_{i})$ is a 
dimensionless function that approaches the constant $H_{i}$ given in 
Eq. (\ref{binaryChik=0Diverge}) in the limit $k \rightarrow 0$. 
The corresponding expression for $\barchia(k)$ in a diblock 
copolymer melt, which is analyzed in Sec. \ref{sec:Diblock}, is of 
the form
\begin{equation}
   \barchia(k) = \barchieff
   + \frac{H(kR)}{N}\Lambda 
   + \delta \barchi^{*}(k)
   \label{diblockChikDiverge}
\end{equation}
where $N$ is the length of the diblock. Both calculations confirm 
that the contributions to $S^{-1}(k=0)$ that Wang absorbed into 
$\barchieff$ do indeed correspond, for $k \neq 0$, to quantities 
that are independent of $k$ and chain architecture, consistent 
with his physical interpretation. 

Our results for the UV divergent part of $\barchia(k)$ 
simplify considerably in the limit $l_{1} = l_{2} = \bar{l}$ of 
equal statistical segment lengths. In this case, the coefficient 
$A$ vanishes in Eq.  (\ref{chieffdef}) for $\barchieff$. In 
addition, we find that the coefficients $H_{i}$ (in a binary 
blend) and $H$ (in a diblock copolymer melt) also vanish in
this limit, for all $k$. In this case, we thus obtain 
\begin{equation}
    \barchia(\kv) 
    = \barchi \left [ 1 - \frac{6 }{\pi^{3}} l \Lambda \right ]
    + \delta \barchi^{*}({\bf k}) \quad,
\end{equation}
where $l = l_{1} = l_{2}$. 
Here, $\delta\barchi^{*}({\bf k})$ is a UV convergent part. 
The term proportional to $\Lambda$ is the UV divergent correction 
to $\barchi$ that was originally identified by de la Cruz 
{\it et al.} \cite{delaCruz1988}.

The corrections to SCFT that are of physical interest arise from the
smaller UV-convergent contribution $\delta\barchi^{*}(k)$. The exact 
one-loop expression for this quantity is given by a UV convergent 
integral, that may be expressed, in either blends or diblock copolymer 
melts, in the non-dimensional form 
\begin{equation}
   N\delta \barchia^{*}(k) =
   \frac{1}{\bar{N}^{1/2}}
   \delta \hat{\chi}^{*}(kR,\chi N, \ldots, )
   \quad, \label{dchi_nondim}
\end{equation}
where $\delta \hat{\chi}^{*}$ is a dimensionless function of the same 
variables as those used in Eq. (\ref{dSsh_nondim}). The $k \rightarrow
0$ limit of our result for this quantity agrees with that obtained by
examining the composition dependence of the free energy density. 

\subsection{End-Effects and Square-Gradient Terms}
The UV divergent terms in Eqs. (\ref{binaryChikDiverge}) and 
(\ref{diblockChikDiverge}) that are of order $\Lambda/N$ cannot be absorbed 
into the renormalized SCFT parameter $\barchieff$, because they depend on 
molecular weight $N$, wavenumber $k$, and chain architecture. To explain 
the physical origin of these terms, we consider a slightly generalized form 
of SCFT in which we allow for two physical effects that are not included in 
the standard form of the theory.  These are:
\begin{enumerate}
\item 
Excess free energies associated with chain ends and junctions between 
blocks in block copoymers.
\item A square-gradient contribution to the one-loop interaction free 
energy.
\end{enumerate}
In Sec. \ref{sec:Defects}, we thus compare our one loop results for 
$\chi_{a}(k)$ to the predictions of a phenomenological model in which 
the UV divergent part of the one-loop contribution to $\Gm_{\rm int}$ is 
assumed to be of the form
\begin{equation}
   \delta \Gm_{\rm int} = \int d\rv [ \;
   \delta f_{\rm local} + \sum_{\alpha}d_{\alpha}\psi_{\alpha} 
   + \frac{1}{2} D( \nabla \phi_{1} )^{2} \; ]\quad,
   \label{FGlocal}
\end{equation}
where $\phi_{1}(\rv) = v \cmon_{1}(\rv)$ is a local volume fraction 
for one of the components in an incompressible liquid.  Here, 
$\delta f_{\rm local}(\rv)$ is a local free energy density, $d_{\alpha}(\rv)$ 
is a local concentration of chain end or junction `defects', and 
$\psi_{\alpha} (\rv)$ is an excess free energy arising from the presence 
of a `defect' of type $\alpha$.  In a binary homopolymer blend in a binary 
homopolymer blend, $d_{1}(\rv)$ and $d_{2}(\rv)$ are the concentrations of 
chain ends for chains of type $1$ and $2$, respectively. In a diblock 
copolymer melt, the index $\alpha$ can take values $1$, $2$, or $J$, where 
$d_{J}(\rv)$ is the local concentration of junctions between the blocks. 
The free energy density $\delta f_{\rm local}(\rv)$, the defect free 
energy $\psi_{\alpha}(\rv)$ for each type of defect, and the coefficient
$D$ are all assumed to be sensitive to local fluid structure, and so may 
depend upon the cutoff $\Lambda$, the statistical segment lengths, and 
the local composition $\phi_{1}(\rv)$.

Excess free energies for chain ends and junctions can arise in the
one-loop approximation, even in a model in which the end and junction 
monomers are assumed to be identical to other monomers of the same 
type, simply because the local environment of a chain end or junction 
is different from that of a monomer in the middle of a long chain. 
This difference is captured at a crude level even by a one-loop theory. 
In any real polymer liquid, the excess free energy associated with chain 
ends would also be sensitive to any differences between the actual 
chemical structure of the terminal units and the chemical repeat unit. 

Previous analyses of the one-loop approximation have ignored these
${\cal O}(\Lambda/N)$ contributions to $\barchia(\kv)$. It was tempting 
for us to do the same, on the grounds that these terms are smaller by 
a factor of $1/N$ than those absorbed into $\barchieff$.  Actually, 
however, these contributions are the same order in an expansion in 
$1/N$ as the term proportional to $\barchiZero$, since $\barchiZero$ 
must be less than a binodal value of of ${\cal O}(1/N)$ in order for 
the homogeneous state of interest to remain stable. We also found 
that we needed to analyze and subtract these ${\cal O}(\Lambda/N)$
divergences in order to carry out the numerical renormalization 
procedure that we are now using to calculate UV-convergent predictions 
of the theory. This procedure is described briefly in Sec. 
\ref{sec:Conclusions}.

We discuss contributions to $\delta\barchia(k)$ that arise from a 
postulated square-gradient contribution to $F_{\rm int}$ simultaneously 
with those that arise from end and junction defects because we find 
that these contributions are otherwise difficult to disentangle. The 
square-gradient contribution to $\delta F_{\rm int}$ in the above model 
simply adds a contribution 
$\delta\barchia(k) = -\frac{v}{2} D k^{2}$ to $\barchia(k)$. 
This, however, can also be written as 
$-\frac{v}{2} D \; (kR)^{2}/Nb^{2}$, where 
$R = \sqrt{N}b$, and so can be absorbed into a contribution of the 
more general functional form $H(kR)\Lambda/N$, if $D \propto \Lambda$. 
The only way for us to identify a squared-gradient contribution is 
thus to explicitly calculate the wavenumber dependence that 
would be produced by end and junction defects alone, and then see 
if our one loop results for $\delta\barchia(k)$ can be expressed 
as the sum of this defects contribution plus an additional square
gradient contribution. 

\subsubsection{Free Energy of Homogeneous Liquids}
Expressions for $\delta f_{\rm local}$ and $\psi_{\alpha}$ in this
generalized SCFT may be obtained by examining the one-loop contribution 
to the free energy density of a homogeneous liquid. 

The quantity $\delta f_{\rm local}$ may be obtained by considering
the $N \rightarrow \infty$ limit of the UV divergent part of the 
free energy density of either a homogeneous binary blends or a
disordered diblock copolymer melts. In either case, we obtain
\begin{eqnarray}
   \delta f_{\rm local} & = & \frac{1}{12\pi^{3}}\left [ 
   \ln \left ( 12 \barBZero \frac{\bar{l}}{v\Lambda^{2}} \right )
   +\frac{2}{3} \right ]\Lambda^{3}
   \nonumber \\
   & - & 
   \frac{ 6 }{ \pi^{2}v }
   \frac{l_{1} l_{2} }{ \bar{l} } \barchiZero \phi_{1}\phi_{2}\Lambda
   \quad, \label{dflocal}
\end{eqnarray}
where $\phi_{i}$ is the macroscopic volume fraction of $i$ monomers.
The one-loop contribution to the parameter $\barchieff$ defined in 
Eq. (\ref{chieffdef}) is related to $\delta f_{\rm local}$ by a 
second derivative with respect to $\phi_{1}$, as in Eq. (\ref{chia_k=0}).

In a homogeneous binary blend, we find that the total UV divergent part 
of the one-loop free energy density $\delta f$ of a liquid of finite
chains may be expressed as a sum
\begin{equation}
   \delta f = \delta f_{\rm local} + 
   \frac{2\phi_{1}}{N_{1}v}\psi_{1} +
   \frac{2\phi_{2}}{N_{2}v}\psi_{2}
   \quad,
\end{equation}
where $2\phi_{\alpha}/N_{\alpha}v$ is the concentration of chain ends for
chains of type $\alpha$, and 
\begin{equation}
   \psi_{\alpha} = 
   -\frac{3}{4\pi^{2}\bar{l}} \; l_{\alpha}^{2} \Lambda 
   \quad. \label{psiend}
\end{equation}
The quantity $\psi_{\alpha}$ is thus tentatively identified as the excess 
free energy of an $\alpha$ chain end.

The corresponding free energy for a disordered diblock copolymer melt can 
be written as a sum
\begin{equation}
   \delta f = 
   \delta f_{\rm local} + \frac{1}{Nv}
   \left ( \psi_{1} + \psi_{2} + \psi_{J}  \right )
\end{equation}
where $\psi_{1}$ and $\psi_{2}$ are excess free energies for the ends
of the $1$ and $2$ blocks, as given by Eq. (\ref{psiend}). The remaining
energy $\psi_{J}$ is given by
\begin{equation}
   \psi_{J} = -\frac{3}{4\pi^{2}\bar{l}}(l_1 - l_2)^{2}\Lambda
   \label{psijct}
   \quad.
\end{equation}
This is assumed to be the excess free energy arising from the junction 
in a single diblock copolymer. The fact that this expression for $\psi_{J}$ 
vanishes in the limit $l_1 = l_2$ is consistent with the fact that in 
the limit $\barchiZero=0$ and 
$l_1 = l_2$ the two blocks of a diblock copolymer become indistinguishable, 
so that the excess free energy associated with the junction must thus 
vanish in this limit. (This argument would allow $\psi_{J} \neq 0$ for 
$l_1 = l_2$ and $\chi_0 \neq 0$, but this does occur to first order in 
a loop expansion.)

\subsubsection{Composition Fluctuations}
The main evidence for our interpretation of the ${\cal O}(\Lambda/N)$
contributions to $S^{-1}(k)$ as a combination of end and junction effects
and square-gradient contributions is a demonstration that the wavenumber 
and parameter dependence of these terms can be explained by this
interpretation.

In Sec.\ref{sec:Defects} we present an RPA calculation for $S^{-1}(k)$ 
in a system with an additional free energy of the form given in Eq. 
(\ref{FGlocal}), using the explicit expressions for $\psi_{\alpha}$ given 
in Eqs. (\ref{psiend}) and (\ref{psijct}). 
For a binary blend, the required calculation is the same as the RPA 
calculation for a blend of polymers with chemically distinct end-groups, 
or a mixture of CAC and DBD triblocks with very short C and D end 
groups. The calculation for diblock copolymers is similar to that
for a pentablock copolymer with very short end and middle blocks. The
inclusion of end effects in the RPA generally yields contributions to 
$S^{-1}(k)$ that are proportional to $1/N$, with a nontrivial 
$k$-dependence that is different in binary blends and diblock copolymer 
melts.

The excess free energies associated with chain ends and junctions 
affect the collective correlation function $S(k)$ if and only if 
the defect free energies $\psi_{\alpha}(\rv)$ depends upon the
composition $\phi_{1}(\rv)$ of its immediate environment. If the 
free energy per defect depends upon the composition of its 
environment, defects will tend to cluster where their free energies 
are lowest, and to favor collective composition fluctuations that 
lower the total defect free energy.  In systems with $l_{1} \neq
l_{2}$, the above expressions for $\psi_{\alpha}(\rv)$ generally do 
depend upon $\phi_{1}(\rv)$, because of the composition dependence 
of the average packing length $\bar{l}(\phi_{1})$. Both end and 
junction defect free energies are independent of $\phi_{1}$ when
$l_1 = l_2$, implying (correctly) that we should find no "defect"
corrections to $S^{-1}(k)$ in this special case.


We find that the results of this generalized RPA calculation exactly 
reproduce the structure of the UV-divergent ${\cal O}(1/N)$ terms 
obtained in our one-loop calculation of $S^{-1}(k)$, if we allow 
for both end effects and a square-gradient contribution to 
$F_{\rm int}[\langle c \rangle]$.  The contributions to $S^{-1}(k)$ 
that arise from the defects vanish in the limit $l_{1}=l_{2}$, for
the reason discussed above.  The value of the coefficient $D$ that 
we infer by this method (which is the same in binary blend and 
diblock copolymer melts of the same composition) is given by
\begin{equation}
   D = - \frac{(l_1-l_2)^{2}}{3\pi^{2}\bar{l}^{2}}\Lambda
   \quad. \label{D_scalar}
\end{equation}
Because both the defect contributions and this square-gradient
coefficient vanish when $l_1 = l_2$, the ${\cal O}(\Lambda/N)$
contribution to $\delta S^{-1}(\kv)$ vanishes in this case.

The success of this approach strongly suggests that the UV divergent 
part of the one-loop contribution to the free energy functional 
$\Gm_{\rm int}[\langle c \rangle]$ is of the form assumed in Eq. 
(\ref{FGlocal}), even for strongly inhomogeneous liquids. We thus
hope that this result will also provide a basis for removing the UV 
divergence of the one-loop approximation for the free energy of 
{\it ordered} phases of block copolymer melts.


\section{Auxiliary Field Method}
\label{sec:Functional}
The Edwards' functional integral representation of $\Zp$ may be obtained 
from the identity
\begin{equation}
     e^{-\frac{1}{2}\cmon*\Uh*\cmon} 
      = \CGint^{-1} \int \! D[J] \;
      e^{-\frac{1}{2}\Jr*\Uh^{-1}*\Jr + i\Jr*\cmon}
      \quad.
\end{equation}
Here, $\int \! D[J]$ represents a functional integral with respect to 
an auxiliary chemical potential field, where $\Jr_{i}(\rv)$ is a field
component that couples to $\cmon_{i}(\rv)$. The constant $\CGint$ is 
given by the integral
\begin{equation}
   \CGint \equiv \int \! D[J] \; e^{-\frac{1}{2}\Jr*\Uh^{-1}*\Jr}
\end{equation}
Substituting this representation of $e^{-U_{\rm int}}$ into the definition
of the grand-canonical partition function $\Zp$ yields a functional 
integral 
\begin{eqnarray}
   \Zp[\hr]   & = & \CGint^{-1} \int\! D[J] \; e^{\Lf[\hr,\Jr]}
   \label{XJint} 
   \\
   \Lf[\hr,\Jr] & \equiv & 
   \ln \Zid[\hr+i\Jr] -\frac{1}{2}\Jr*\Uh^{-1}*\Jr
   \eqsp, \label{Ldef}
\end{eqnarray}
Here, we have introduced the notation $\Zid[\hir]$ for the partition 
function of an ideal gas of molecules subjected to an applied field 
$\hir$, for which the total potential energy is $U_{0}+U_{\rm ext}[\hir]$.
In Eq. (\ref{Ldef}), $\Zid[\hr+i\Jr]$ is the grand-canonical partition 
function for an ideal gas in which monomers of type $i$ are subjected to a 
fluctuating complex field
\begin{equation}
   \hir_{i}(\rv) \equiv  \hr_{i}(\rv) + iJ_{i}(\rv)
   \eqsp. \label{hirdef}
\end{equation}
Hereafter, quantities such as $\Zid$ and $\hir$ that are defined in 
this ideal gas reference state will be denoted with a tilde.

The functional derivative of $\ln \Zid[\hir]$ with respect to $\hir$ is 
the average monomer concentration
\begin{equation}
   \la \cmon_{i}(\rv) \ra = 
   \frac{\delta \ln \Zid[\hir]}{\delta\hir_{i}(\rv)}
   \eqsp. \label{dlnZidhir}
\end{equation}
Here, $\la \cdots \ra$ denotes an average taken in the ideal gas 
subjected to a field $\hir$. Higher derivatives yield
\begin{equation}
   \Sshi^{(n)}_{i_1 i_2 \ldots i_n} (\rv_1,\ldots,\rv_{n}) 
   = \frac{\delta^{(n)} \ln \Zid[\hir]}
          {\delta\hir_{i_1}(\rv_1)\delta\hir_{i_2}(\rv_2)
           \cdots \delta\hir_{i_n}(\rv_n)}
   \label{SshidlnZiddhir}
\end{equation}
where 
\begin{equation}
    \Sshi^{(n)}_{i_1 i_2 \ldots i_n}(\rv_1,\ldots,\rv_n) 
    \equiv \sum_{a m} \sum_{s_1 \cdots s_n} \la 
    \cmon_{a m i_{1}}(\rv_{1}) \cdots \cmon_{a m i_{n}}(\rv_{n}) 
    \ra
\end{equation}
is an $n$-point intramolecular correlation function in the ideal 
gas reference state for monomers of types $i_1, i_2, \ldots, i_n$ 
on the same molecule. 

\subsection{Gaussian Fluctuations}
As already noted, a very simple form of mean-field theory is obtained 
by applying a saddle-point approximation to functional integral 
(\ref{XJint}).  Requiring that 
\begin{equation}
   \frac{\delta L[h,\Jr]}{\delta\Jr_{i}(\rv)} = 0
\end{equation}
and using Eq. (\ref{dlnZidhir}) yields the saddle-point condition
given in Eq. (\ref{hSCF}), in which
$\hir_{i}(\rv)$ denotes the saddle-point value of the field 
$\hir = h + i\Jr$ defined in Eq. (\ref{hirdef}), and in which
$\la \cmon_{j}(\rv) \ra$ is the corresponding ideal-gas monomer 
concentration. 

A Gaussian approximation for $\Zp$ is obtained by the approximating
the deviation $\delta \Lf$ of $\Lf$ from its saddle point value by an 
expansion to second order in the deviation $\delta \Jr_{i}(\rv) \equiv
\Jr_{i}(\rv) - \Jrs_{i}(\rv)$ of $\Jr$ from the saddle-point field 
$\Jrs$. We thus approximate
\begin{equation}
    \delta \Lf \simeq
    - \frac{1}{2}\sum_{ij} \int \! d\rv \int \! d\rv' \;
    \Ghi^{-1}_{ij}(\rv,\rv')
    \delta \Jh_{i}(\rv) \, \delta \Jh_{j}(\rv')
    \quad.  \label{LGauss}
\end{equation}
where 
\begin{eqnarray}
   \Ghi^{-1}_{ij}(\rv,\rv') & = & -
   \frac{\delta \Lf[\Jr,\hr]}
             {\delta \Jr_{i}(\rv) \delta \Jr_{i}(\rv')}
   \nonumber \\
   & = &
   \Sshi_{ij}(\rv,\rv') + \Uh^{-1}_{ij}(\rv,\rv')
   \eqsp,  \label{Gidef}
\end{eqnarray}
or, in the Fourier space representation for a homogeneous fluid,
\begin{equation}
   \Ghi^{-1}_{ij}(\kv) = \Sshi_{ij}(\kv) + \Uh^{-1}_{ij}(\kv)
   \quad. 
\end{equation}
The ``propagator" $\Ghi$ obtained by inverting $\Ghi^{-1}$ is 
the screened interaction potential identified by Edwards 
\cite{Edwards1965,Edwards1966}, which is closely analogous to 
the electrostatic screened interaction in the Debye-H\"{u}ckel 
theory of electrolytes. It was noted by Edwards that results 
of the one-loop theory could be obtained from a perturbation 
theory in which monomers interact via this screened interaction. 
It was shown more systematically by one of us \cite{Morse2006} 
that the diagrammatic perturbation theory that arises naturally 
from the auxiliary field method is equivalent to a type of 
molecular cluster expansion in which interaction lines (or 
``bonds") represent factors of the screened interaction. 

In the Gaussian approximation 
\begin{equation}
   \Zp \simeq \Zp_{s} \Zp_{G}
   \quad,
\end{equation}
where $\Zp_{s} = e^{L[\hr,\Jrs]}$ is the saddle-point value, and 
$\Zp_{G}$ is a factor arising from Gaussian fluctuations. The Gaussian
contribution to the free energy in a homogeneous liquid is 
\begin{eqnarray}
   -\ln \Zp_{G} & = & 
   \frac{1}{2}V\int_{\kv}
   \ln \det [ \Ghib^{-1}(\kv)\Uhb(\kv) ]
   \nonumber \\
    & = & \frac{1}{2}V\int_{\kv}
   \ln \det [ \Iv + \Sshib(\kv)\Uhb(\kv) ]
   \eqsp, \label{FGauss}
\end{eqnarray}
where $\det[ \cdots ]$ denotes the determinant of a $2 \times 2$ 
matrix. 

\subsection{Incompressibilty and Regularization}
In the limit $k \ll \Lambda$, an explicit expression may be given for 
$\Ghib(\kv)$ for a nearly incompressible homogeneous liquid.  By taking 
the incompressible limit $B_{0} \rightarrow \infty$ , while 
approximating 
$F(k/\Lambda) = 1$ for $k \ll \Lambda$, we obtain
\begin{equation}
   \Gh_{ij}(\kv) = 
   \frac{ 1 - 2 v \barchiZero |\Sshi(\kv)| \Sshi^{-1}_{ij}(\kv) }
        {\Sshi_{+}(\kv) - 2 v \barchiZero |\Sshi(\kv)|}
   \quad. \label{GhSsInc} 
\end{equation}
The `1' in the numerator denotes a contribution of unity to every
matrix element. 

In the opposite high-wavenumber limit $k \gg \Lambda$, the bare 
potential $\Uhb(\kv)$ in any liquid with a large but finite bare 
compression modulus $\barBZero$ is assumed to become vanishingly small, 
as a result of the decay of the crossover function $F(k/\Lambda)$. 
This decay of the bare potential will also cause the screened 
potential $\Ghib(\kv)$ to become very small for $k \gg \Lambda$.

In Eq. (\ref{FGauss}), the Gaussian contribution to the free energy
is expressed as a Fourier integral in which the integrand depends on 
$\Ghi(\kv)$. This is a generic feature of the auxiliary field theory: 
Fluctuation corrections to mean-field results for all quantities of 
interest may be expressed as Fourier integrals involving factors of 
the screened interaction $\Ghi(\kv)$.  Because both $\Uh(\kv)$ and 
$\Ghi(\kv)$ vanish for $k \gg \Lambda$, the use of any model with a 
nonzero range of interaction $\Lambda^{-1}$, rather than a point-like 
interaction, thus naturally introduces a cutoff length. 

The effects of a nonzero range of interaction $\Lambda^{-1}$ may 
thus be crudely mimicked by treating $\Lambda$ as a cutoff wavenumber, 
and simply suppressing contributions from wavevectors $k \gg \Lambda$ 
in all Fourier integrals. This regularization scheme, which we will
adopt, is equivalent to the use of a model in which $\Uhb(\kv)$ is 
independent of $k$ for all $k < \Lambda$, and zero for all $k > 
\Lambda$. In what follows, we will also restrict ourselves to the 
nearly-incompressible limit, in which $\Ghib(\kv)$ is given for all 
$k < \Lambda$ by Eq. (\ref{GhSsInc}). 

We consider a Gaussian approximation for the free energy density of 
a homogeneous liquid ({\it e.i.}, either a binary homopolymer blend 
or a disordered diblock copolymer melt) in which the integral in
Eq. (\ref{FGauss}) is restricted to $k < \Lambda$, and in which the 
integrand is evaluated for $k < \Lambda$ by considering the limit 
of large $\barBZero$. This yields a one-loop correction to the free 
energy density
\begin{equation}
  \delta f \equiv - \ln \Zp_{G}/V
\end{equation}
given by
\begin{equation}
   \delta f = \frac{1}{2}\int\limits_{|\qv|< \Lambda}
   \ln \left \{
   \left [ \Sshi_{+}(\qv) - 2v\barchiZero|\Sshi(\qv)| \right ] 
   \barBZero v \right \}
   \quad. \label{FGauss_inc}
\end{equation}
Here, $\Sshi_{+}(\qv)$ and $|\Sshi(\qv)|$ are the sum of elements and 
determinant, respectively, of the ideal-gas correlation function matrix
$\Sshi_{ij}(\qv)$, defined by analogy to Eq. (\ref{SshPlusDet}) for the 
related quantities $\Ssh_{+}(\qv)$ and $|\Ssh(\qv)|$.  

\section{Free Energy in Binary Blends}
\label{sec:BinaryChi}

Several authors 
\cite{delaCruz1988,Fredrickson1994,Fredrickson1995,Wang2002} have 
obtained a UV divergent contribution to $\chi_{a}(\kv=0)$ in a 
binary homopolymer blend by calculating the one-loop free energy
$\delta f$ for a homogeneous mixture, as a function of composition, 
and then using Eq. (\ref{chia_k=0}) to extract $\barchia(\kv=0)$. 
In this section, we review and extend this approach.

\subsection{High-$q$ Behavior of $\Sshi$ and $\Ghi$}
To analyze the UV divergence of $\delta f$, we will need an asymptotic 
expansion of the high-$q$ behavior of the intramolecular two-point 
function $\Sshi_{ij}(\qv)$, as an expansion in increasing powers of 
$1/q$. In a binary blend of two Gaussian homopolymers, the function 
$\Sshi_{ij}(\qv)$ is a diagonal matrix with elements
\begin{equation}
   \Sshi_{ij}(\qv) = \delta_{ij}\cmon_{i}D_{i}(\qv)
\end{equation}
with
\begin{equation}
   \Dh_{i}(\qv) = N_{i}g(Q_{i}^{2})
\end{equation}
where $Q_{i}^{2} \equiv q^{2} N b_{i}^{2}/6$, and where
\begin{equation}
   g(x) \equiv 2 ( e^{-x} - 1 + x )/ x^{2}
\end{equation}
is the Debye function.  The required high-$q$ expansion 
of $\Dh_{i}(\qv)$ may be obtained by simply dropping the 
exponentially decaying term in the Debye function (which
is not an analytic function of $1/q$). This yields an
approximation
\begin{equation}
   \Dh_{i}(\qv) \simeq 
   \Dh_{i}^{(0)}(\qv) + \Dh_{i}^{(1)}(\qv) 
\end{equation}
\begin{eqnarray}
   \Dh_{i}^{(0)}(\qv) & = & \frac{2N_i}{Q_{i}^{2}}
   = \frac{12 l_{i}}{v} \frac{1}{q^{2}} \\
   \Dh_{i}^{(1)}(\qv) & = & -\frac{2N_i}{Q_{i}^{4}} 
   = \frac{-72 l_{i}^{2}}{v^{2}N_{i}} 
   \; \frac{1}{q^{4}}
   \label{Dh_asymp_blend}
\end{eqnarray}
where $v$ is a monomer reference volume, and $l_{i} = v/b_{i}^{2}$.

A corresponding expansion of the screened interaction to the same
order yields 
\begin{equation}
   \Ghi_{ij}(\qv) \simeq 
   \Ghi^{(0)}(\qv) + \Ghi^{(1)}(\qv)
   + \Ghi_{ij}^{(\chi)}(\qv)
   \label{Ghiasymp}
\end{equation}
where
\begin{eqnarray}
   \Ghi^{(0)}(\qv) & = & \frac{v^2}{12 \bar{l} } \; q^{2} 
   \label{Ghiasymp0} \nonumber \\
   \Ghi^{(1)}(\qv) & = & 
   \frac{v}{2\bar{l}^{2}} \left
   ( \frac{\phi_{1}l_{1}^{2}}{N_{1}} + \frac{\phi_{2}l_{2}^{2}}{N_{2}} 
   \right ) 
   \nonumber \\
   \Ghi_{ij}^{(\chi)}(\qv) & = &  
   \frac{ 2v }{ \bar{l}^{2} } Z_{ij} \barchiZero
   \label{Ghiasymp1} 
\end{eqnarray}
with $\bar{l} \equiv \phi_{1} l_{1} + \phi_{2} l_{2}$, and
\begin{eqnarray}
   Z & \equiv &
   \left [ \begin{array}{cc}
   - \phi_{2}^{2}l_{2}^{2} & \phi_{1}\phi_{2}l_{1}l_{2} \\
   \phi_{1}\phi_{2}l_{1}l_{2} & - \phi_{1}^{2}l_{1}^{2} 
   \end{array} \right ] 
   \quad.
\end{eqnarray}
In Eq. (\ref{Ghiasymp}) and (\ref{Ghiasymp1}), the quantities 
$\Ghi^{(0)}(\qv)$ and $\Ghi^{(1)}(\qv)$ are written with no 
monomer type indices $i$ and $j$ to indicate that the values 
of these quantities are actually independent of $i$ and $j$. 

\subsection{Free Energy Density}
\label{sub:FreeEnergyBlend}
In a binary homopolymer blend, Eq. (\ref{FGauss_inc}) for $\delta f$ 
reduces to
\begin{equation}
   \delta f \simeq \frac{1}{2} 
   \int\limits_{\qv} \!\!
   \ln [( \Sshi_{1} + \Sshi_{2} - 2 v \barchiZero \Sshi_{1}\Sshi_{2} ) 
        \barBZero v ]
   \quad, \label{FGauss_blend}
\end{equation}
The dominant contribution to the argument of the logarithm in 
Eq.  (\ref{FGauss_blend}) in the high-$k$ limit arises from the 
leading order contribution to 
$\Sshi_{+} \equiv \Sshi_{1} + \Sshi_{2}$, which is
\begin{equation}
   \Sshi_{+}^{(0)}(\qv) \equiv 
   \sum_{i}\cmon_{i} D_{i}^{(0)}(\qv)
   = \frac{12 \bar{l}}{v^{2}} \frac{1}{q^{2}} 
\end{equation}
Note that $\Sshi_{+}^{(0)}(\qv) = 1/G^{(0)}(\qv)$.
Factoring $\Sshi_{+}^{(0)}\barBZero$ out of the argument of 
the logarithm yields an expression
\begin{equation}
   \delta f  = 
   \delta f^{(0)} + \frac{1}{2}\int\limits_{\qv}
   \ln \left \{  1 + \Gh^{(0)}
   \left [ \Delta \Sshi_{+} - 2 v \barchiZero|\Sshi| \right ] \right \}
   \label{dfLnFactor}
\end{equation}
in which 
$\Delta \Sshi_{+}(\qv) \equiv \Sshi_{+}(\kv) - \Sshi_{+}^{(0)}(\kv)$,
and in which
\begin{eqnarray}
   \delta f^{(0)} & = & 
   \frac{1}{2} \int\limits_{\qv}
   \ln \left ( \Sshi_{+}^{(0)}\barBZero v \right )
   \nonumber  \\
   & = & \frac{1}{12\pi^{2}}\left [ 
   \ln \left ( 12 \barBZero \frac{\bar{l}}{v\Lambda^{2}} \right )
   +\frac{2}{3} \right ]\Lambda^{3}
   \label{df0Blend}
\end{eqnarray}
is the most strongly UV divergent contribution.

Upon expanding the integrand of the remaining integral in Eq. 
(\ref{dfLnFactor}) in powers of $1/q$, we find that the leading 
order terms are of $1/q^{2}$, and yield UV divergent contributions of 
${\cal O}(\Lambda)$ to the integral, but that all subsequent terms in
the expansion are UV convergent. The total UV divergent contribution 
to $\delta f$ is given by a sum
\begin{equation}
    \delta f \simeq 
    \delta f^{(0)} + \delta f^{(\chi)} + \delta f^{(1)}
\end{equation}
where
\begin{eqnarray}
    \delta f^{(\chi)}
    & =  & -\barchiZero\int\limits_{\qv}
    \Gh^{(0)}|\Sshi^{(0)}| 
    \\
    \delta f^{(1)}
    & = & \frac{1}{2} \int\limits_{\qv}
    \Gh^{(0)}\Sshi^{(1)}_{+}
\end{eqnarray}
where 
$|\Sshi^{(0)}(\qv)| = \cmon_{1}\cmon_{2}D^{(0)}_{1}(\qv)D^{(0)}_{2}(\qv)$
and 
$\Sshi^{(1)}_{+} = \sum_{i}\cmon_{i}\Dh^{(1)}_{i}(\qv)$. 
Completing the integrals yields
\begin{eqnarray}
    \delta f^{(\chi)}
    & = &
    -\frac{6\barchiZero}{\pi^{2}v}
    \frac{l_{1}l_{2}}{\bar{l}}\phi_{1}\phi_{2}\Lambda
    \label{dfChiBlend}
    \\
    \delta f^{(1)}
    & = & -\frac{3}{2\pi^{2}v\bar{l}}
    \left ( \frac{\phi_{1}l_{1}^{2}}{N_{1}} + 
     \frac{\phi_{2}l_{2}^{2}}{N_{2}} \right ) \Lambda
    \label{df1Blend}
\end{eqnarray}
The quantity $\delta f^{(0)}$, which is the one-loop contribution to 
the free energy density of a nearly incompressible blend of infinite 
chains with $\chi = 0$, arises solely from the strong repulsive 
interactions that suppress overall concentration fluctuations. This
quantity generally depends on composition, because of the composition
dependence of $\bar{l}$, and thus can contributes to $\chi_{a}(0)$.
It is independent of composition only in the special case $l_{1} = 
l_{2} = \bar{l}$. The contribution $\delta f^{(\chi)}$ is a negative 
contribution that reflects a reduction in the average interaction 
energy from its mean-field value by correlations.  We show in Sec. 
\ref{sec:Defects} that $\delta f^{(1)}$ arises from changes in 
packing near the chain ends. 

\subsection{Direct Correlation Function}
Once the UV divergent contribution $\delta f$ is known, the 
corresponding contribution $\delta \Ch_{ij}(\kv=0)$ to the direct 
correlation function may be obtained from the relation
\begin{equation} 
   \delta \Ch_{ij}(0) = \frac{\partial^{2} ( \delta f )}{\partial \cmon_{i} \partial \cmon_{j}}
\end{equation} 
Differentiation of Eqs. (\ref{df0Blend}), (\ref{dfChiBlend}), and 
(\ref{df1Blend}) yields a UV divergent contribution 
\begin{eqnarray}
   \delta \Ch_{ij}(0) & \simeq & \delta \Ch_{ij}^{(0)}    
                     + \delta \Ch_{ij}^{(\chi)} + \delta \Ch_{ij}^{(1)}   
   \nonumber \\
   \delta \Ch_{ij}^{(0)}(0) & = &
   \frac{ v^2 l_{i} l_{j} }{12\pi^2\bar{l}^2} \Lambda^{3}
   \\
   \delta \Ch_{ij}^{(\chi)}(0) & = & 
   \frac{12 v}{ \pi^2 } \frac{l_i l_j }{ \bar{l}^3} 
   Z_{ij} \barchiZero \Lambda 
   \\
   \delta \Ch_{ij}^{(1)}(0) & = & 
   \frac{6v}{2\pi^{2}}\frac{l_i l_j}{\bar{l}^{3}}
   \left (  \frac{\phi_{1} l_{1}^{2}}{N_{1}}  +
   \frac{\phi_{2} l_{2}^{2}}{N_{2}} \right ) \Lambda
   \nonumber \\
   & - & \frac{6v}{4\pi^{2}}\frac{l_i l_j}{\bar{l}^{2}}
   \left (  \frac{l_{i}}{N_{i}} +
   \frac{l_{j}}{N_{j}} \right ) \Lambda
\end{eqnarray}
By applying Eq. (\ref{chiadef}), we obtain a corresponding UV divergent 
contribution to $\barchia(0)$,
\begin{eqnarray}
   \delta \barchia (0) & \simeq &
   \frac{ (l_1- l_2)^2  v}{24\pi^2\bar{l}^2} \Lambda^{3}
   -\frac{6 l_1^2  l_2^2 }{ \pi^2 \bar{l}^3}\barchiZero\Lambda
   \nonumber \\
   & - &
   \frac{3}{ 2\pi^2 } 
   \frac{ l_1 l_2 (l_1-l_2) }{ \bar{l}^3 }   
   \left ( \frac{l_{1}}{N_{1}} - \frac{l_{2}}{N_{2}} \right )
   \Lambda
   \quad,
\end{eqnarray} 
of the form given in Eq. (\ref{binaryChik=0Diverge}).
Note that the coefficient $A$ of the $\Lambda^{3}$ contribution
and coefficients $H_{1}$ and $H_{2}$ of the terms linear in 
$\Lambda/N_{i}$ all vanish in the case $l_{1}=l_{2}$ of two 
polymers with equal statistical segment lengths.

An alternative method of deriving $\delta \Ch_{ij}(0)$, which is
useful for comparison to the subsequent calculation of $\delta\Ch_{ij}
(\qv)$ at $\qv \neq 0$, is to apply Eq. (\ref{Chk=0}) to Eq. 
(\ref{FGauss}) before evaluating the Fourier integral. 
A straightforward differentiation yields an integral 
\begin{eqnarray} 
   \delta \Ch_{ij}(0) & = & -\frac{1}{2}
   \frac{\partial^{2}}{\partial \cmon_{i} \partial \cmon_{j}}
   \int_{\qv}
   \ln \det [ \Ghib^{-1}(\qv)\Uhb(\qv) ] \nonumber \\
   & = & 
   \frac{1}{2} \int_{\qv}
   \Dh_{i}(\qv)
   \Ghi_{ij}(\qv) \Ghi_{ij}(\qv)
   \Dh_{j}(\qv)
   \eqsp. \label{Chk=0Gauss}
\end{eqnarray}
We show in what follows that this expression can be recovered
by taking the $\kv \rightarrow 0$ limit of our expression for 
$\delta \Ch_{ij}(\kv)$. 

\section{Gaussian Approximation for $S(k)$}
\label{sec:GaussianSq}
In this section, we derive an Gaussian approximation for the 
inverse correlation function $S^{-1}_{ij}(k)$ at $k \neq 0$.
The derivation is based on a calculation of the second functional 
derivatives of $\ln \Zp_{G}[h]$ with respect to an external field 
$h_{i}$. 

For this purpose, we now adopt a compact notation for position
and monomer type indices in which $S(1,2)$ is used as shorthand
for a function $S_{i_1 i_2}(\rv_{1}, \rv_{2})$. In this notation,
an integer label $I$ is used as shorthand for a position $r_{I}$ 
and a monomer type index $i_I$. For example, identity 
(\ref{Sccd2lnZ}) becomes 
\begin{equation}
  S(1,2) = 
  \frac{\delta^{2} \ln \Zp}{\delta \hr(1) \delta \hr(2)} 
  \quad,
\end{equation}
where $\delta /\delta h(2)$ denotes a functional derivative 
with respect to $\hr_{i_2}(\rv_{2})$. 

The one-loop contribution to $S(1,2)$ is given by a derivative
\begin{equation}
  \delta S(1,2) = \left .
  \frac{\delta^{2} \ln \Zp_{G}}{\delta \hr(1) \delta \hr(2)} 
  \right |_{\mu}
\end{equation}
in which $\Zp_{G}$ is the Gaussian contribution to the partition
function of a system in an applied field $\hr$.  
The Gaussian contribution to the free energy of a system that
is subjected to an inhomogeneous field $\hir$ is a functional
\begin{equation}
   \ln \Zp_{G} = -\frac{1}{2}{\rm Tr} \ln \Ghi^{-1} \Uh
   \label{lnZGinhomo}
\end{equation}
Here, $\Ghi^{-1}(1,2)$ is a integral operator, which is generally 
not translationally invariant in a fluid that is subjected to an 
inhomogeneous external field $\hr$. The symbols $\ln$ and ${\rm Tr}$ 
denote the generalized logarithm and trace for such an operator. 
The function 
\begin{equation}
   \Ghi^{-1}(1,2;[\hir]) \equiv 
   \Sshi^{(2)}(1,2;[\hir]) + \Uh^{-1}(1,2)
\end{equation}
is a functional of the field $\hir$, which is the saddle point of 
$\hr + i\Jr$ obtained in the presence of an external field $\hr$.  
The Gaussian integral $\Zp_{G}$ is thus also a functional of $\hir$.

A straightforward differentiation of Eq. (\ref{lnZGinhomo}) twice 
with respect to the external field $h$ yields
\begin{eqnarray}
  \delta S(1,2) 
  & = & 
  \int_{1'}\int_{2'}
  \frac{\delta^{2} \ln \Zp_{G}[\hir]}{\delta \hir(1') \delta \hir(2')} 
  \frac{\delta \hir(1')}{\delta h(1)}
  \frac{\delta \hir(2')}{\delta h(2)} 
  \nonumber \\
  & + &
  \int_{1'}
  \frac{\delta \ln \Zp_{G}[\hir]}{\delta \hir(1')} 
  \frac{\delta \hir(1')}{\delta h(1)\delta h(2)}
  \eqsp. \label{d2lnZGdh1dh2}
\end{eqnarray}
Here, we have introduced the shorthand
\begin{equation}
   \int_{I} \equiv \sum_{i_{I}} \int d\rv_{I}
\end{equation}
for integration over a position $\rv_{I}$ and summation over 
allowed values of a corresponding monomer type index $i_{I}$.
To evaluate Eq. (\ref{d2lnZGdh1dh2}), we must evaluate both the 
first and second derivatives of $\Zp_{G}[\hir]$ with respect to 
the saddle point field $\hir$, and the first and second
derivatives of $\hir$ with respect to the external field $\hr$.

To evaluate the required derivative of $\Zp_{G}[\hir]$, we apply the 
identities\cite{Morse2006}
\begin{eqnarray}
    \frac{\delta {\rm Tr} \ln \Ghi^{-1}}{\delta \hir(1)}
    & = & \int_{1} \Ghi(2,3)
    \frac{\delta \Ghi^{-1}(2,3)}{\delta \hir(1)} 
    \nonumber \\
    \frac{\delta \Sshi^{(n)}(1,\ldots,n)}{\delta \hir(n+1)}
    & = & \Sshi^{(n+1)}(1,\ldots,n,n+1) 
    \label{dSshidhir} \\
    \frac{\delta \Ghi(1,2)}{\delta \hir(3)} & = & 
    - \int_{1'} \int_{2'}\Ghi(1,1')\Sshi^{(3)}(1',2',3)\Ghi(2',2)
    \nonumber \label{dGhidhir} 
\end{eqnarray}
to obtain
\begin{equation}
  \frac{\delta \ln \Zp_{G} }{\delta \hir(1)} 
   =  - \frac{1}{2} \int_{2}\int_{3}
  \Sshi^{(3)}(1,2,3) \Ghi(2,3)
  \label{dlnZGdhir1}
\end{equation}
and
\begin{equation}
  \frac{\delta^{2} \ln \Zp_{G}}{\delta \hir(1) \delta \hir(2)} 
   = 
  - \frac{1}{2} \int_{3}\int_{4}
  \Sshi^{(4)}(1,2,3,4) 
  \Ghi(3,4) \quad
  \label{d2lnZGdhir1dhir2}
\end{equation}
\vspace{-8pt}
$$
  \quad\quad
  + \; \frac{1}{2} \int_{3}\int_{4}\int_{5}\int_{6}
  \Sshi^{(3)}(1,3,4)
  \Ghi(3,5)
  \Sshi^{(3)}(2,5,6)
  \Ghi(6,4)\quad
$$

To evaluate the derivatives of $\hir$ with respect to the
external field $\hr$, we differentiate the saddle saddle-point 
condition \begin{equation}
  \hir(1) = \hr(1) - \int_{2} \Ur(1,2)\Sshi^{(1)}(2) 
  \quad,
\end{equation}
with respect to $\hr$. 
Note that $\Sshi^{(1)}(2)$ is a functional of $\hir(1)$, with
derivatives given by Eq. (\ref{dSshidhir}).  A single functional 
derivative yields
\begin{equation}
   \int_{2}
   \frac{\delta \hir(2)}{\delta \hr(1)}
   \left[ \delta(2,4) + \int_{3}\Sshi^{(2)}(2,3)\Ur(3,4)\right] 
   = \delta(1,4)
   \label{del_saddlepoint}
\end{equation}
Let
\begin{equation}
  \Scci^{-1}(1,2) \equiv \Sshi^{-1}(1,2) + \Uh(1,2)
  \label{SccinvRPA}
\end{equation}
denote the inverse correlation function obtained in the
mean-field approximation, in which $C(1,2) = - U(1,2)$.
Using this definition for $\Scci^{-1}$ and Eq. (\ref{Gidef}) 
for $\Ghi^{-1}$, we may rewrite Eq. (\ref{del_saddlepoint})
in either of the equivalent forms 
\begin{eqnarray}
   \frac{\delta \hir(2)}{\delta \hr(1)} 
    & = & \int_{3} \Scci(1,3)\Sshi^{-1}(3,2) 
    \label{dhirdh1} 
    \\
    & = & \int_{3} U^{-1}(1,3)\Ghi(3,2)
    \quad. \label{dhirdh2}
\end{eqnarray}
A second functional derivative yields
\begin{eqnarray}
   \frac{\delta^{2}\hir(3)}{\delta \hr(1) \delta \hr(2)} 
   & = & - 
   \int_{1'}\int_{2'}\int_{3'}\int_{4'}
   \frac{\delta \hir(1')}{\delta \hr(1)} 
   \frac{\delta \hir(2')}{\delta \hr(2)} 
   \nonumber \\
   & & \times
   \Sshi^{(3)}(1',2',4')\Ur(4',3')
   \frac{\delta \hir(3)}{\delta \hr(3')} 
   \label{d2hirdh1dh2}
\end{eqnarray}

An explicit expression for $\delta \Scc(1,2)$ may then be 
obtained by combining Eqs. (\ref{d2lnZGdh1dh2}), (\ref{dlnZGdhir1}),
(\ref{dhirdh1}) or (\ref{dhirdh2}), and 
(\ref{d2hirdh1dh2}).  We are more interested, however, in obtaining 
a one-loop contribution to the inverse correlation function 
$\Scc^{-1}(1,2)$ that appears in the OZ relation. To the order 
required here, this is related to $\delta \Scc(1,2)$ by 
\begin{equation}
  \delta \Scc^{-1}(1,4) = -\int_{2} \int_{3}
  \Scci(1,2) \delta \Scc(2,3) \Scci(3,4)
  \label{deltaSinv}
\end{equation}
Combining Eq. (\ref{deltaSinv}) with our expression for 
$\delta\Scc(1,2)$ yields a one-loop contribution
\begin{equation}
  \delta \Scc^{-1}(1,2) =  
  -\int_{1'} \int_{2'}
  \Sshi^{-1}(1,1') 
  \delta \Lambda(1',2')
  \Sshi^{-1}(2',2)
\end{equation}
in which
\begin{eqnarray}
  \delta \Lambda(1,2)
  & \equiv & 
  \frac{\delta^{2} \ln \Zp_{G}}{\delta \hir(1) \delta \hir(2)} 
  \label{dLmGaussian}
  \\
  & - & \int_{3} \int_{4} 
  \Sshi^{(3)}(1,2,3)\Ghi(3,4) 
  \frac{\delta \ln \Zp_{G}}{\delta \hir(4)} 
  \quad.  \nonumber 
\end{eqnarray}
The required functional derivatives of $\ln \Zp_{G}$ are given in 
Eqs. (\ref{dlnZGdhir1}) and (\ref{d2lnZGdhir1dhir2}).

\section{Diagrammatic Approach}
\label{sec:Diagrams}
\begin{figure}
\centering
\includegraphics[width = 3.0in, height=!]{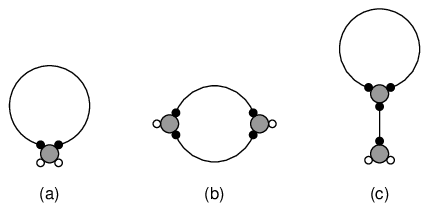}
\caption{
The three one-loop diagrams of $\Sshi$ vertices and $-\Ghi$ 
bonds that contribute to the expression for $\delta \Lm(1,2)$ 
given in Eq. (\ref{dLmGaussian}).  Diagrams (a) and (b) 
represent cluster integrals arising from the first and second 
terms of the r.h.s. of Eq. (\ref{d2lnZGdhir1dhir2}) for 
$\delta^{2}\ln \Zp /\delta \hir(1)\delta \hir(2)$, respectively, 
in the first line of Eq. (\ref{dLmGaussian}). Diagram (c) 
represents the second line of Eq. (\ref{dLmGaussian}), in
which $\delta \ln \Zp /\delta \hir(4)$ is given by Eq. 
(\ref{dlnZGdhir1}).
}
\label{fig:Lm1loop}
\end{figure}

It is convenient to introduce a diagrammatic representation of 
cluster integrals such as those obtained in Eqs. (\ref{dlnZGdhir1}), 
(\ref{d2lnZGdhir1dhir2}), and (\ref{dLmGaussian}). We adopt the 
diagrammar used previously in Ref. [13].
Some examples of the type of diagrams used here are shown in Fig. 
\ref{fig:Lm1loop}.  
Each integral is represented as a diagram containing vertices and bonds. 
An $n$-point vertex, shown as a shaded circle with $n$ smaller circles 
around its circumference, represents a function of $n$ coordinates and 
type indices, such as $\Sshi^{(n)}(1,\ldots,n)$. Each of the small 
circles around the perimeter of a vertex is either a field circle (shown 
blackened), which represents an argument of the corresponding function 
that must be integrated over, or a root circle (white), which represents 
a fixed parameter, rather than a integration variable.  Each bond 
represents a function of two coordinates and type indices, such as the 
bare interaction $\Ur(1,2)$ or the screened interaction $\Ghi(1,2)$. 
Each bond must be connected at each end to either a vertex field 
site, or a free root site. (A free root site is small white circle 
that is not associated with a vertex, which is used simply to 
indicate the arguments associated with the free end of a bond are 
known parameters).  The value of a diagram is the value of the integral 
obtained by integrating over the coordinates associated with all of the 
black circles, divided by a combinatorical prefactor that is given by 
the order of the group of permutational symmetries of the diagram.
The diagrams discussed here are all diagrams of $\Sshi$-vertices and 
$-\Ghi$ bonds, in which a factor of $\Sshi^{(n)}$ is associated with 
each $n$-point vertex and a factor of $-\Ghi$ is associated with each 
bond. In diagrams with $-\Ghi$ bonds, vertices with no white circles 
(i.e., with no root circles) must each have three or more black 
(field) circles, representing factors of $\Sshi^{(n)}$ with $n \geq 3$. 

Fig. \ref{fig:Lm1loop} shows a diagrammatic representation of Eq. 
(\ref{dLmGaussian}), for $\delta \Lambda(1,2)$, in which we have 
used Eqs. (\ref{dlnZGdhir1}) and (\ref{d2lnZGdhir1dhir2}) for the 
required functional derivatives of $\ln \Zp_{G}$.  The correspondence
between diagrams and integral expressions is discussed in the figure 
caption. 

It was shown in Ref. [13]
how $\ln \Zp$ and various 
correlation functions could be expressed to any order in perturbation 
theory as sums of well-defined infinite sets of such cluster diagrams.  
It was found that the two-point correlation function $S(1,2)$ can be 
expressed as an infinite sum
\begin{equation}
 \Scc(1,2) = 
  \left \{ \begin{tabular}{l}
  Sum of all connected diagrams \\
  of $\Sshi$ vertices and $-\Ghi$ bonds\\
  with $2$ roots circles labelled \\
  $1$ and $2$
\end{tabular} \right \} 
\label{SccSsU}
\end{equation}
The intramolecular correlation function $\Ssh(1,2)$ in an 
interacting fluid is given by the subset of these diagrams in which 
both of the root sites are one the same vertex (the root vertex):
\begin{equation}
 \Ssh(1,2) = 
  \left \{ \begin{tabular}{l}
  Sum of connected diagrams \\
  of $\Sshi$ vertices and $-\Ghi$ bonds\\
  with $2$ roots circles labelled \\
  $1$ and $2$ on the same vertex 
  \end{tabular} \right \}
\label{SsaSsSmG}
\end{equation}
Among the diagrams described in Eq. (\ref{SsaSsSmG}) is one 
consisting of a single two-point $\Sshi$ vertex, with no bonds,
which represents the ideal gas contribution $\Sshi(1,2)$. 
The corresponding expansion of the intramolecular correlation 
function $\Ssh_{a}(1,2)$ for molecule of a specific type in a 
mixture may be obtained by replacing the $\Sshi$ root vertex 
in each diagram of Eq. (\ref{SsaSsSmG}) (i.e., the vertex
with two white circles) by a corresponding $\Sshi_{a}$ vertex, 
representing a factor of the intramolecular correlation 
function $\Sshi_{a}$ for the specified molecule type, while 
using $\Sshi$ vertices for all other vertices in the diagram.  

\subsection{Diagrammatic Resummations}
The screened interaction $-\Ghi$ used in the cluster expansions 
described above can be expressed algebraically as an infinite 
geometrical series
\begin{eqnarray}
   -\Ghi & = & -[ 1 + \Uh*\Sshi]^{-1}*\Uh 
        \nonumber \\
        & = &  - \Uh + \Uh*\Sshi*\Uh - \cdots
\end{eqnarray}
where $\Sshi$ denotes $\Sshi^{(2)}$.
This can also be expressed diagrammatically as the sum of an 
infinite series of all possible chain diagrams of alternating 
$-\Uh$ bonds and $\Sshi^{(2)}$ vertices. 
By substituting this diagrammatic expansion of $-\Ghi$ into 
the above expansions of $\Scc$ and $\Ssh$, we may obtain a 
formally equivalent expansions of these quantities in terms of 
diagrams of $\Sshi$ vertices and $-\Uh$ bonds.  That is, the 
perturbation theory may be expressed in terms of either the 
screened interaction $\Ghi$ or the underlying bare interaction 
$\Uh$. The descriptions of the infinite sums of diagrams of 
$\Sshi$ vertices and $-\Uh$ bonds required to construct $\Scc$ 
and $\Ssh$ are identical to those given in Eqs. (\ref{SccSsU}) 
and (\ref{SsaSsSmG}) for diagrams of $\Sshi$ vertices and 
$-\Ghi$ bonds, except for a replacement of $-\Ghi$ bonds by 
$-\Uh$ bonds, and a change in the rule for the nature of the 
allowed $\Ssh$ vertices: Two-point $\Sshi^{(2)}$ vertices with 
two field circles and no root circles are allowed in diagrams 
of $-\Uh$ bonds, but are prohibited in diagrams of $-\Ghi$ 
bonds. 

A Dyson equation for $S(1,2)$ may be obtained by defining a function
\begin{equation}
  \Lm(1,2) = 
  \left \{
\begin{tabular}{l}
  Sum of bond irreducible diagrams \\
  of $\Sshi$ vertices and $-\Ghi$ bonds with\\
  $2$ roots circles labelled $1$ and $2$. 
\end{tabular} \right \} 
\label{LSsR}
\end{equation}
A ``bond irreducible" diagram is one that cannot be divided into 
two disconnected pieces that each contain one of the two root 
circles by cutting or removing only one bond. Eq. (\ref{LSsR}) 
can also be expressed as a sum of all bond-irreducible diagrams 
of $\Sshi$ diagrams and $-\Uh$ bonds, if $\Sshi^{(2)}$ vertices 
with two field sites and no root sites are allowed. This set 
of diagrams includes the trivial diagram consisting only of a 
$\Sshi^{(2)}$ vertex with two root sites and no bonds, which 
represents the function $\Sshi^{(2)}(1,2)$. Thus, to a first 
approximation (or ``tree level"), $\Lm(1,2) \simeq \Sshi(1,2)$.  
The only one-loop diagrams that contribute to $\Lm(1,2)$ are the 
three diagrams shown in Fig. \ref{fig:Lm1loop}. The expression 
for $\delta \Lm(1,2)$ given explicitly in Eq. (\ref{dLmGaussian})
is thus the one-loop contribution to the quantity $\Lm(1,2)$
defined in Eq. (\ref{LSsR}).

The function $S$ can be expressed in terms of $\Lm$ as a geometric 
series
\begin{eqnarray}
    \Scc & = & \Lm -\Lm*\Uh*\Lm + \cdots 
     \nonumber \\
         & = & \Lm -\Lm*\Uh*\Scc
     \eqsp,
\end{eqnarray}
or diagrammatically as a sum of chain diagrams of $\Lm$ 
vertices connected by $-\Uh$ bonds. Resumming this series, 
or solving the recursion relation, yields 
\begin{equation}
   \Scc^{-1}(1,2) = \Lm^{-1} (1,2) + \Uh(1,2)
   \label{SL}
\end{equation}
Approximating $\Lm(1,2)$ by $\Sshi^{(2)}(1,2)$ yields the mean
field approximation for $\Scc$.  Note that Eq. (\ref{SL}) for 
$\Scc^{-1}(1,2)$ is {\it not} the same as the generalized OZ 
equation.  

\subsection{One-Loop Contributions}
To calculate one-loop corrections to $\Scc^{-1}$, it is useful to
define
\begin{equation}
   \Lm(1,2) = \Sshi(1,2) + \delta \Lm(1,2)
   \eqsp, \label{LmSshdLm}
\end{equation}
where $\delta \Lm(1,2)$ is a sum of all contributions to $\Lm(1,2)$ 
other than the tree-level contribution $\Sshi(1,2)$. To first order
in a loop expansion $\delta \Lm(1,2)$ is given by the sum of the three 
one-loop diagrams shown in Fig. \ref{fig:Lm1loop}. Substituting 
Eq. (\ref{LmSshdLm}) into Eq. (\ref{SL}) yields a geometrical series 
in which the first few terms are
\begin{eqnarray}
   \Scc^{-1} 
          &  = & U + \Sshi^{-1}
               - \Sshi^{-1}*\delta \Lm*\Sshi^{-1} + \cdots
   \label{SccInvExpand}
\end{eqnarray}
The one-loop contribution to $\Scc^{-1}(1,2)$ is thus given by
\begin{equation}
   \delta\Scc^{-1} =
               - \Sshi^{-1}*\delta \Lm*\Sshi^{-1}
   \label{dSccInvdLm}
\end{equation}
with $\delta \Lm$ approximated by the sum of the three one-loop 
diagrams for $\Lm$. 

The one-loop contribution to the OZ expression for $\Scc^{-1}$ may 
also be expressed as a sum
\begin{equation}
    \delta \Scc^{-1} =
    \delta \Ssh^{-1} - \delta \Ch
\end{equation}
where $\delta\Ssh^{-1}$ and $\delta \Ch$ represent one loop
contributions to $\Ssh^{-1}$ and $\Ch$, respectively. The one-loop 
correction to $\Ssh^{-1}$ is given by the convolution
\begin{equation}
    \delta \Ssh^{-1} \simeq 
     - \Sshi^{-1}*\delta\Ssh*\Sshi^{-1}
\end{equation}
where $\delta \Ssh$ represents the one-loop correction to $\Ssh$.
According to the diagrammatic rule given in Eq. (\ref{SsaSsSmG}), 
the one-loop contribution to $\Ssh(1,2)$ is given by the sum of 
diagrams (a) and (c) in Fig. \ref{fig:Lm1loop}.  It follows that 
we may identify $-\delta \Ch$ with the remaining contribution to
Eq. \ref{dSccInvdLm} that arises from the contribution of diagram 
(b) to $\delta\Lm$. That is, to first order in a loop expansion,
\begin{equation}
    \delta \Ch \simeq \Sshi^{-1}*\Sgh*\Sshi^{-1} 
    \quad,
    \label{dChofSgh}
\end{equation}
where $\Sgh$ is the value of diagram (b). This expression was
obtained previously\cite{Morse2006} as the one-loop contribution 
to a general diagrammatic expansion of the direct correlation 
function. 

\subsubsection{Intramolecular Correlations}
The quantity $\delta\Ssh_{ij}(\kv)$ may be expressed as a sum
\begin{equation}
   \delta \Ssh_{ij}(\kv) = \sum_{a} \delta \Ssh_{a,ij}(\kv) 
\end{equation}
where $\delta \Ssh_{a,ij}(\kv)$ is a one loop correction to the 
intramolecular correlation function $\Ssh_{a,ij}(\kv)$ for molecules 
of species $a$. Using the diagrammatic rules discussed above, the 
one-loop contribution $\delta \Ssh_{a,ij}(\kv)$ may be expressed 
as a Fourier integral
\begin{equation}
   \delta \Ssh_{a,ij}(\kv) = \cpoli_{a} \; [ \;
   I_{a,ij}^{(2)}(\kv) -
   \sshi_{a,ij}(\kv) \overline{G}_{ab}\cpoli_{b}
   I^{(0)}_{b}
   \; ] 
   \label{dSshk1}
\end{equation}
Here,
\begin{eqnarray}
   I_{a}^{(0)} & \equiv & -\frac{1}{2}\int_{\qv}
   \sshi_{a,ij}(\qv)\Ghi_{ij}(\qv)
   \nonumber \\
   I_{a,ij}^{(2)}(\kv) & \equiv & -\frac{1}{2}\int_{\qv}
   \sshi_{a,ijkl}^{(4)}(\kv,-\kv,\qv,-\qv)\Ghi_{kl}(\qv)
   \label{Ia02def}
\end{eqnarray}
where $\sshi_{a,ij}(\kv) \equiv \sshi^{(2)}_{a,ij}(\kv,-\kv)$, and
\begin{equation}
   \overline{G}_{ab} \equiv \sum_{kl}N_{ak}\Gh_{kl}(\kv=0)N_{bl}
   \label{Gbardef}
\end{equation}
where $N_{ak}$ is the number of monomers of type $k$ on a molecule 
of type $a$. 
Here and in what follows, we use the notation
\begin{equation}
    \int_{\qv} \equiv \int \frac{d^{3}q}{(2\pi)^{3}}
\end{equation}
for Fourier integrals. To obtain the above expression, we have used 
the identity $\sshi^{(3)}_{a,ijk}(\kv,-\kv,0) = N_{ak} \sshi^{(2)}(\kv)$. 
The first term in square brackets in Eq. (\ref{dSshk1}) corresponds 
to the diagram (a) in \ref{fig:Lm1loop}, while the second corresponds to 
diagram (c). To obtain the above expressions for $\delta\Ssh_{a,ij}(\kv)$, 
the root vertices in these diagrams must be taken to be $\Sshi_{a}$ 
vertices. A corresponding contribution for the sum $\delta\Ssh_{ij}(\kv)$
may be obtained either by adding the results for $\delta\Ssh_{a,ij}$ for 
different species in a mixture, or by taking the root vertices in these 
diagrams to be $\Sshi$ vertices (with no species index).

\subsubsection{Direct Correlation Function}
Eq. (\ref{dChofSgh}) for $\delta \Ch$ may be expressed more explicitly 
in Fourier space as a product
\begin{equation}
    \delta \Ch_{ij}(\kv) \simeq 
    \Sshi^{-1}_{ik}(\kv) \Sgh_{kl}(\kv)\,
    \Sshi^{-1}_{lj}(\kv)
    \eqsp, \label{dC1loop}
\end{equation}
in which
\begin{eqnarray}
   \Sgh_{ij}(\kv) & = &
   \frac{1}{2} \int_{\qv}
   \Sshi_{ikm}^{(3)}(\kv,\qmv,-\qplv) \Ghi_{kl}(\qplv)
   \label{Sg1loop} 
   \\
   & &  \quad \times \;
   \Sshi_{jln}^{(3)}(-\kv,-\qmv,\qplv)\Ghi_{ln}(\qmv)
   \eqsp, \nonumber
\end{eqnarray}
is the Fourier transform of diagram (b) of Fig. 1, where
$\qv_{\pm} \equiv \qv \pm \frac{\kv}{2}$.

\section{Canonical vs. Grand-Canonical Ensemble}
\label{sec:Concentration}
Because we have worked thus far in grand-canonical ensemble, the 
one loop contribution to $\Scc^{-1}$ derived above is a correction
to the mean-field result for a system with a fixed set of chemical 
potentials, rather than fixed set of molecular concentrations. In 
grand-canonical ensemble, a one-loop approximation for the free 
energy will generally yield a slightly different concentration for 
each type of molecule than that obtained at the same chemical 
potentials from the saddle-point approximation. In this section, we 
first calculate the difference between the molecular concentrations 
obtained in the mean-field and one-loop approximation at equal
chemical potentials,  and then use this to obtain a one-loop 
approximation for $\Scc^{-1}$ for a system with fixed molecular 
concentrations.

\subsection{Concentration at Fixed Chemical Potential}
\begin{figure}
\centering
\includegraphics[width = 2.5in, height=!]{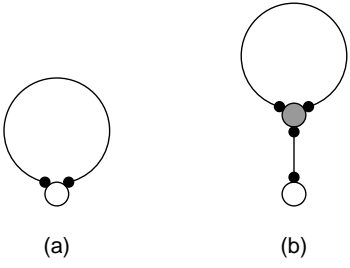}
\caption{
The two one-loop diagrams of $\Sshi$ vertices and $-\Ghi$ bonds
that contribute to the molecular number density $\cpol_{a}$. 
Diagrams (a) and (b) correspond to the first and second lines 
of Eq. (\ref{dcpol1}), respectively.  Here, whitened vertices 
(larger circles) are used to represent factors of the ideal-gas 
intramolecular correlation function $\Sshi_{a}$ for a specific 
molecular species $a$, while the gray vertex in diagram (b) 
represents a factor of $\Sshi$, as defined in 
Eq. (\ref{Ssh2sumdef}).}
\label{fig:rho1loop}
\end{figure}

In grand-canonical ensemble, molecular concentration is given by
a derivative
\begin{equation}
  \cpol_{a} = \frac{1}{V}\frac{\partial \ln \Zp}{\partial \mu_{a}}
  \quad.
\end{equation}
The concentration obtained in a Gaussian or one-loop approximation 
thus differs from that obtained in mean-field theory at equal
chemical potential by an amount 
\begin{equation}
  \delta \cpol_{a} = \frac{1}{V}
  \frac{\partial \ln \Zp_{G}}{\partial \mu_{a}} 
  \quad.
\end{equation}
Here, the derivative is evaluated at fixed temperautre and fixed 
values of the chemical potentials of species other than $a$. Because 
the saddle-point field $\hir$ will shift in response to changes 
in chemical potential, however, and $\ln \Zp_{G}$ is expressed as
a functional of the saddle-point field, this derivative is a sum
\begin{equation}
   V\delta \cpol_{a} = 
   \left .  \frac{\partial \ln \Zp_{G}}{\partial \mu_{a}} \right |_{\hir}
   + \int_{1} \left . 
   \frac{\partial \hir(1)}{\partial \mu_{a}}
   \frac{\delta \ln \Zp_{G}}{\delta \hir(1)} \right |_{\mu}
   \label{dcpol0}
\end{equation}
where $\delta \ln \Zp_{G}/\delta \hir(1)$ is given in Eq. 
(\ref{dlnZGdhir1}), and where 
\begin{equation}
   \left . \frac{\partial \ln \Zp_{G}}{\partial \mu_{a}} \right |_{\hir}
    = - \frac{1}{2}\int_{1} \int_{2}\Sshi_{a}(1,2)\Ghi(2,1)
    \label{dlnZGdmu}
\end{equation}
A straightforward functional derivative yields the identity
\begin{equation}
   \frac{\partial \hir(1)}{\partial \mu_{a}} = \int_{2}
   \Ghi(1,2)\Sshi^{(1)}_{a}(2)
\end{equation}
where $\Sshi^{(1)}_{ai}(\rv)$ is the contribution of molecules 
of type $a$ to the concentration of monomers of type $i$ in the
reference ideal gas. This is given in a homogeneous fluid by a 
constant
\begin{equation}
   \Sshi^{(1)}_{ai}(\rv) = N_{ai}\cpoli_{a}
\end{equation}
for all $\rv$, where $N_{ai}$ is the number of monomers of type 
$i$ on a molecule of type $a$. By combining Eqs. (\ref{dcpol0})
and (\ref{dlnZGdmu}), we obtain
\begin{eqnarray}
   V\delta \cpol_{a} & = &
   - \frac{1}{2} \int_{1} \int_{2} \Sshi_{a}(1,2) \Ghi(2,1)  
   \label{dcpol1}
   \\
   & & + \frac{1}{2} 
   \int_{1} \int_{2} \int_{3} \int_{4}
   \Sshi_{a}^{(1)}(1)\Ghi(1,2)\Sshi(2,3,4)\Ghi(3,4)
   \nonumber
\end{eqnarray}
This expression is shown diagrammatically in Fig. \ref{fig:rho1loop}.
Evaluating the Fourier representation of these diagrams yields
\begin{equation}
   \delta \cpol_{a} = \cpoli_{a} [ \;
   I_{a}^{(0)} 
   - \overline{G}_{ab} \cpoli_{b} I^{(0)}_{b}
   \; ]
   \label{dcpol2}
\end{equation}
where $I_{a}^{(0)}$ and $\overline{G}_{ab}$ are defined in Eqs
(\ref{Ia02def}) and (\ref{Gbardef}). 

\subsection{$S(k)$ at Fixed Concentration}
We would like to obtain an expression for the one-loop correction
to $S(k)$ in an interacting liquid with a fixed concentration 
$\cpol_{a}$ for each species of molecules. To do so, we consider
a one-loop approximation in which each chemical potential is
shifted by an amount $\delta \mu_{a}$ from the value used in the 
mean-field calculation, where $\delta\mu_{a}$ is chosen so as to 
produce a shift $-\delta \cpol_{a}$ that exactly cancels the 
the one-loop contributions to $\cpol_{a}$ at fixed chemical 
potential. To first order in a loop expansion, the only effect of 
this shift in chemical potentials is to shift the intramolecular 
correlation function by an amount
\begin{equation}
   \delta \Ssh_{a,ij}(\kv) = -\delta \cpol_{a} \sshi_{a,ij}(\kv)
   \label{dSshdmu}
\end{equation}
where $\delta \cpol_{a}$ is the one-loop contribution to $\rho_{a}$ 
at fixed chemical potential. 
There is no corresponding correction to $\Ch_{ij}(\kv)$, because 
this is given to zeroth order by a quantity $-\Uh_{ij}(\kv)$ that 
is independent of $\mu$.

Upon combining Eq. (\ref{dSshdmu}) with Eq. (\ref{dcpol2}) for 
$\delta\rho_{a}$, we find that the contribution of diagram (c) of 
Fig. \ref{fig:Lm1loop} to $\delta \Ssh_{a}$ is precisely cancelled 
by the contribution of diagram (b) of Fig. \ref{fig:rho1loop} to 
$\delta\cpol_{a}$. The contributions arising from diagram (a) of 
Fig. \ref{fig:Lm1loop} 
and diagram (a) of Fig. \ref{fig:rho1loop} yield a correction 
$\delta \Ssh_{a,ij}(\kv) = \cpol_{a}\delta \ssh_{a,ij}(\kv)$, 
where
\begin{equation}
   \delta \ssh_{a,ij}(\kv) = 
   I_{a,ij}^{(2)}(\kv) - \sshi_{a,ij}(\kv) I^{(0)}_{a}
\end{equation}
This one-loop contribution to $\ssh_{a,ij}(\kv)$ at fixed 
molecular concentration may also be expressed as an integral
\begin{equation}
   \delta \ssh_{a,ij}^{(2)}(\kv) \simeq 
    - \frac{1}{2}\sum_{ij} \int_{\qv}
   \sschi_{a,ijkl}^{(4)}(\kv,-\kv,\qv,-\qv)\Ghi_{kl}(\qv)
   \eqsp, \label{ssh1loop2} 
\end{equation}
where 
\begin{eqnarray}
   \sschi_{a,ijkl}^{(4)}(\kv,-\kv,\qv,-\qv) 
   & \equiv & 
   \sshi_{a,ijkl}^{(4)}(\kv,-\kv,\qv,-\qv) 
   \nonumber \\ &  &
   -\sshi_{a,ij}^{(2)}(\kv)\sshi_{a,kl}^{(2)}(\qv)
\end{eqnarray}
This expression for $\delta \ssh_{ij}^{(2)}(\kv)$ was obtained 
in Ref. [13]
by similar reasoning. It was obtained 
previously by Barrat and Fredrickson by calculating a one-loop
correction to $\ssh_{ij}(\kv)$ in canonical, rather than 
grand-canonical, ensemble. 

\section{Asymptotic Analysis of Single-Chain Correlations}
\label{sec:SingleChain}
To show that the one-loop approximation for $\Ssh_{ij}^{-1}(\kv)$ 
is renormalizable, we must show that the UV divergent parts of the 
one-loop contributions to $\ssh_{ij}(\kv)$ and $\Ch_{ij}(\kv)$ each 
have a specific wavenumber dependence. In both cases, the proof 
relies critically upon an analysis of the dominant asymptotic high-$q$ 
behavior of a 3- or 4-point intramolecular correlation function.
To show that dominant UV divergent part of Eq. (\ref{dC1loop}) for 
$\delta\Ch_{ij}(\kv)$ is independent of $\kv$, as required by our 
criteria for renormalizability, we must examine the asymptotic 
high-$q$ behavior of the 3-point function 
$\Sshi^{(3)}_{ijk}(\kv,\qv_{-},-\qv_{+})$ that appears in Eq
(\ref{Sg1loop}).  Similarly, to show that the UV divergent part 
of Eq. (\ref{ssh1loop2}) for $\delta \ssh_{a,ij}(\kv)$ has the 
non-trivial wavenumber dependence implied by Eq. (\ref{SshRenorm}), 
we must examine the asymptotic behavior of the four-point function
$\sschi_{ijkl}^{(4)}(\kv,-\kv,\qv,-\qv)$. In both cases, we need
the asymptotic behavior of a multi-point correlation function in 
the limit of large integration wavevector $\qv$, for arbitrary 
$\kv$. The correlation functions required here are closely 
related to, but distinct from, the 3- and 4-point vertex functions 
introduced by Leibler \cite{Leibler1980}, which are defined as 
the 3rd and 4th functional derivatives of the SCFT free energy 
$F[\langle \cmon\rangle]$.  For simplicity, we limit ourselves in 
this section to the identification of the dominant high-$q$ behavior 
for homopolymers. A more systematic asymptotic expansion for 
homopolymers and the generalization to diblock copolymers are 
given in the appendix.  

\subsection{Multi-Point Correlations}
\label{sub:MultiPoint}
To begin, we consider some general features of the $n$-point 
intramolecular correlation function $\sshi^{(n)}(\kv_1,\ldots,\kv_n)$ 
for a Gaussian homopolymer. This quantity is given by an integral
\begin{equation}
   \sshi^{(n)}(\kv_{1},\ldots,\kv_{n})  = 
   \int\limits_{0}^{N} d^{n}s 
   \left \langle e^{ i \kv_{j} \cdot \Rv (s_{j}) } \right \rangle
   \quad, \label{cint1}
\end{equation}
where $\langle\cdots\rangle$ indicates an average over conformations 
of a single Gaussian chain. Here, we have introduced the notation 
\begin{equation}
   \int\limits_{0}^{N} d^{n}s \equiv 
   \int\limits_{0}^{N} \! ds_{n} \cdots 
   \int\limits_{0}^{N} ds_{2} \int\limits_{0}^{N} \! ds_{1} \;
\end{equation}
for an integral over $n$ contour variables.  Because we consider only 
homopolymers in this section, no monomer species indices are needed, 
or used. 

To calculate $\sshi^{(n)}(\{\kv\})$, we divide the integral over 
$0 < s_{1}, s_{2},\ldots, s_{n} < N$ into $n!$ contributions arising 
from different ways of ordering the values of $s_{1}$, $s_{2}$, \ldots, 
$s_{n}$, each corresponding to a different ways of ordering labelled 
monomers along a chain. We first consider a restricted $n$ dimensional 
integral in which the variables $s_{1},\ldots,s_{n}$ are restricted to 
the ordered subpace $s_{1} < s_{2} < \cdots < s_{n}$, and then reconstruct 
the original integral by summing over all $n!$ permutations of the index 
labels $i=1,\ldots,n$.  By this method, Eq. (\ref{cint1}) may be 
rewritten as a sum 
\begin{equation}
   \sshi^{(n)}_{i}( \kv_{1},\ldots,\kv_{n} ) = 
   \sum_{\{ P \}} 
   \ssho^{(n)}_{i}(\kv_{1}^{P},\ldots,\kv_{n}^{P}) 
   \label{sshi_permutation_sum}
\end{equation}
in which the list $\kv_{1}^{P},\ldots,\kv^{P}_{n}$ is a permutation 
$P$ of the original list of wavevectors $\kv_{1},\ldots,\kv_{n}$,
and $\sum_{\{P\}}$ denotes a sum over all possible permutations of 
the $n$ wavevectors. Here, 
\begin{equation}
   \ssho^{(n)}(\kv_{1},\ldots,\kv_{n}) \equiv
   \int\limits_{0}^{N} d^{n}\underline{s} \;
   \langle e^{i \sum_{j}\kv_{j} \cdot \Rv(s_{j}) } 
   \rangle \label{cint2}
\end{equation}
is an ordered integral, in which we introduce the notation
\begin{equation}
   \int\limits_{0}^{N} d^{n}\underline{s} \equiv 
   \int\limits_{0}^{N} \! ds_{n} \cdots 
   \int\limits_{0}^{s_{3}} ds_{2} \int\limits_{0}^{s_{2}} \! ds_{1} \;
   \quad,
\end{equation}
to indicate an ordered integral over the space 
$0 < s_{1} < s_{2} < s_{3} < \cdots < s_{n} < N$.

To obtain an expression for the ordered integral $\ssho^{(n)}_{i}$, 
we first rewrite the sum of dot products that appears in the exponential 
in Eq. (\ref{cint2}) as a sum
\begin{equation}
  \sum_{j=1}^{n} \kv_{j} \cdot \Rv(s_{j}) = 
  \ev_{n}\cdot\Rv(s_{n}) -
  \sum_{j=1}^{n-1} \ev \cdot \Delta \Rv_{j+1,j}
\end{equation}
in which 
\begin{equation}
   \Delta \Rv_{ij} \equiv \Rv(s_{i}) - \Rv(s_{j})
\end{equation}
and in which
\begin{equation}
    \ev_{i} \equiv \sum_{j=1}^{i} \kv_{j} 
    \quad, \label{qsum_def}
\end{equation}
where $\kv_{1},\ldots,\kv_{n}$ denotes the ordered list of arguments 
of the ordered integral $\ssho^{(n)}_{i}(\kv_{1},\ldots,\kv_{n})$. 
Using the property 
\begin{equation}
   \langle e^{ i \ev \cdot [ \Rv (s) - \Rv (s') ] } \rangle
    = e^{ - e^{2} b^{2}  |s-s'|/6 }
\end{equation}
of a Gaussian chain, where $e^{2} \equiv |\ev|^{2}$, we then obtain 
an explicit expression
\begin{equation}
   \ssho^{(n)}_{i}(\kv_{1},\ldots,\kv_{n})  
    =  
   \int\limits_{0}^{N} d^{n}\underline{s} \;
    e^{-\sum_{j=1}^{n-1}|\ev_{j}|^{2} b^{2} s_{j+1,j}/6 } 
   \label{ssho_explicit}
\end{equation}
where 
\begin{equation}
    s_{ij} \equiv s_{i} - s_{j}
\end{equation}
is a difference in monomer contour variables, for any set of wavevectors 
$\kv_{1},\ldots,\kv_{n}$ for which $\ev_{n}=\sum_{i=1}^{n}\kv_{i}=0$, as 
required by translational invariance.

By changing variables in the above integrals to contour variables 
$\sd_{i} \equiv s_{i}/N$, for which $0 < \sd_{i} < 1$, we find that
\begin{eqnarray}
   \sshi^{(n)}(\kv_{1},\ldots,\kv_{n};N,b) & = &
   N^{n} \sshid^{(n)}(\Kv_{1},\ldots,\Kv_{n}) 
   \nonumber \\
   \ssho^{(n)}(\kv_{1},\ldots,\kv_{n};N,b) & = &
   N^{n} \sshod^{(n)}(\Kv_{1},\ldots,\Kv_{n}) 
\end{eqnarray}
where the functions
$\sshid^{(n)}$ and $\sshod^{(n)}$ depend only upon the re-scaled 
wavevectors $\Kv_{j} \equiv \kv_{j}\sqrt{N}b/6$, and 
\begin{equation}
   \sshod^{(n)}_{i}(\Kv_{1},\ldots,\Kv_{n})  
    =  
   \int\limits_{0}^{1} d^{n}\underline{\sd} \;
   e^{-\sum_{j=1}^{n-1}E_{j}^{2}\sd_{j+1,j} } 
   \quad. \label{sshod_explicit}
\end{equation}
Here, $E_{j}^{2} \equiv |\Kv_{1}+\cdots+\Kv_{j}|^{2}$, and the
integral in Eq. (\ref{sshod_explicit}) is taken over 
$0 < \sd_{1} < \sd_{2} < \ldots < \sd_{n} < 1$. The function 
$\sshid^{(n)}$ is an $n$-point generalization of the Debye 
function, which is related to $\sshod^{(n)}$ by a sum over 
permutations analogous to Eq. (\ref{sshi_permutation_sum})

\subsection{Three Point Function}
Consider the asymptotic high-$q$ behavior of the function 
$\ssh^{(3)}(\Kv,\Qv_{-},-\Qv_{+})$, which is needed in Eq. (\ref{Sg1loop}) 
to calculate $\delta\Ch_{ij}(\Kv)$.  Adding the $3!=6$ permutations of the 
monomers or wavevector labels yields 
\begin{eqnarray}
   \sshid^{(3)}(\Kv,\Qv_{-},-\Qv_{+}) & = &
   2\sshod^{(3)}(\Kv,\Qv_{-},-\Qv_{+})  
   \nonumber \\
   & + &
   2\sshod^{(3)}(\Kv,-\Qv_{+},\Qv_{-})
   \nonumber \\
   & + &
   2\sshod^{(3)}(\Qv_{-},\Kv,-\Qv_{+})
   \label{ssh3_homo_sum} 
\end{eqnarray}
where $\Qv_{\pm} \equiv \Qv \pm \Kv/2$. 
Equal contributions to this sum are made by permutations that are related 
by a reversal of the order of the wavevector arguments, so that, 
{\it e.g.}, $\sshod^{(3)}(\Kv,\Qv_{-},-\Qv_{+}) = \sshod^{(3)}(-\Qv_{+},\Qv_{-},\Kv)$. 

We are interested here in the asymptotic behavior of 
$\sshid^{(3)}(\Kv,\Qv_{-},-\Qv_{+})$ in the limit $Q^{2} \gg 1$ and $Q^{2} \gg K^{2}$, 
for otherwise arbitrary $K$.  First, consider the first two ordered integrals in 
Eq.  (\ref{ssh3_homo_sum}), in which $\Kv$ is the first of the three arguments. 
These may be related by taking $\Qv \rightarrow -\Qv$, so we need consider only 
$\sshod^{(3)}(\Kv,\Qv_{-},-\Qv_{+})$. This is given by an integral 
\begin{equation}
   \sshod^{(3)}(\Kv,\Qv_{-},-\Qv_{+}) = 
   \int\limits_{0}^{1} d^{3}\underline{\sd} \;
   e^{-Q_{+}^{2} s_{32}-K^{2} s_{21}} 
   \quad.
\end{equation}
In the limit $Q^{2} \gg 1$ of interest, this integral is dominated by 
contributions in which the separation $\sd_{32}$ is of order $1/Q^{2}$, 
yielding $\sd_{32} \ll 1$. The dominant asymptotic behavior in this 
limit may thus be obtained by replacing the integral with respect to 
$\sd_{3}$ over the domain $\sd_{2} < \sd_{3} < 1$ by an integral over 
the seminfinite domain $\sd_{2} < \sd_{3} < \infty$, while still 
requiring that $0 < \sd_1 < \sd_2 < 1$. To leading order in an 
expansion in powers of $1/Q$, this approximation yields
\begin{equation}
   \sshod^{(3)}(\Kv,\Qv_{-},-\Qv_{+}) \simeq 
   \frac{1}{2Q^{2}} \gf(K^{2})
   + {\cal O} \left ( Q^{-4} \right )
   \quad, \label{sshod3_leading}
\end{equation}
where
\begin{equation}
   \gf(K^{2}) = 
   2 \int\limits_{0}^{1}ds_{2}\int\limits_{0}^{s_{2}}ds_{1}\;
   e^{-K^{2}s_{21}} 
   \label{DebyeInt}
\end{equation}
is the Debye function.  Because the result is invariant under
$\Qv \rightarrow -\Qv$, this approximation yields
$\sshod^{(3)}(\Kv,-\Qv_{+},\Qv_{-}) \simeq \sshod^{(3)}(\Kv,\Qv_{-},-\Qv_{+})$.
Next, we consider the remaining ordered integral,
\begin{equation}
   \sshod^{(3)}(\Qv_{-}, \Kv, -\Qv_{+}) = 
   \int\limits_{0}^{1} d^{3}\underline{\sd} \;
   e^{-Q_{+}^{2} s_{32}-Q_{-}^{2} s_{21}} 
   \quad.
\end{equation}
By reasoning similar to that discussed above, we may approximate
this integral in the limit $Q^{2}_{\pm} \gg 1$ by an integral over 
a domain $-\infty < \sd_{1} < \sd_{2}$, $\sd_{2} < \sd_{3} < \infty$, 
and $0 < \sd_{2} < 1$. This yields a leading order approximation
\begin{equation}
   \sshod^{(3)}(\Kv, \Qv_{-},-\Qv_{+}) 
   \simeq \frac{1}{Q^{4}} 
   + {\cal O}(Q^{-6}) \quad,
\end{equation}
which does not contribute to the leading order ${\cal O}(Q^{-2})$ 
term in $\sshid^{(3)}$. The leading order contribution to 
$\sshid^{(3)}$ thus arises from the four permutations in which 
$\Kv$ is either the first or last argument of $\sshod^{(3)}$, and 
is equal to $4$ times the r.h.s. of Eq. (\ref{sshod3_leading}).

A more systematic asymptotic expansion, which is outlined in the 
appendix, yields 
\begin{equation}
   \sshid^{(3)}(\Kv,\Qv_{-},-\Qv_{+})  = 
   \sshid^{(3,0)} + \sshid^{(3,1)} + \cdots
   \label{sshid3_expand}
\end{equation}
with 
\begin{eqnarray}
   \sshid^{(3,0)} & = &
   \frac{2}{Q^{2}} \gf(K^{2})
   \label{ssh3_0_homo}
   \\
   \sshid^{(3,1)} & = &
   \frac{1}{2 Q^{4} } 
   \left [ K^2 (3+4\alpha^2) g(K^{2}) - 4 \right ]
   \label{ssh3_1_homo}
   \quad. 
\end{eqnarray}
where $\alpha \equiv \Qv\cdot\Kv/QK$. Further terms in the expansion 
are not needed to analyze the UV divergence of $\delta\Ch_{ij}(\kv)$. 
The observation that the leading order contribution $\sshid^{(3,0)}$ 
has the same $K$-dependence as the two-point function $\sshid^{(2)} = g(K^{2})$ 
plays an essential role in our analysis of $\delta \Ch_{ij}(\kv)$. 

\subsection{Four Point Function}
The function $\psi^{(4)} \equiv \psi^{(4)}(-\kv,\kv,-\qv,\qv)$ may be 
expressed as an integral
\begin{equation}
   \psi^{(4)}  = 
   \int\limits_{0}^{N}d^{n}s
   \left ( \langle 
   e^{i\kv \cdot \Delta \Rv_{21}} \,
   e^{i\qv \cdot \Delta \Rv_{43}} \rangle 
   - 
   \langle e^{i\kv\cdot \Delta \Rv_{21}} \rangle
   \langle e^{i\qv\cdot \Delta \Rv_{43}} \rangle
   \right ) 
   \nonumber
   \label{psi4def}
\end{equation}
in which the integral ranges over $0 < s_{i} < N$ for all $i=1,\ldots,4$.  
Like the integral for $\sshi^{(4)}$, this integral may be divided into 
contributions arising from each of the $4!=24$ different possible 
permutations of the contour variables $\sd_{1},\ldots,\sd_{4}$.  
We thus define functions $\psio$ analogous to the ordered integrals 
$\ssho^{(4)}$, in which arguments $\pm\kv$ and $\pm\qv$ are listed in the 
same order as order of the values of the associated contour variables.
For example,
\begin{eqnarray}
   & & \psio^{(4)}(-\kv,-\qv,\qv,\kv)  
   \nonumber \\
   & & \equiv  
   \int\limits_{0}^{N}d^{n}\underline{s}
   \left ( \langle 
   e^{i\kv \cdot \Delta \Rv_{41}} \,
   e^{i\qv \cdot \Delta \Rv_{32}} \rangle 
   - 
   \langle e^{i\kv\cdot \Delta \Rv_{41}} \rangle
   \langle e^{i\qv\cdot \Delta \Rv_{32}} \rangle
   \right )
   \nonumber
   \label{psio4def}
\end{eqnarray}
where the integration is over a subspace 
$0 < s_{1} < s_{2} < s_{3} < s_{4} < N$.  
We also define functions
\begin{eqnarray}
   \psid^{(4)}(\Kv,-\Kv,\Qv,-\Qv) & \equiv &
   N^{-4} \psi^{(4)}(\kv,-\kv,\qv,-\qv)
   \nonumber \\
   \psiod^{(4)}(\Kv_{1},\Kv_{2},\Kv_{3},\Kv_{4}) & \equiv &
   N^{-4} \psio^{(4)}(\kv_{1},\kv_{2},\kv_{3},\kv_{4})
   \quad
\end{eqnarray}
analogous to $\sshid$ and $\sshod$, which depend only upon the 
dimensionless wavevectors $\Kv \equiv \kv \sqrt{Nb^{2}/6}$ and 
$\Qv \equiv \qv \sqrt{Nb^{2}/6}$. 

The function $\psiod^{(4)}$ is unchanged by permutations of its 
arguments that reverse the order 
({\it i.e.}, $\Kv_{1},\ldots,\Kv_{4} \rightarrow \Kv_{4},\ldots,\Kv_{1}$) 
or that reverse all of their signs
({\it i.e.}, $\Kv_{1},\ldots,\Kv_{4} \rightarrow -\Kv_{1},\ldots,-\Kv_{4}$).
Using these symmetries, we find that 
\begin{eqnarray}
  \psid^{(4)} & = & 
  4\psiod(\Kv,\Qv,-\Kv,-\Qv) + 4\psiod(\Kv,-\Qv,-\Kv,\Qv)
  \nonumber \\
  & + & 2\psiod(\Kv,\Qv,-\Qv,-\Kv) + 2\psiod(\Kv,-\Qv,\Qv,-\Kv)
  \nonumber \\
  & + & 2\psiod(\Qv,\Kv,-\Kv,-\Qv) + 2\psiod(-\Qv,\Kv,-\Kv,\Qv) 
  \quad
\end{eqnarray}
The values of $\psiod^{(4)}$ arising from the 8 permutations 
in which $\pm\Kv$ are the first two or the last two arguments, 
such as $\psio^{(4)}(\Kv,-\Kv,\Qv,-\Qv)$, vanish because of a 
cancellation of the first and second terms in the integrand of 
Eq. (\ref{psi4def}). 

In the limit $Q^{2} \gg 1$ and $Q^{2} \gg K^{2}$ of interest, the 
dominant contributions to $\psid^{(4)}$ arise from the 4 permutations 
in which $\pm\Kv$ are the 1st and last arguments of $\psid$. (The 
analysis required to show this is given in the appendix). The sum
of these permutations yields a dominant contribution of 
${\cal O}(Q^{-4})$ in an expansion in powers of $1/Q$.  For example,
the function $\psiod^{(4)}(\kv,\qv,-\qv,-\kv)$ is given by an integral
\begin{eqnarray}
   && \psiod^{(4)}(\Kv,\Qv,-\Qv,-\Kv) 
   \label{psio4kqmqmkdef} \\
   && \quad = 
   \int\limits_{0}^{1}d^{4}\underline{\sd} \left (
    e^{-K^{2}\sd_{43}-Q_{+}^{2}\sd_{32}-K^{2}\sd_{21}} 
    - e^{-K^{2}\sd_{41}-Q^{2}\sd_{32}}  \right )
    \nonumber
\end{eqnarray}
Throughout this subsection, $Q_{\pm}^{2} \equiv |\Qv \pm \Kv|^{2}$.
In the limit $Q^{2} \gg 1$ of interest, the dominant behavior of 
this integral is
\begin{eqnarray}
    \psiod^{(4)}(\Kv,\Qv,-\Qv,-\Kv) & \simeq & 
    \left [
    \frac{-2K\alpha}{Q^{3}} + \frac{4(K\alpha)^{2}}{Q^{4}} 
    \cdots \right ] \nonumber 
    \\
    & \times &
    \int\limits_{0}^{1}d\sd_{4}\int\limits_{0}^{s_{4}}d\sd_{1}\;
    \sd_{41} \; e^{-K^{2}\sd_{41}}
    \quad\quad \label{psio4asymp0}
\end{eqnarray}
where $\alpha \equiv \Kv\cdot\Qv/QK$. The integral in the second 
line may be expressed as a derivative of Eq. (\ref{DebyeInt}) for 
the Debye function, since
\begin{equation}
   \frac{dg(K^{2})}{d(K^{2})} = - 2
   \int\limits_{0}^{1}d\sd_{2}\int\limits_{0}^{\sd_{2}}d\sd_{1}\;
   \sd_{21} \; e^{-K^{2}\sd_{21}}
\end{equation}
Upon adding the 4 permutations in which $\pm\Kv$ are the first and 
last arguments of $\psiod^{(4)}$, the terms linear in $\alpha$ cancel.
This yields a leading order contribution
\begin{equation}
    \psid^{(4)}(\Kv,-\Kv,\Qv,-\Qv) \simeq  -
    \frac{8 K^{2} \alpha^{2}}{Q^{4}}
    \frac{d \gf(K^{2}) }{d(K^{2})}
    \quad.
\end{equation}
The dominant contribution to the function $\psi^{(4)} = N^{4}\psid^{(4)}$, 
may be expressed in terms of a partial derivative
\begin{equation}
   \psi^{(4,0)}(\kv,-\kv,\qv,-\qv) =
   -\frac{ 288 \alpha^{2} }{ q^{4} b^{2} }
   \frac{ \partial \sshi^{(2)}(\kv) }{ \partial (b^{2})}
   \label{psi4asymp0}
\end{equation}
of $\sshi^{(2)}$ respect to $b^{2}$. Here, we have introduced the 
notation $\psi^{(4,0)}$ to denote the leading order term in an 
expansion of $\psi^{(4)}$ in powers of $1/q$. This asymtotic 
expression is used in Sec.  \ref{sec:Blend} to relate the one-loop 
correction to $\sshi^{(2)}$ to a renormalization of the statistical 
segment length.

\section{Correlations in Binary Blends}
\label{sec:Blend}
We can now analyze the UV divergent contributions to $\Scc^{-1}_{ij}(\kv)$ 
in a binary homopolymer blend. We consider the intramolecular and direct
correlation functions separately.

\subsection{Intramolecular Correlations}
\label{sec:IntramolecularBlend}
The one-loop contribution to $\ssh ^{(2)}_{a,ij}(\kv)$ in either a blend 
or a copolymer melt is given by Eq. (\ref{ssh1loop2}).  In case of a binary 
homopolymer blend this simplifies to
\begin{equation}
     \delta \ssh ^{(2)}_i(\kv) = 
   - \frac{1}{2}\int_{\qv} 
     \psi^{(4)}_i(\kv,-\kv,\qv,-\qv)\Ghi_{ii}(\qv)
     \label{dssh_blend}
\end{equation}
where the integral is constrained to $|\qv| < \Lambda$, and where we 
have introduced the notation $\ssh^{(2)}_i \equiv \ssh^{(2)}_{ii}$ and 
$\psi^{(4)}_{i} \equiv \psi^{(4)}_{iiii}$ appropriate for a homopolymer 
mixture. 

To determine the order of the dominant ultraviolet divergence of 
$\delta\ssh ^{(2)}_i(\kv)$, we note that Eq. (\ref{Ghiasymp0}) yields 
$\Ghi(\qv) \propto q^2$ and that Eq. (\ref{psi4asymp0}) yields $\psi^{(4)} 
\propto q^{-4}$ for large $q$. We thus expect 
\begin{eqnarray}
    \int\limits_{q < \Lambda} d^{3}q \; \psi^{(4)}_{i}\Ghi_{ii} 
    \sim \Lambda \quad.
\end{eqnarray}
All subdominant terms in an expansion of the integrand in powers of 
$1/q^{2}$ yield contributions to the integrand that are smaller by
by factors of ${\cal O}(1/q^{2})$, which lead to UV convergent 
contributions to the integral.  

The ultraviolet divergent part of the one-loop approximation for 
$\delta \ssh^{(2)}_{i}$ is thus given by a cutoff integral
\begin{equation}
     \delta \ssh ^{(2)}_i(\kv) = 
   - \frac{1}{2}\int_{\qv} 
     \psi^{(4,0)}_i(\kv,-\kv,\qv,-\qv)\Ghi^{(0)}(\qv)
     \label{dssh_blend_UV1}
\end{equation}
in which $\psi^{(4,0)}$ denotes the leading order approximation for
$\psi^{(4)}$ given in Eq. (\ref{psi4asymp0}), and $\Ghi^{(0)}(\qv)$
denotes the leading order asymptotic approximations to $\Gh_{ij}(\qv)$ 
given in Eq. (\ref{Ghiasymp0}) (which is the same for all $i$ and $j$).
A straightforward integration yields
\begin{eqnarray}
   \delta\ssh_i^{(2)}(\kv) \simeq
   \frac{ \partial\sshi_{i}^{(2)}(\kv) }{ \partial(b_i^2) }\delta(b_i^2)
   \label{dssh_blend_UV2}
\end{eqnarray}
where
\begin{eqnarray}
     \delta(b_i^2) 
     & = & b_{i}^{2} \frac{2 l_{i}^{2}}{\pi^{2}\bar{l}} \;\Lambda
     \label{deltabi2}
\end{eqnarray}
is a shift in the value of $b_{i}^{2}$.

The UV convergent correction to $\ssh(\kv)$ is given by the difference
\begin{equation}
     \delta \ssh^{*}_i(\kv) = 
   - \frac{1}{2}\int_{\qv} \left [
     \psi^{(4)}_i \Ghi_{ii}
     - \psi^{(4,0)}_i \Ghi^{(0)} \right ]
     \label{dssh_star_blend}
\end{equation}
between the one-loop integral expression and the UV divergent part.
Here, $\psi^{(4)}$ and $\psi^{(4,0)}$ represent the exact and leading
asymptotic expression for $\psi^{(4)}(\kv,-\kv,\qv,-\qv)$, respectively,
while $\Ghi_{ii}$ and $\Gh^{(0)}$ represent the exact and leading 
asymptotic expressions for $\Ghi_{ii}(\qv)$.  This is a convergent
integral, which we evaluate by taking $\Lambda \rightarrow \infty$.
It is straightforward to show that both terms in this integrand may 
be expressed as $v N^{3}$ (where $N$ is $N_{1}$ or $N_{2}$) times a 
function of a dimensionless wavenumber $qR$ (where $R = \sqrt{N}b$ 
and $b$ is $b_{1}$ or $b_{2}$) and of the dimensionless parameters 
$\chi_{0} N$, $N_{1}/N_{2}$, and $b_{1}/b_{2}$. By non-dimensionalizing
the measure in the wavevector integral by length scale $R$, we may 
express $\delta \ssh^{*}$ as a convergent dimensionless integral
times a prefactor of $v N^{3}/R^{3} = N^{2}/\bar{N}^{1/2}$. The
corresponding non-dimensional form of $\delta \Ssh_{i}^{*}$ is
given in Eq. (\ref{dSsh_nondim}).

\subsection{Direct Correlation Function}
\label{sec:DirectBlend}
The one-loop contribution to the direct correlation function is given 
by Eqs. (\ref{dC1loop}) and (\ref{Sg1loop}).  In case of a binary 
homopolymer blend, for which all of the monomer type indices of the 
functions $\Sshi^{(2)}_{ij}$ or $\Sshi^{(3)}_{ijk}$ must be the same, 
Eqs.  (\ref{dC1loop}) and (\ref{Sg1loop}) may be combined and simplified 
to obtain
\begin{equation}
   \delta \Ch_{ij}(\kv) \simeq 
   \frac{1}{2} \int_{\qv}
   \Dh_{i}(\kv,\qv) \Ghi_{ij}(\qplv) 
   \Dh_{j}(-\kv,-\qv)\Ghi_{ij}(\qmv)
   \eqsp, 
   \label{dC1loop_blend} 
\end{equation}
where 
\begin{equation}
   \Dh_{i}(\kv,\qv) \equiv
   \frac{\Sshi^{(3)}_{i}(\kv,\qmv,-\qplv)}{\Sshi_{i}(\kv)}
   \quad.
\end{equation}
Here, $\Sshi_{i} \equiv \Sshi_{ii}^{(2)}$ and $\Sshi_{i}^{(3)}
\equiv \Sshi_{iii}^{(3)}$ are the intramolecular two- and 
three-point functions for homopolymers of type $i$, respectively. 

Consider the $\kv \rightarrow 0$ limit of Eq. (\ref{dC1loop_blend}).
By using the long-wavelength limits
\begin{eqnarray}
   \Sshi_{i}^{(2)}(0) & = & N_{i}\cmon_{i} \nonumber \\
   \Sshi^{(3)}_{i}(0,\qv,-\qv) & = & N_{i}\Sshi_{i}(\qv)
\end{eqnarray}
we find that
\begin{equation}
   \lim_{\kv \rightarrow 0}
   \Dh_{i}(\kv,\qv) \equiv \frac{\Sshi_{i}(\qv)}{\cmon_{i}}
   \quad. \label{Dkq_def_blend}
\end{equation}
By substituting this into Eq. (\ref{dC1loop_blend}), we may 
immediately confirm that $\kv \rightarrow 0$ limit of Eq. 
(\ref{dC1loop_blend}) for $\delta \Ch_{ij}(\kv)$ is equivalent 
to Eq. (\ref{Chk=0Gauss}) for $\delta \Ch_{ij}(\kv=0)$, which was
obtained by considering the composition dependence of the free
energy density.

To find the divergent part of Eq. (\ref{dC1loop_blend}) for 
$\Ch_{ij}(\kv)$, we will need high-$q$ asymptotic expansions for 
the screened interaction $\Ghi_{ij}(\qplv)$, and the function 
$\Dh_{i}(\kv,\qmv,-\qplv)$. To the required order in an expansion 
in powers of $1/q$, 
\begin{eqnarray}
  G_{ij}(\qv_{\pm}) &=& 
  G^{(0)} + G^{(\pm)} + G^{(1)} + G^{(\chi)}_{ij} 
  \nonumber \\
  G^{(0)} &=& 
  \frac{ v^2 }{12 \bar{l} }q^{2}
  \nonumber \\
  G^{(\pm)} &=& \pm
  \frac{v^2}{12 \bar{l}} \; \kv\cdot\qv
  \nonumber \\
  G^{(1)} &=& 
  \frac{v}{2 \bar{l}^2}
  \left ( 
     \frac{\phi_{1}l_{1}^{2}}{N_{1}} +
     \frac{\phi_{2}l_{2}^{2}}{N_{2}} 
  \right )
    + \frac{k^2  v^2}{48 \bar{l}}
  \nonumber \\
  G^{(\chi)}_{ij} &=& 
  \frac{2 v}{\bar{l}^2}Z_{ij} \barchiZero
  \label{prop_homo}
\end{eqnarray}
To obtain a corresponding asymptotic expansion of $\Dh_{i}$ in 
the high-$q$ limit, for fixed $\kv$, we note that 
\begin{equation}
\Dh_{i}(\kv,\qv) = 
\frac{N_{i}\sshid^{(3)}_{i}(\Kv,\Qv_{-},-\Qv_{+})}{g(K^{2}_{i})}
\end{equation}
and use expansion (\ref{sshid3_expand}) of $\sshid^{(3)}$.
This yields
\begin{eqnarray}
   \Dh_{i}(\kv,\qv) & \simeq &
   \Dh^{(0)}_{i} + \Dh^{(1)}_{i} + \cdots
   \nonumber \\
   \Dh^{(0)}_{i} & = & \frac{12 l_{i}}{v q^{2}}
   \\
   \Dh^{(1)}_{i} & = & \left [
   \frac{3 l_{i}}{v} k^{2} (3+4\alpha^2) 
  -\frac{72 l_{i}^{2}}{v^{2}N_{i}g(K_{i}^{2})} \right ]
   \frac{1}{q^{4}} 
   \nonumber
\end{eqnarray}
where $K_i^2=k^2 N_i b_{i}^{2}/6 = k^2  N_i v/(6 l_i)$.
The key simplifying feature of this expansion is the fact that the 
leading order contribution $\Dh^{(0)}_{i}$ to $\Dh_{i}(\kv,\qv)$ is 
independent of $\kv$.

By substituting these expressions into Eq. (\ref{dC1loop_blend}), 
and keeping only terms of ${\cal O}(1)$ and ${\cal O}(1/q^{2})$ 
in the integrand that lead to UV divergent integrals, we obtain 
a UV divergent contribution to $\delta\Ch_{ij}(\kv)$ as a sum:
\begin{equation}
   \delta \Ch_{ij}(\kv) \simeq 
   \delta \Ch_{ij}^{(0)} + 
   \delta \Ch_{ij}^{(\chi)} +
   \delta \Ch_{ij}^{(1)} 
   \label{DeltaCh_int_blend_sum}
\end{equation}
where 
\begin{eqnarray}
   \delta \Ch_{ij}^{(0)} & \simeq &
   \frac{1}{2} \int_{\qv}
   \Dh_{i}^{(0)} \Dh_{j}^{(0)} 
   \Ghi^{(0)}\Ghi^{(0)}
   \nonumber \\
   \delta \Ch_{ij}^{(\chi)} & \simeq & 
   \int_{\qv}
   \Dh_{i}^{(0)} \Dh_{j}^{(0)}  
   \Ghi_{ij}^{(\chi)} \Ghi^{(0)}
   \label{DeltaCh_int_blend} \\
   \delta \Ch_{ij}^{(1)} & \simeq & 
   \frac{1}{2} \int_{\qv}
   \Dh_{i}^{(0)} \Dh_{j}^{(0)}  
   [ \Ghi^{(+)} \Ghi^{(-)} + 
     2\Ghi^{(1)} \Ghi^{(0)} ]
   \nonumber \\
   & + & 
   \frac{1}{2} \int_{\qv}
   \left [
   \Dh_{i}^{(1)}\Dh_{j}^{(0)} +
   \Dh_{i}^{(0)}\Dh_{j}^{(1)}
   \right ] 
   \Ghi^{(0)} \Ghi^{(0)} 
   \nonumber
\end{eqnarray}
Evaluating the integrals with respect to $\qv$ yields
\begin{eqnarray}
  \delta \Ch_{ij}^{(0)} & = &
  \frac{ l_i l_j  v^2}{12\pi^2\bar{l}^2} \Lambda^3 
  \nonumber \\
  \delta \Ch_{ij}^{(\chi)} & = &  
  \frac{12 v}{\pi^2} \frac{l_{i}l_{j} }{\bar{l}^3} Z_{ij} \barchiZero \Lambda 
  \nonumber \\
  \delta \Ch_{ij}^{(1)} & = & 
  \frac{7 k^2  l_i l_j   v^2}{12  \pi^2\bar{l}^2} \Lambda
  \label{dC1loop_blend_explicit} \\
  & + & 
  \frac{3  l_i l_j   v } {\pi^2\bar{l}^3 }
  \left ( \frac{\phi_1 l_1^2}{N_1} + \frac{\phi_2 l_2^2}{N_2} \right ) 
  \Lambda
  \nonumber \\
  & -  &
  \frac{3  l_i l_j   v }{ 2\pi^2\bar{l}^2 }
  \left (
  \frac{ l_i }{ N_i \gf(K_{i}^{2})} + 
  \frac{ l_j }{ N_j \gf(K_{j}^{2})} 
  \right ) \Lambda
  \quad. \nonumber
\end{eqnarray}
The divergent part of the corresponding apparent $\chi$ parameter 
is a sum of the three terms
\begin{eqnarray}
\delta \barchi^{(0)}
& = & \frac{ (l_1- l_2)^2  v}{8\pi^2\bar{l}^2}\frac{\Lambda^3}{3}
\nonumber \\
\delta \barchi^{(\chi)} 
& = &-\frac{6 l_1^2  l_2^2 }{ \pi^2 \bar{l}^3} 
      \barchiZero \Lambda
\nonumber \\
\delta \barchi^{(1)} & = &
     \frac{7 k^2  (l_1- l_2)^2   v}{24  \pi^2\bar{l}^2} \Lambda
     \\
     & + & 
     \frac{3 (l_1- l_2)^2 } { 2 \pi^2\bar{l}^3 }
     \left ( 
          \frac{\phi_{1}l_{1}^{2}}{N_{1}} +
          \frac{\phi_{2}l_{2}^{2}}{N_{2}} 
     \right ) \Lambda
     \nonumber \\
     & + &  
     \frac{3 (l_1- l_2)}
     { 2\pi^2\bar{l}^2 }
     \left ( 
     \frac{l_{2}^{2}}{N_2 \gf(K_{2}^{2})} -
     \frac{l_{1}^{2}}{N_1 \gf(K_{1}^{2})}
     \right ) \Lambda
     \quad. \nonumber
\end{eqnarray}
Note that both $\chi^{(0)}$ and $\chi^{(1)}$ vanish in the case 
$l_{1} = l_{2}$ of equal statistical segment lengths.

The UV convergent contribution $\delta \Ch_{ij}^{*}(\kv)$ is given
by the difference between the exact integral expression of Eq.
(\ref{dC1loop_blend}) and the sum of the UV divergent integrals given 
in Eqs.  (\ref{DeltaCh_int_blend_sum}) and (\ref{DeltaCh_int_blend}).
The resulting UV convergent integral can be non-dimensionalized by
arguments similar to those used to non-dimensionalize 
$\delta\Ssh_{ij}^{*}(k)$, which in this case yield a non-dimensional
integral times a prefactor of $v/(N\bar{N}^{1/2})$. The 
corresponding non-dimensional form of $\delta\barchia^{*}(k)$ is 
given in Eq. (\ref{dchi_nondim}).

\section{Correlations in Diblock Copolymer Melts}
\label{sec:Diblock}

We now consider the calculation of $\delta S_{ij}(q)$ for a diblock 
copolymer melt. The calculation is closely analogous to that given
above for a blend. 

\subsection{Two-Point Functions}
The two-point intramolecular function $\Sshi_{ij}(\qv)$ for a Gaussian diblock 
copolymer of length $N$ with blocks of length $f_{1}N$ and $f_{2}N$, is a matrix
\begin{equation}
  \Sshi_{ij}(\qv) = \frac{N}{v}\gf_{ij}(\qv)
\end{equation}
where
\begin{eqnarray}
   \gf_{ij} & = &
   \left [ \begin{array}{cc}
   \gf_{1} & \ef_{1} \ef_{2} \\
   \ef_{1} \ef_{2} & \gf_{2}
   \end{array} \right ]
   \nonumber \\
   \gf_{i} & \equiv &
   2 ( e^{-f_{i}Q_{i}^{2}} - 1 + f_{i} Q_{i}^{2} ) / Q^{4}_{i}
   \nonumber \\
   \ef_{i} & \equiv & ( 1 - e^{-f_{i}Q_{i}^{2}} )  / Q_{i}^{2}
   \label{gf_def_diblock}
\end{eqnarray}
with $Q_{i}^{2} \equiv q^{2} N b_{i}^{2}/6$. The high-$q$ asymptotic behavior 
of these function may be obtained by dropping all terms that contain factors
of $e^{-f_i Q_i^2}$.

The high-$q$ behavior of the propagator $\Ghi_{ij}(\qv)$ may be 
approximated to the required accuracy by an expansion
\begin{eqnarray}\nonumber
    \Ghi_{ij}(\qv) & \simeq &
    \frac{ v^2 }{12 \bar{l} }|\qv|^2
    \nonumber \\
    & + & 
    \frac{(l_1^2-l_1 l_2 +l_2^2)v}{2\bar{l}^2N}
    + \frac{2v}{\bar{l}^2}\barchiZero Z_{ij}
   \nonumber
   \label{Ghasymp_diblock}
\end{eqnarray}
Here, $\bar{l}$ and $Z_{ij}$ have the same values as in a 
homopolymer blend of the same composition, with $\phi_{i} 
= f_{i}$. That is,
\begin{eqnarray}
  \bar{l} & \equiv & f_{1}l_{1} + f_{2}l_{2}
  \nonumber \\
  Z & \equiv & \left [ \begin{array}{cc}
    -f_2^2 l_2^2  &  f_1f_2l_1l_2 \\
     f_1f_2l_1l_2 &   -f_1^2 l_1^2  \end{array} 
    \right ] \quad.
\end{eqnarray}
The only difference between this expansion and the corresponding 
expansion of $\Ghi_{ij}(\qv)$ in a homopolymer blend of equal 
composition is the nature of the term $\Ghi^{(1)}_{ij}(\kv)$ that 
is proportional to $1/N$ and independent of $\chi$, which is the 
first term in the second line of Eq. (\ref{Ghasymp_diblock}).

\subsection{Free Energy Density}
The one-loop contribution $\delta f$ to the free energy density of 
a homogeneous diblock copolymer melt can be calculated by a procedure 
closely analogous to that given in subsection \ref{sub:FreeEnergyBlend} 
for a binary homopolymer blend. The UV divergent contribution is of
the form
\begin{equation}
    \delta f \simeq 
    \delta f^{(0)} + \delta f^{(\chi)} + \delta f^{(1)}
    \quad. \label{dfDiblock}
\end{equation}
The expressions for $\delta f^{(0)}$ and $\delta f^{(\chi)}$ are 
identical to those given in Eqs. (\ref{df0Blend}) and 
(\ref{dfChiBlend}), respectively, for a blend of the same composition.
As for a homopolymer blend, the sum of these two terms yields the 
quantity $\delta f_{\rm local} = \delta f^{(0)} + \delta f^{(\chi)}$.
The remaining ${\cal O}(1/N)$ contribution is given in a diblock
melt by
\begin{eqnarray}
    \delta f^{(1)} & = & 
    \frac{1}{2} \int\limits_{\qv}
    \Gh^{(0)}\Sshi^{(1)}_{+}
    \nonumber \\
    & = & -\frac{3}{2\pi^{2}v\bar{l}}
    \frac{l_{1}^{2} - l_{1}l_{2} + l^{2}_{2}}{N}
    \Lambda
    \label{df1Diblock}
\end{eqnarray}
Here, $\Sshi^{(1)}_{+} = \sum_{ij} \Sshi^{(2,1)}_{ij}$, where
$\Sshi^{(2,1)}_{ij}$ is the ${\cal O}(N/Q^{4})$ contribution to
$\Sshi^{(2)}_{ij}(\qv)$. 

\subsection{Intramolecular Correlations}
In the diblock case, we calculate $\delta \ssh_{ij}^{(2)}(\kv)$ using
Eq. (\ref{ssh1loop2}). To analyze the divergence of this expression 
we need to identify the high-$\qv$ behavior of its components. As in
the blend case, this integral has a divergent part proportional to 
$\Lambda$, which may be obtained by using the dominant contributions
to $\psi^{(4)}_{ijkl}(\kv,-\kv,\qv,-\qv)$ and $\Ghi_{kl}^{(2)}(\kv)$.
As in a blend, the dominant contribution to $\Ghi_{kl}^{(2)}(\kv)$ 
is independent of the values of the indices, and is given by Eq. 
(\ref{Ghiasymp0}). 

The high-$q$ behavior of $\psi^{(4)}_{ijkl}$ is analyzed in the
appendix.  As for the case of the blend, we find a simple expression 
for the dominant asymptotic behavior $\psi^{(4)}$ which largely 
determines the form of the result: The dominant contribution 
to the function $\psi^{(4)}_{ijkl}(\kv,-\kv,\qv,-\qv)$ arises only
for elements with $k=l$, and is given by 
\begin{equation}
   \psi^{(4)}_{ijkk}(\kv,-\kv,\qv,-\qv) \simeq
   -\frac{288\alpha^{2}}{q^4b_{k}^{2}}
   \frac{\partial \sshi^{(2)}_{ij}(\kv)}{\partial (b_{k}^{2})}
\end{equation}
For $i=j$, the derivative with respect to $b_{k}^{2}$ is nonzero 
only for $i=j=k$, while for $i \neq j$, the derivative is nonzero
for both $k=1$ and $k=2$.  The dominant contributions to elements 
of $\psi^{(4)}_{ijkl}(\kv,-\kv,\qv,-\qv)$ with $k \neq l$ are all 
found to be ${\cal O}(1/q^{6})$ or smaller, and thus do not 
contribute to the divergent part of Eq. (\ref{ssh1loop2}) for 
$\delta\sshi^{(2)}_{ij}(\kv)$. 

Using this asymptotic result for $\psi^{(4)}$, the UV divergent 
contribution to Eq. (\ref{ssh1loop2}) can be written as a sum
\begin{eqnarray}
   \delta\ssh_{ij}^{(2)}(\kv) \simeq 
   \sum_{k}
   \frac{ \partial\sshi_{ij}^{(2)}(\kv) }{ \partial(b_k^2) }
   \delta( b_k^2)
\end{eqnarray}
where $\delta (b_{k}^{2})$ is again given by Eq. (\ref{deltabi2}).
For the diblock melt, as for the blend, we thus find that the 
divergent contributions to $\delta\ssh_{ij}^{(2)}(\kv)$ can be
absorbed into a renormalization of statistical segment lengths.

\subsection{Direct Correlation Function}
The one-loop contribution to the direct correlation function in
a diblock copolymer melt can be written as 
\begin{equation}
   \delta \Ch_{ij}(\kv) \simeq 
   \frac{1}{2} \int_{\qv}
   \Dh_{ikm}(\kv,\qv) \Ghi_{kl}(\qplv) 
   \Dh_{jln}^{(3)}(\kv,\qv)\Ghi_{mn}(\qmv)
   \eqsp, 
   \label{dC1loop_diblock} 
\end{equation}
where, in this context, we define
\begin{equation}
    \Dh_{ikl}(\kv,\qv) \equiv
    \Sshi^{-1}_{ij}\Sshi^{(3)}_{jkl}(\kv,\qmv,-\qv_{+})
    \quad, \label{Dh_ikl_diblock_def}
\end{equation}
with $\Dh_{ikl}(\kv,\qv) = \Dh_{ikl}(-\kv,-\qv)$. 
As for the binary blend, we expand the three point function 
$\Sshi^{(3)}_{jkl} = \Sshi^{(3)}_{jkl}(\kv,\qmv,-\qplv)$ in the 
high-$q$ limit as a sum
\begin{equation}
   \Sshi^{(3)}_{jkl} = 
   \Sshi^{(3,0)}_{jkl} + \Sshi^{(3,1)}_{jkl}
\end{equation}
and $\Dh_{ikl}(\kv,\qv)$ as a corresponding sum
\begin{equation}
   \Dh_{ikl} \simeq \Dh^{(0)}_{ikl} + \Dh^{(1)}_{ikl}
\end{equation}
where 
$\Sshi^{(3,0)}_{jkl}$ and $\Dh_{ikl}^{(0)}$ are ${\cal O}(1/q^{2})$ 
and
$\Sshi^{(3,1)}_{jkl}$ and $\Dh_{ikl}^{(1)}$ are ${\cal O}(1/q^{4})$.

A detailed analysis of the high-$q$ behavior of $\Sshi^{(3)}_{jkl}$,
which is outlined in the appendix, shows that the leading order 
contribution $\Sshi^{(3,0)}_{jkl}$, which is of ${\cal O}(1/q^{2})$,
is nonzero only for $k=l$, and has four elements $\Sshi^{(3,0)}_{jkk}$ 
that are all of the form
\begin{equation}
   \Sshi^{(3,0)}_{jkk}(\kv,\qmv,-\qplv) =
   \Sshi^{(2)}_{jk}(\kv)\frac{12 l_{k}}{vq^{2}}
\end{equation}
It follows from definition Eq. (\ref{Dh_ikl_diblock_def}) and this
expression that the corresponding leading order contribution to 
$\Dh_{ikl}$ is 
\begin{equation}
    \Dh_{ikl}^{(0)}(\kv,\qv) = 
    \delta_{ik} \delta_{kl} \Dh^{(0)}_{i}(\qv)
\end{equation}
where
\begin{equation}
    \Dh^{(0)}_{i}(\qv) \equiv \frac{12 l_{i}}{v q^{2}}
\end{equation}
is independent of $\kv$, and is the same function as that found in
the binary blend. 

The only nonzero elements of the subdominant contribution
$\Sshi^{(3,1)}_{jkl} = \Sshi^{(3,1)}_{jkl}(\kv,\qmv,-\qplv)$ are:
\begin{eqnarray}
  \Sshi^{(3,1)}_{jjj}  & = &
  \frac{18 l_j^2}{v^{3}}
  [ K_{j}^2(3+4\alpha^2) \gf_{j} -4f_{j} ] \frac{1}{q^4} 
  \nonumber \\
  \Sshi^{(3,1)}_{jkk} & = & 
  \frac{18 l_j^2}{v^{3}}  
  \ef_{j} [ K_{k}^2(3+4\alpha^2)\ef_{k}-4 ]
  \frac{1}{q^{4}}
  \nonumber \\
  \Sshi^{(3,1)}_{jjk} & = & 
  \frac{36 l_j l_k \ef_{j}}{v^3} \frac{1}{q^4}
  \nonumber
\end{eqnarray}
where $k \neq j$. In order to calculate the one-loop contribution
to $\delta \Ch_{ij}(\kv)$, we will need a quantity
\begin{equation}
   \Dh^{(1)}_{i}(\kv,\qv) \equiv 
   \sum_{kl} \Dh^{(1)}_{ikl}(\kv,\qv)
   \quad.
\end{equation}
This is given by
\begin{equation}
   \Dh^{(1)}_{i} = \left [
   \frac{3 l_{i}}{v} k^{2} (3+4\alpha^2) 
    -\frac{72}{Nv^{2}} \gf_{ij}^{-1}L_{j}
   \right ]
   \frac{1}{q^{4}} 
   \label{Dh1_diblock}
\end{equation}
where we have defined a vector $L_{i}$ with components
\begin{eqnarray}
   L_{1} & \equiv & f_{1}l_{1}^{2} + \ef_{1}l_{2}^{2} - \ef_{1}l_{1}l_{2}
   \nonumber \\
   L_{2} & \equiv & f_{2}l_{2}^{2} + \ef_{2}l_{1}^{2} - \ef_{2}l_{1}l_{2}
   \label{Hk_def}
\end{eqnarray}
The expression for $\Dh_{i}^{(1)}$ in a homopolymer blend given 
in Eq. (\ref{Dh_asymp_blend}) may be recovered by the 
replacements $L_{i} \rightarrow l_{i}^{2}$ and 
$g^{-1}_{ij}/N \rightarrow \delta_{ij}/(\gf_{i}N_{i})$.

When expressed in terms of $\Gh_{ij}$ and the quantity $\Dh_{i}$ 
defined above, the integral expression for the divergent part of 
$\delta \Ch_{ij}(\kv)$ in a diblock copolymer melt is identical 
to that given in Eq. (\ref{DeltaCh_int_blend}) for a homopolymer 
blend. The only differences between the expressions obtained 
for $\delta\Ch_{ij}$ in a diblock copolymer melt and that in 
a homopolymer blend of the same composition arise from the use 
of different expressions for the the $k$-independent part of
$\Gh^{(1)}_{ij}$ [given for a diblock by the first term of the 
second line of Eq.  \ref{Ghasymp_diblock}], and for $\Dh^{(1)}_{i}$
[given by Eq. (\ref{Dh1_diblock})].  Moreover, the integral
expressions for $\delta \Ch_{ij}^{(0)}(\kv)$ and 
$\delta \Ch_{ij}^{(\chi)}(\kv)$ are identical to those obtained 
for a homopolymer blend of the same composition.  Only the 
contribution $\delta \Ch_{ij}^{(1)}(\kv)$ differs from that 
obtained for a corresponding blend. This is given by 
\begin{eqnarray}
  \delta \Ch_{ij}^{(1)} & = & 
  \frac{7 k^2  l_i l_j v^2}{12 \pi^2\bar{l}^2} \Lambda
  \nonumber \\
  & + & 
  \frac{3 v l_{i}l_{j}( l_{1}^{2} - l_1 l_2 + l_2^{2} ) } {\pi^2 N \bar{l}^3 } 
  \Lambda
  \nonumber \\
  & - &
  \frac{3 v }{ 2N\pi^2\bar{l}^2 }
  ( \gf_{ik}^{-1}L_{k}l_{j} + l_{i} L_{k} \gf_{kj}^{-1} ) 
  \Lambda
  \label{dCh1_diblock}
\end{eqnarray}
The first line, the term proportional to $k^{2}$, is identical to 
the corresponding expression for a blend of the same composition. 
The second line, which arises from the integral involving the 
$k$-independent part of $\Ghi^{(1)}$, is different because of the 
use of different expression for $\Ghi^{(1)}$. The third line is 
different because of the use of a different expression for 
$\Dh^{(1)}_{i}$. 

The corresponding expression for the UV divergent contribution to
$\delta \barchi(\kv)$ is similar to that obtained for a
blend of the same composition. The expressions for $\delta \chi^{(0)}$ 
and $\delta \barchi^{(\chi)}$ terms are identical to those 
obtained in a homopolymer blend. The expression for 
$\delta \barchi^{(1)}(\kv)$ is different, but is retains the
property that $\delta \chi^{(1)} = 0$ in the limit $l_1 = l_2$ of 
equal statistical segment lengths.

\section{End Effects}
\label{sec:Defects}
We show in this section that the form of our results for the UV 
divergent part of $\delta C_{ij}(k)$ are consistent with the existence 
of a UV divergent one-loop contribution to the free energy of the form 
proposed in Eq. (\ref{FGlocal}). Here, it is convenient to start from 
the more general expression
\begin{equation}
   \delta \Gm_{\rm int} = \int d\rv [ \;
   \delta f_{\rm local} + \sum_{\alpha}d_{\alpha}\psi_{\alpha} 
   + \frac{1}{2} D_{ij}\nabla \cmon_{i} \nabla \cmon_{j}\; ]
   \label{FGlocal2} \quad,
\end{equation}
in which the gradient-squared term is expressed in terms of monomer
concentrations, as would be required to describe a slightly compressible
liquid.  The postulated free energy is the sum of a Ginzburg-Landau 
like functional of the monomer concentrations plus additional free 
energies arising from chain ends and (for a diblock) from the junction
that connects the two blocks.  

The excess free energies arising from chain ends and junctions are 
assumed to depend linearly on the concentrations of these defects 
because the defect concentrations are dilute. Direct defect-defect 
interactions, which would yield contributions to $S^{-1}(k)$ of 
${\cal O}(1/N^{2})$, could become important in systems of relatively 
short chains with strongly interacting end-groups, but do not appear 
in our model within the one-loop approximation.

\subsection{Monomer and Defect Chemical Potentials}
We compare our one-loop results to a form of SCFT in which the average 
monomer concentrations are calculated from those of ideal gas reference 
system with a self-consistent field Hamiltonian
\begin{equation}
   \tilde{U} = U_{\rm chain} - \sum_{i}\hir_{i}*\cmon_{i} 
     + \sum_{\alpha}d_{\alpha}*\psi_{\alpha}
\end{equation}
in which a chain end of type $\alpha=1$ or $\alpha=2$, or a junction 
($\alpha = J$) at position $\rv$ is penalized by a free energy 
$\psi_{\alpha}(\rv)$.  In the absence of any external field, monomers of 
type $i$ are subjected to a field
\begin{equation}
   -\hir_{i} = \Uh_{ij}*\cmon_{j}
             + \frac{\delta (\delta \Gm_{\rm int})}{\delta \cmon_{i}}
    \quad,
\end{equation}
in which $\delta \Gm_{\rm int}$ is given by Eq. (\ref{FGlocal2}), and 
$\delta/\delta\cmon_{i}$ represents a functional derivative.

To construct an RPA calculation of $S(k)$,  we consider the deviations from 
a homogeneous reference state induced by a small external field $\hr_{i}$ 
that couples only to the monomer density. This external perturbation will 
induce deviations $\delta \cmon_{i}$ and $\delta d_{\alpha}$ in monomer and 
defect concentrations, respectively, and deviations $\delta\hir_{i}$ and 
$\delta\psi_{\alpha}$ in the conjugate fields.  These deviations in chemical 
potential fields are given, to linear order in the deviations in the 
concentrations, by
\begin{eqnarray}
    - \delta \hir_{i} 
     & = & U_{ij}' * \delta \cmon_{j} + V_{i\alpha} * \delta d_{\alpha}
       - \hr_{i}
     \nonumber \\
     \delta \psi_{\alpha} & = & V_{\alpha i}^{T}*\delta \cmon_{i}
     \label{d_fields}
\end{eqnarray}
where
\begin{eqnarray}
   \Uh_{ij}' & \equiv & -\frac{\delta\hir_{j}}{\delta \cmon_{i}}
   = \Uh_{ij} + 
   \frac{\delta^2 (\delta \Gm_{\rm int})}{\delta \cmon_{i}\delta \cmon_{j}}
   \nonumber \\
   \Vh_{i\alpha} & \equiv & 
   \frac{\delta \psi_{\alpha}}{\delta \cmon_{i}}
   \quad.
\end{eqnarray}
Here, $\Uh_{ij}'$ is an effective monomer-monomer interaction, and 
$V_{i\alpha}$ where is an effective monomer-defect interaction. 
Here, and throughout this section, we use `*' to represent spatial 
convolution only, and display the monomer and defect type indices 
explicitly.

In Fourier space, the effective interaction $\Uh_{ij}'(\kv)$ is 
given by 
\begin{equation}
   \Uh_{ij}'(\kv)  =  \Uh_{ij}(\kv) 
   + \frac{\partial^2 (\delta f_{\rm local})}{\partial c_{i}\partial c_{j}}
   + W_{ij} + D_{ij}k^{2} 
\end{equation}
where
\begin{equation}
   W_{ij} \equiv
   \sum_{\alpha}d_{\alpha}
   \frac{\partial^2 \psi_{\alpha}}{\partial \cmon_{i}\partial \cmon_{j}}
\end{equation}
Here $d_{\alpha}$ is the defect density in the homogeneous reference
state: For hopolymers, the ends densities are
$d_{\alpha} = 2 \phi_{\alpha}/(vN_{\alpha})$; for diblock copolymers,
both ends and junction densities are $d_{\alpha} = 1/(vN)$.
All derivatives are evaluated in this homogeneous reference state. Note 
that our definition of $\Uh_{ij}'$ as a second functional derivative 
of the interaction free energy yields a contribution $D_{ij}k^{2}$ 
that arises from the (postulated) gradient-squared contribution to 
$\delta \Gm_{\rm int}$.  


\subsection{Generalized RPA}
To complete the RPA linear response calculation, we must combine the
above with a description of the linear response of an ideal gas of
polymers. The linear response of the monomer and defect concentrations 
to deviations in the above combination of fields is given by
\begin{eqnarray}
   \delta\cmon_{i} & = & 
   \Sshi_{ij}*\delta \hir_{j} 
   - \Rhi_{i \beta} * \delta \psi_{\beta}
   \nonumber \\
   \delta d_{\alpha} & = & 
     \Rhi_{\alpha j}^{T} * \delta \hir_{j}
   - \Ehi_{\alpha \beta} * \delta \psi_{\beta} 
     \label{d_concentrations}
\end{eqnarray}
Here, $\Sshi_{ij}$ is the ideal-gas intramolecular correlation 
between monomers of types $i$ and $j$, $\Rhi_{j \alpha}$ is the 
intramolecular correlation between $\alpha$ defects and $j$ 
monomers, and $\Ehi_{\alpha\beta}$ is the intramolecular correlation 
between defects of type $\alpha$ and $\beta$. All of these functions 
are diagonal in a homopolymer blend (i.e., are nonzero only for 
$i=j$, $i=\alpha$ or $\alpha = \beta$), and all become non-diagonal 
in a diblock copolymer melt. 

By combining linear response equations (\ref{d_concentrations}) with
self-consistency conditions (\ref{d_fields}), it is straightforward
to show that, in a Fourier representation, $\delta \cmon_{i}(\kv)$ 
is given by
\begin{equation}
   \delta \cmon_{i} = 
   A_{ij}[ - B_{jk}\delta \cmon_{k} + \hr_{j} ]
\end{equation}
where all quantities are implicitly functions of a wavenumber $\kv$, 
and 
\begin{eqnarray}
   A_{ij} & \equiv & 
   P^{-1}_{ik}\Sshi_{kl}[P^{T}]^{-1}_{lj}
   \nonumber \\
   P_{ij} & \equiv & \delta_{ij} 
                   + \Rhi_{i\alpha}\Vh_{\alpha j}^{T}
   \nonumber \\
   B_{jk} & \equiv & \Uh'_{jk} - 
                     \Vh_{j\alpha}\Ehi_{\alpha\beta}\Vh_{\beta k}^{T}
   \quad.
\end{eqnarray}
Solving for the inverse response/correlation function 
$S_{ij}^{-1}(\kv) = \delta \hr_{j}(\kv)/ \delta \cmon_{i}(\kv)$ 
yields
\begin{eqnarray}
   S^{-1}_{ij} & = & A^{-1}_{ij} + B_{ij} 
   \nonumber \\
   & = &
   \Sshi^{-1}_{ij} + \Uh_{ij} 
   + \delta S^{-1}_{ij} 
\end{eqnarray}
in which 
\begin{equation}
   \delta S^{-1}_{ij} = 
   \frac{\partial(\delta f_{\rm local})}{\partial \cmon_{i}\partial \cmon_{j}}
   + D_{ij}k^{2} - \delta C_{ij}^{(d)}
\end{equation}
is the correction to mean-field theory, and where 
$\delta C_{ij}^{(d)}$ is a contribution arising from the interaction 
of the end and junction `defects' with the monomer concentration.  This 
quantity is given by a sum
\begin{eqnarray}
   -\delta C_{ij}^{(d)}  & \equiv & 
   -\delta C_{ij}^{(d1)} - \delta C_{ij}^{(d2)} 
   \nonumber \\
   - \delta C_{ij}^{(d1)} & \equiv & W_{ij} +
       \Vh_{i\alpha}\Rhi_{\alpha k}^{T}\Sshi^{-1}_{kj}
     + \Sshi^{-1}_{ik}\Rhi_{k\alpha}\Vh_{\alpha j}^{T}
   \nonumber \\
   - \delta C_{ij}^{(d2)} & \equiv &
         \Vh_{i\alpha}\left [ 
         \Rhi_{\alpha k}^{T}\Sshi^{-1}_{kl}\Rhi_{l\beta}
          - \Ehi_{\alpha\beta} \right] \Vh_{\beta j}^{T}
   \label{Cd_defect} 
\end{eqnarray}
Note that the quantity $\delta C_{ij}^{(d1)}$ depends linearly upon 
the effective interaction $V_{i\alpha}$, while $\delta C_{ij}^{(d2)}$ 
is second-order in $V_{i\alpha}$. 

\subsection{Comparison to One-Loop Results}
To test our phenomenological model, we compare our explicit one loop
calculations for the $\delta C^{(1)}_{ij}(\kv)$ to the results of the
above generalized RPA calculation, while using the explicit expressions 
for $\psi_{\alpha}$ given in Eqs. (\ref{psiend}) and (\ref{psijct}), 
and treating $D_{ij}$ as a free parameter. We find that, in both 
blends and diblock copolymer melts, the quantity $\delta C^{(1)}_{ij}$ 
obtained in the one-loop approximation for $\delta \Ch_{ij}(\kv)$ can 
be written as a sum of the form
\begin{equation}
   -\delta \Ch_{ij}^{(1)}(\kv) = 
    D_{ij}k^{2} - \delta \Ch_{ij}^{(d1)}(\kv)
   \label{Ch1_end}
\end{equation}
where $\delta\Ch_{ij}^{(d1)}$ is the first-order defect contribution
given in Eq. (\ref{Cd_defect}), and
\begin{equation}
   D_{ij} = -\frac{v^{2} l_{i}l_{j}}{3\pi^{2} \bar{l}^{2}} \Lambda
   \quad. \label{D_value}
\end{equation}
The same value is obtained for $D_{ij}$ in blends and diblock copolymer 
melts of the same composition. The scalar coefficient given in Eq. 
(\ref{D_scalar}) is obtained by requiring that $\nabla \cmon_{1} 
= - \nabla \cmon_{2}$ in an incompressible liquid.  

In both blends and diblock copolymer melts, the result from the one 
loop approximation is thus identical to that obtained from the 
generalized RPA, {\it except} for the absence in our one-loop results 
of the contribution $\delta \Ch_{ij}^{(d2)}$ predicted by RPA result. 
Note that the contribution $\delta \Ch_{ij}^{(d2)}$ that is 
``missing" from the results of the one-loop approximation has a 
qualitatively different dependence on both $V_{i\alpha}$ and 
$\Lambda$ than terms that are found in the one-loop calculation: 
The quantity $\delta \Ch_{ij}^{(2d)}$ is second order in the 
strength of the coupling $V_{i\alpha}$, and so would yield a
contribution $\delta \Ch_{ij}^{(2d)} \propto \Lambda^{2}$ in a 
theory in which $V_{i\alpha} \propto \Lambda$, while the terms
that {\it do} appear in our one-loop results are linear in 
$V_{i\alpha}$, and proportional to $\Lambda$. 
The one-loop calculation of $\delta \Ch_{ij}$ yields only terms 
proportional to $\Lambda$ and $\Lambda^{3}$, so it appears that 
the missing contribution predicted by the RPA could not possibly 
be generated by a one-loop approximation.  We assume that the 
discrepancy occurs because terms that are quadratic in the 
magnitude of the one-loop contributions to $V_{i\alpha}$ appear 
only at second order in a loop expansion of $S^{-1}(k)$. The 
$\Lambda$-dependence of the missing contribution appears to be 
consistent with this conjecture: The power counting analysis 
of the loop expansion given in Sec. \ref{sec:PowerCounting} 
and in Ref. [13] 
indicates that the dominant UV divergence of the two-loop contribution 
to $\delta\Ch$ will be ${\cal O}(\Lambda^{4})$, to which we expect 
to find subdominant corrections of ${\cal O}(\Lambda^{2})$ and 
${\cal O}(\ln \Lambda)$.  While this conjecture could be proved 
only by analyzing the two-loop theory, it seems clear that the 
missing terms cannot be produced by a one-loop approximation, 
and so may be neglected when making this comparison.

Below, we present some details of the application of the 
generalized RPA to binary blends and diblock copolymer melts, 
respectively. 

\subsubsection{Binary Blends}
For a binary blend of homopolymers, 
\begin{eqnarray}
   V_{i\alpha} & = &  
   \frac{3 v l_{i} }{4\pi^{2}\bar{l}^{2}} \; l_{\alpha}^{2} \Lambda
   \nonumber \\
   W_{ij} & = & -
   \frac{3 v l_{i}l_{j}}{\pi^{2}\bar{l}^{3}}
   \left [
   \frac{\phi_{1}l_{1}^{2}}{N_{1}} +
   \frac{\phi_{2}l_{2}^{2}}{N_{2}} \right ] \Lambda
   \nonumber \\
   \Sshi_{ij}     & = & \delta_{ij} v^{-1}\phi_i N_{i}\gf_{i}
   \nonumber \\
   \Rhi_{i\alpha} & = & \delta_{i\alpha} 2v^{-1}\phi_i \ef_{i}
\end{eqnarray}
where $\gf_{i} \equiv g(K_{i}^{2})$ is the Debye function, and 
where $\ef_{i} \equiv (1 - e^{-K_{i}^{2}})/K_{i}^{2}$ for a 
homopolymer.  This yields
\begin{equation}
     - \delta C_{ij}^{(d1)}  \equiv  
     W_{ij}
     + \frac{3vl_{i}l_{j}}{2\pi^{2} \bar{l}^{2}} 
     \left [ \frac{l_{i}\ef_{i}}{N_{i}\gf_{i}} +
     \frac{l_{j}\ef_{j}}{N_{j}\gf_{j}} \right ] \Lambda
\end{equation}
By using the identity $\ef_{i} = 1 - \gf_{i}K_{i}^{2}/2$ and the 
definition of $K_{i}^{2} = k^{2} v /(6 l_{i})$, we obtain the 
alternative expression 
\begin{equation}
     -\delta C_{ij}^{(d1)}  \equiv 
      W_{ij}
     + \frac{3vl_{i}l_{j}}{2\pi^{2} \bar{l}^{2}} 
     \left [ 
     \frac{l_{i}}{N_{i}\gf_{i}} + \frac{l_{j}}{N_{j}\gf_{j}} 
     \right ] \Lambda
     - \frac{v^{2} l_{i}l_{j}}{4\pi^{2} \bar{l}^{2}} \Lambda
\end{equation}
By substituting this expression for $\delta \Ch_{ij}^{(d1)}$ into 
Eq. (\ref{Ch1_end}), and using Eq. (\ref{D_value}) for $D_{ij}$, we 
reproduce the one-loop result for $\delta C^{(1)}_{ij}$ given in Eq. 
(\ref{dC1loop_blend_explicit}), 

\subsubsection{Diblock Copolymer Melts}
For a diblock copolymer melt, we obtain
\begin{eqnarray}
   V_{iJ} & = &  
   \frac{3 v l_{i} }{4\pi^{2}\bar{l}^{2}} \; (l_{1}-l_{2})^{2} \Lambda
   \nonumber \\
   W_{ij} & = & 
   - \frac{3v l_{i}l_{j}(l_{1}^{2} - l_{1}l_{2} + l_{2}^{2})}
        {\pi^{2}N\bar{l}^{3}} \Lambda
\end{eqnarray}
and
\begin{eqnarray}
    \Sshi_{ij}& = & \delta_{ij} v^{-1}N\gf_{ij}
    \nonumber \\
    \Rhi_{i\alpha} & = & v^{-1}\ef_{i}  
    \quad (\alpha = i)
    \nonumber \\
    \Rhi_{i\alpha} & = & v^{-1}\ef_{i}e^{-f_{\alpha}K_{\alpha}^{2}}
    \quad (\alpha \neq i, J)
    \nonumber \\
    \Rhi_{iJ} & = & v^{-1}\ef_{i}
\end{eqnarray}
where $\gf_{ij}$, $\gf_{i}$, $\ef_{i}$ are defined for a diblock 
copolymer by Eq. (\ref{gf_def_diblock}). This yields
\begin{equation}
     - \delta C_{ij}^{(d1)}  \equiv 
     W_{ij} + \frac{3v l_{i}l_{j}}{2\pi^{2}N\bar{l}^{2}} 
     \left [ g^{-1}_{ik}\ef_{i}l_{j} + l_{i}\ef_{k}\gf_{kj}^{-1} \right ] 
     \Lambda
\end{equation}
By using the identity $\ef_{i} = f_{i} - \gf_{i}K_{i}^{2}/2$, we obtain 
the alternative expression 
\begin{equation}
     - \delta C_{ij}^{(d1)}  \equiv 
     W_{ij} + \frac{3vl_{i}l_{j}}{2\pi^{2}N\bar{l}^{2}} 
     \left [ g^{-1}_{ik}L_{k}l_{j} + l_{i}L_{k}\gf_{kj}^{-1} \right ] 
     \Lambda
     - \frac{v^{2} l_{i}l_{j}}{4\pi^{2} \bar{l}^{2}} \Lambda
\end{equation}
where $L_{k}$ is given by Eq. (\ref{Hk_def}). 
By using this expression for $\delta C_{ij}^{(d1)}$ and Eq. 
(\ref{D_value}) for $D_{ij}$ in Eq. (\ref{Ch1_end}), we reproduce Eq. 
(\ref{dCh1_diblock}).

\section{Beyond One-Loop}
\label{sec:PowerCounting}
In this section, we briefly look beyond the one-loop theory. We
consider the structure of UV divergences of an arbitrary diagram
in the unrenormalized diagrammatic perturbation theory, at the 
level of naive power counting. We also consider the structure 
of a renormalized loop expansion, and argue that this should 
yield an asymptotic expansion of corrections to SCFT in powers 
of $1/\sqrt{\bar{N}}$.

For this purpose, we use the diagrammatic formalism presented 
in Ref. [13].
There, it was shown that corrections to the Gaussian or one-loop 
approximation for the grand potential 
$\ln Z$ for a polymer liquid could be expressed as an infinite sum 
of connected diagrams of $-\Ghi$ bonds and $\Ssh$ vertices, with 
no root sites. Here, we consider the dominant UV divergence of 
an arbitrary diagram in the expansion of the free energy density 
$\ln Z/V$ for a model with a wavenumber cutoff $\Lambda$.  
For generality, we consider a model of continuous Gaussian 
polymers in a space of arbitrary spatial dimension $d$. We
consider a generic connected diagram with $B$ bonds, each 
representing a factor of $-\Ghi$, connecting $V$ vertices, each 
representing a factor of $\Sshi^{(n)}$, where $n$ is the number 
of attached bonds.  The number $L$ of loops, or independent 
wavevector integrals, in such a diagram is $L = V - B + 1$. For 
simplicity, we consider only the dominant UV divergent to the 
free energy density of a one-component liquid (or of a blend 
with $\chi = 0$ and $b_{1}=b_{2}$), without attempting to analyze 
the dependence of the free energy density on $\chi$, or on the 
composition of a blend.  

The analyze the UV divergence of an arbitrary diagram, we must
characterize the asymptotic high-$q$ behavior of $\Ghi(q)$, and
of $\Sshi^{(n)}(\qv_{1}, \ldots,\qv_{n})$ for arbitrary $n$. The 
dominant $q$-dependence of the screened interaction $\Ghi(q)$, 
given in Eq. (\ref{Ghiasymp1}), is
\begin{equation}
    \Ghi(q) \sim (qb)^{2}/c \quad.
    \label{Ghi_scaling}
\end{equation}
where $c=1/v$ is the monomer concentration.
The function $\Sshi^{(n)}$ can be expressed, for a Gaussian 
homopolymer in any spatial dimension $d$, as a product of 
the form $\Sshi^{(n)}(\qv_{1},\ldots,\qv_{n}) = c N^{n-1} 
\sshid^{(n)}(\Qv_{1},\ldots,\Qv_{n})$, where $c/N$ is the 
number concentration of polymers, $\Qv_{i} \equiv \qv_{i}R$, and 
$R = b\sqrt{N/6}$. To characterize the high-$q$ limit, it is useful
to consider the limit of infinitely long chains, $N \rightarrow 
\infty$, at constant monomer concentration $c$. In this limit, 
all nonzero wavevectors become "large" compared to $1/R$.  The 
function $\Sshi^{(n)}(\qv_{1},\ldots,\qv_{n})$ generally approaches 
a nonzero limit as $N \rightarrow \infty$. For $\Sshi^{(n)}$ to 
approach a value that is independent of $N$ in this limit, 
$\sshid^{(n)}(\Qv_{1},\ldots,\Qv_{n}R)$ must approach a 
homogeneous scaling function
\begin{equation}
   \lim_{ N \rightarrow \infty}
   \sshid^{(n)}(\lambda \Qv_{1}, \ldots, \lambda \Qv_{n} ) =
   \lambda^{2(n-1)}\sshid^{(n)}(\Qv_{1},\ldots,\Qv_{n})
\end{equation}
when $|\Qv| \gg 1$ for all $n$ arguments.  At a power counting 
level, this implies that the function $\sshid^{(n)}$ is of order 
$Q^{2(n-1)}$ when all of its arguments are of order $Q$.
Corresponding, $\Sshi^{(n)}$ must be of order
\begin{equation}
   \Sshi^{(n)} \sim \frac{c (qb)^{2}} {(qb)^{2n}}
   \label{Sshi_n_scaling}
\end{equation}
when all of its arguments are of order $q \gg 1/R$.

To count powers of $q$ in the Fourier integral associated with an 
arbitrary diagram in the expansion of $\ln Z/V$, it useful to 
associated one of the $n$ factors of $(qb)^{2}$ in the denominator 
of Eq. (\ref{Sshi_n_scaling}) for $\Sshi^{(n)}$ with one end of one 
of the $n$ bonds that must be attached to the associated vertex. This 
method of counting counting leaves an overall factor of $c(qb)^{2}$ 
for each vertex, and a factor of $c^{-1}(qb)^{-2}$ for each bond
in an incompressible liquid.  In a diagram with $L$ loops, in $d$ 
dimensions, we obtain an additional factor of order $q^{dL}$ from 
the integration over $L$ wavevectors. By combining these factors, 
we find that the contribution to $\ln Z/ V$ from a generic 
connected $L$-loop diagram with a cutoff wavenumber $\Lambda$ is 
of order
\begin{equation}
   \Lambda^{d} (\Lambda^{d-2}c/b^{2})^{L-1} 
   \label{LoopExpand_UV1}
\end{equation}
In the physically relevant case $d=3$, we may define a packing 
length $p=1/(c b^{2})$, to rewrite this as
\begin{equation}
   \Lambda^{3} (\Lambda p)^{L-1}
   \label{LoopExpand_UV2}
\end{equation}
The order of UV divergence is thus expected to increases by one 
factor of $\Lambda p$ at each order in the loop expansion. 

In the one-loop theory considered here, with $d=3$ and $L=1$, 
this argument yields a UV divergent contribution of order 
$\Lambda^{3}$. This agrees with the results of our explicit
one-loop calculation of free energy density of an incompressible 
liquid. Corresponding $L$-loop contributions to the functional 
derivatives of $F[\langle c \rangle]$, such as the second 
derivative $S^{-1}(k)$, are expected to have the same dependence 
on $\Lambda$ as the underlying free energy contributions, at
eac order in the loop expansion. Thus, for example, the 
dominant UV divergent one-loop contribution to $S^{-1}(k)$ 
in our explicit one-loop calculation is also of order 
$\Lambda^{3}$. 

We have shown that this UV divergence of the one-loop theory
can be removed by renormalization.  Let us assume, for the moment, 
that this procedure can be extended to arbitrary order in a loop 
expansion.  We imagine that the contribution of an $L$-loop 
diagram to the renormalized perturbation theory can be obtained 
by subtracting an asymptotic approximation for the integrand of 
the corresponding Fourier integral that is accurate at high
wavenumbers, $q \gg 1/R$. This will generally leave a UV 
convergent contribution that arises primarily from wavenumbers
of order $1/R$, due to deviations of the integrand from the
asymptotic approximation at low wavenumbers.  The resulting
contribution to the renormalized perturbation theory is expected 
to be similar in magnitude to the value of the unrenormalized 
integral evaluated with a cutoff $\Lambda \sim 1/R$.  This 
suggests that the renormalized perturbation theory (if one exists) 
will yield a loop expansion in which the $L$-loop correction 
to the SCFT free energy functional will have a prefactor of 
order
\begin{equation}
  \frac{1}{R^{3} } \left ( \frac{p}{R} \right )^{L-1}
  \sim \frac{1}{vN\bar{N}^{L/2}}
  \label{Renormal_LoopExpand_Nbar}
\end{equation}
We have used the fact that $R/p = N^{1/2}b^{3}/v = \bar{N}^{1/2}$ 
to obtain the second expression in the above.
The value of each diagram will be given by this prefactor times
a non-dimensionalized convergent integral whose value is a
dimensionless function of $\chi N$ and the other SCFT dimensionless 
paramters.  If our renormalization procedure can be extended 
beyond the one-loop level, the resulting renormalized loop 
expansion is thus expected to yield an asymptotic expansion 
of the free energy in powers of $1/\sqrt{\bar{N}}$.

Eq. (\ref{Renormal_LoopExpand_Nbar}) may also be obtained by 
dimensional analysis: If we non-dimensionalize all lengths 
in units of $R = \sqrt{N}b$ in the functional Taylor expansion 
of the statistical weight $L$ that appears in the Edwards 
functional integral, we obtain an expression for $L$ as the 
product of a non-dimensionalized functional times a large 
parameter $\sqrt{\bar{N}}$. The existence of such a large
prefactor is known to imply \cite{Amit1984} that the loop 
expansion (or, more precisely, the renormalized loop 
expansion) will yield an expansion in powers of 
$\bar{N}^{-1/2}$.

\section{Polymer Solutions}
\label{sec:Solutions}
UV divergences also appear in the excluded volume problem for 
polymers in good solvent. The diagrammatic formalism and power 
counting arguments given above for an incompressible liquid can 
also be applied to the excluded volume problem. A comparison of 
the two problems in the same language is instructive. 

In the standard Edwards model \cite{Edwards1965,Edwards1966} 
of a polymer in good solvent, the two-body interaction is 
approximated by a point-like effective interaction
$U(\rv) = a\delta(\rv)$, or $U(q) = a$, where $a$ is an
effective excluded volume. The one-loop theory given here for 
incompressible liquids is closely analogous to Edwards' one-loop
theory of solutions.  The screened interaction in the original 
Edwards theory is given, in the limit $N \rightarrow \infty$, by
\begin{equation}
    G^{-1}(q) = \frac{12c}{q^{2}b^{2}} + a^{-1}
\end{equation}
where the first term is the limit $qR \gg 1$ of $\Sshi(q)$. In 
this theory, we thus obtain an essentially unscreened interaction 
$\Ghi(q) \sim a$ for $k\xi \gg 1$, and a screened interaction 
identical to that given in Eq. (\ref{Ghi_scaling}) for $k\xi \ll 1$,
where $\xi \sim b/\sqrt{ca}$ is the Edwards screening length.

The one-loop theory for a nominally incompressible liquid is 
thus very similar to the Edwards theory with a cutoff wavenumber
$\Lambda \ll \xi^{-1}$.  In studies of non-dilute polymer 
solutions one is generally interested in the dependence of the 
radius of gyration, osmotic pressure etc. upon contributions 
from wavelengths less than $\xi$, which determine the 
concentration dependence of these quantities.  To retain this 
information, one must use a cutoff $\Lambda \gg \xi^{-1}$.  

The UV divergence contributions from the regime $q\xi \gg 1$ 
may be analyzed by repeating the power counting arguments 
given above for an incompressible liquid, while using the 
unscreened interaction $\Ghi(q) = a$, rather than the screened 
interaction $\Ghi(q) \propto q^{2}$. By this method, we find 
that the contribution to the free energy density of an $L$-loop 
diagram with $V$ vertices is of order
\begin{equation}
    \Lambda^{d} 
    (\Lambda b)^{(L-1)(d-4) - 2V} (a/b^{d})^{L-1} (ac)^{V}    
\end{equation}
In the infinite dilution limit, in which one considers only 
the interaction among monomers of a single chain, the only
relevant diagrams have only one vertex, $V=1$. In this limit, 
the above expression becomes equivalent to the known result 
\cite{DesCloizeaux1990} for the naive degree of divergence 
of diagrams in the perturbation theory for swelling of a 
single self-avoiding chain. 

In the physically relevant case $d=3$, the only UV divergent
diagrams in the expansion of the free energy density for a 
solution are the one-loop diagram with one-vertex ($L=V=1$), 
which diverges as $\Lambda$, and the two-loop contribution 
with one vertex ($L=2$ and $V=1$), which exhibits a logarithmic
divergence in $d=3$. The strongly divergent one-loop diagram 
was correctly identified by Edwards with a divergence in the 
free energy per monomer due to interactions between pairs of 
nearby monomers on the same chain. Edwards removed this 
divergence by subtracting the free energy per monomer of a 
single isolated chain from the total free energy. All of the 
diagrams involving more than one vertex, which are necessary 
to calculate, e.g., the second virial coefficient, are UV 
convergent in $d=3$. 

Alexander-Katz {\it et al}.  \cite{Fredrickson2005} have 
recently considered the UV divergence of the chemical potential 
in a stochastic field-theoretic simulation of a polymer 
solutions. In this context, the UV divergence shows up as 
a dependence of the polymer chemical potential $\mu$ upon 
a spatial discretization length $\Delta x$. They found that 
that the UV divergence of $\mu$ found in their simulation 
could be removed by subtracting a one-loop approximation for 
the free energy per monomer of a system of non-interacting 
chains from their simulation results. (The above analysis 
suggests that a remaining logarithmic divergence should 
have remained, but they reported no evidence that suggests
this). Our analysis of the incompressible liquid model 
indicates that this simple subtraction will not be 
sufficient to remove the UV divergences from analogous 
simulations of dense polymer mixtures, in which the 
divergence appears at all orders in the loop expansion. 

The appearance of negative exponents of $\Lambda$ in all
but a few diagrams in the expansion of the dilute solution
free energy in $d=3$ is a symptom of the fact that most 
diagrams in this theory are infrared (IR) rather than 
UV divergent. In the infinite dilution limit, in which 
only diagrams with $V=1$ are relevant, the IR divergence 
is cutoff at a wavelength of order $R \sim \sqrt{N}b$. 
In this limit, a generic diagram diverges with increasing 
chain length $N$ as $N^{d/2} N^{(4-d)(L-1)/2}$ for all $d$ 
less than the upper critical dimension $d_{c}=4$.  In a 
semidilute or concentrated solution, this IR divergence 
is cutoff at the screening length $\xi$.  The quantities 
that we calculate in this paper, other than the free energy 
density (i.e., the direct correlation function, which is 
analogous to the second virial coefficient, and the 
interaction-induced change in the single-chain correlation 
function) are actually UV convergent and IR divergent in 
the dilute solution problem.  It is, of course, the IR 
divergence of the excluded volume problem that makes the 
problem interesting, and that leads to nontrivial scaling 
behavior for a self-avoiding walk. 

The above analysis makes it clear that the excluded volume 
problem is ``renormalizable" in the sense that this word is 
normally used in quantum or statistical field theory, and 
that the theory of incompressible liquids considered here is 
not. A field theory is normally said to have a renormalizable 
UV divergence if the divergence can be absorbed into a finite 
number of measurable parameters, such as the mass and charge 
in quantum electrodynamics. More precisely, it is usually required 
that only a finite number of vertex functions contain a primitive 
UV divergence in spatial dimensions $d$ less than or equal to an 
upper critical dimension $d_{c}$.\cite{Amit1984}  For this to be
true, it must generally be the case that the degree of UV
divergence of all vertex functions decrease with increasing order 
in a loop expansion for all $d < d_{c}$.  For any $d < d_{c}$ 
only a few low order diagrams thus remain UV divergent, but the 
theory instead becomes susceptible to the appearance of IR 
divergences in (in different contexts) the limit $T \rightarrow T_{c}$ 
or $N \rightarrow \infty$, leading to nontrivial critical 
phenomena. The Edwards model for polymers in good solvent fits 
this description, with an upper critical dimension $d_{c} = 4$. 
In the model of incompressible liquids considered here, however, 
we see from Eq.  (\ref{LoopExpand_UV2}) that the degree of UV 
divergence increases with increasing order in a loop expansion 
for all $d > 2$, implying that $d_{c}=2$. This model would thus
normally be said be unrenormalizable in $d=3$.  The difference 
between the upper critical dimensions of the two theories is a 
result of the replacement of the unscreened point-like interaction, 
which is independent of wavenumber, by a strongly wavenumber 
dependent screened interaction, $\Ghi(q) \propto q^{2}$.

The notion of ``renormalizability" used in this paper is 
thus different from its usual meaning in field theory, and 
is specific to the physics of dense polymer mixtures. In
either context, a theory is said to be renormalizable if all 
of its UV divergences can be absorbed into the parameters of 
an appropriate phenomenological model.  In addition, for 
renormalization to be useful, it is required in both contexts
that the functional form of a phenomenological model that is
sufficiently flexible to absorb all UV divergences also be 
sufficiently constrained to allow nontrivial predictions to 
be made. A theory is properly described as unrenormalizable 
if the functional form of the theory required to absorb all 
UV divergent parts of the free energy functional is so 
flexible that it could describe all conceivable behavior. In 
the present context, the appropriate phenomenological model 
is a SCFT of Gaussian chains. The only constraint that we 
place on the functional form of this theory is that interaction 
free energy functional (excluding the contributions of chain 
ends and junctions) be independent of chain length $N$ and 
chain architecture.  We show here that, to first order in a 
loop expansion, a SCFT of this form is recovered as the 
$N \rightarrow \infty$ limit of the true free energy functional, 
and that the sensitivity of the theory to the cutoff (or more 
generally, to monomer scale chemical details) can be absorbed 
into this $N \rightarrow \infty$ limit. The renormalized 
perturbation theory makes nontrivial predictions about 
$N$-dependent, UV convergent corrections to this theory. If 
the same ideas were applied to a liquid of point particles, 
there would be no constraint on the functional form of the 
free energy, and so the theory would have no predictive power. 
It is thus the appearance of $1/\bar{N}$ as a small parameter 
that makes it possible to extract useful information from a 
theory that would be considered unrenormalizable by the usual 
rules of field theory.

\section{Conclusions}
\label{sec:Conclusions}
Several previous calculations have shown that predictions of a
coarse-grained model of polymer liquids, when extended beyond 
the mean-field level, depend very sensitively on the value 
chosen for an arbitrary coarse-graining length.  In light of 
this, it is reasonable to ask whether such models can make 
any unambiguous physical predictions.  We know how to extract 
physical predictions from a UV divergent theory only if we 
can absorb all UV divergences into the values of a finite number 
of phenomenological coefficients.
The question of whether such coarse-grained theories can predict
anything thus inevitably boils down to the question of whether 
they are, in some sense, renormalizable.  The definition of 
"renormalizability" that we introduce here (as discussed above) 
is based upon the assumption that a renormalized SCFT with an
interaction free energy that is independent of $N$ is obtained
as the limit $N \rightarrow \infty$ of the true free energy, and 
that all UV divergences can be absorbed into this asymptotic 
theory.

In this paper, we have explicitly demonstrated the renormalizability 
of the one-loop approximation for $S^{-1}(k)$ in the auxiliary field 
approach, for arbitrary $k$, for both polymer blends and diblock 
copolymer melts. We showed that all UV divergent contributions to 
$S^{-1}(k)$ can be absorbed into changes in the values of the 
parameters of an appropriate form of SCFT.  To make this work, it was 
necessary to allow for renormalization of all of the parameters of the 
standard SCFT, i.e., of the statistical segment lengths as well as the 
local interaction free energy. We also found that it was necessary to 
allow for some free energy contributions that are plausible on physical 
grounds, but not usually considered, i.e., a square-gradient interaction
free energy and excess energies for chain end and for junctions in 
block copolymers.

At a conceptual level, this analysis is important because it provides 
evidence for the logical consistency of the study of coarse-grained 
models, and for our assumption that some form of SCFT is exact in the 
limit $N \rightarrow \infty$.  Because we have analyzed only one 
vertex function, $S^{-1}(k)$, only to first order in a loop expansion, 
our calculation provides a consistency check, but not a proof of 
renormalizability.  This is quite different from the situation in 
quantum and statistical field theory, in which methods were developed
long ago to prove renormalizability to all orders in perturbation 
theory. An analogous proof would require an enormous generalization 
of the explicit analysis given here.

At a practical level, our analysis is important as a necessary step 
in the development of a rigorous renormalized perturbation theory of 
corrections to SCFT: Identification and removal of all UV divergences 
is a prerequisite to the systematic study of the long wavelength 
physics that coarse-grained models are intended to describe.  In 
subsequent work on this subject, we will focus on examination of 
physical predictions of the renormalized one-loop theory for 
corrections to SCFT, and on extending the theory beyond the one-loop 
level.

Our procedure for extracting the UV convergent contributions of
physical interest is, for the moment, to simply subtract the results 
of our analytic calculation of the UV divergent contribution to each 
quantity from the results of a numerical evaluation of the underlying 
Fourier integral, using the same finite cutoff wavenumber $\Lambda$ 
in both calculations. To make this procedure work, we must identify 
and subtract all UV divergent contributions to the unrenormalized 
integrals, including the ${\cal O}(\Lambda/N)$ divergences discussed 
in Sec. \ref{sec:Defects}. We have confirmed numerically that this 
procedure yields results that are nearly independent of $\Lambda$ for 
large values of $\Lambda$, and that converge in the limit $\Lambda 
\rightarrow \infty$. It is worth noting that this numerical procedure 
provides a very stringent test of the correctness of our results: Any 
error in either our analytic calculation of UV divergent contributions 
or in our numerical integration would destroy the required cancellation 
of UV divergent terms.

\appendix

\section{Single-Chain Correlations}
\label{appendix}
In this appendix, we provide further details of the derivation of 
asymptotic high-$q$ expansions of the three- and four-point correlation 
functions for both homopolymers and diblock copolymers.  The leading 
order terms in the required expansions were obtained for homopolymers 
in Sec. \ref{sec:SingleChain}.

\subsection{Generalization to Block Copolymers}
To begin, we generalize to block copolymers the discussion of 
multi-point correlations for a Gaussian homopolymer that was given 
in subsection \ref{sub:MultiPoint}. We consider a $n$-point 
correlation function
\begin{equation}
   \sshi^{(n)}_{i_1,\ldots,i_2}(\kv_{1},\ldots,\kv_{n})  = 
   \int\limits_{0}^{N} d^{n}s 
   \left \langle e^{ i \kv_{j} \cdot \Rv (s_{j}) } \right \rangle
   \quad, \label{cint1_multiblock}
\end{equation}
Here, $d^{n}s$ should be understood to indicate an integral over
all values of $s_{1},\ldots,s_{n}$, subject only to the constraint
that, in a block copolymer, the integral over $s_{j}$ must be 
taken over only the block of the polymer that contains monomers 
of type $i_{j}$. (For simplicity, we will only discuss situations
in which each block is chemically distinct). For example, in a 
diblock copolymer for which monomers with $0 < s < fN$ are of 
type $1$,
\begin{equation}
    \sshi^{(3)}_{122}(\kv_{1},\kv_{2},\kv_{3}) = 
    \int\limits_{fN}^{N}ds_{3}\int\limits _{fN}^{N}ds_{2}
    \int\limits_{0}^{fN}ds_{1} 
   \left \langle e^{ i \kv_{j} \cdot \Rv (s_{j}) } \right \rangle
\end{equation}
We may also write $\sshi^{(n)}$ as sum of ordered integrals 
involving different permutations of the wavevector arguments, 
of the form given in Eq.  (\ref{sshi_permutation_sum}), if it
is understood that:
i) The sum over permutations must be restricted to permutations
that do not change any monomer types, but only that exchange 
the identities of monomers of the same type, and 
ii) The ordered integral $\sshi^{(n)}_{i_1,\ldots,i_n}$ is defined 
by integrating over a subspace defined by the requirement that 
$s_{1} < s_{2} < \ldots < s_{n}$ {\it and} that the integral 
with respect to monomer index $s_{j}$ may not extend beyond the 
block containing monomers of type $i_{j}$. For example,
\begin{equation}
    \sshi^{(3)}_{122}(\kv_{1},\kv_{2},\kv_{3}) = 
    \ssho^{(3)}_{122}(\kv_{1},\kv_{2},\kv_{3}) +
    \ssho^{(3)}_{122}(\kv_{1},\kv_{3},\kv_{2}) 
\end{equation}
where
\begin{equation}
    \ssho^{(3)}_{122}(\kv_{1},\kv_{2},\kv_{3}) =
    \int\limits_{fN}^{N}ds_{3}
    \int\limits_{fN}^{s_{3}}ds_{2}
    \int\limits_{0}^{fN}ds_{1} 
   \left \langle e^{ i \kv_{j} \cdot \Rv (s_{j}) } \right \rangle
\end{equation}
The value of an ordered integral $\ssho^{(n)}_{i_1,\ldots,i_{n}}$ can 
be evaluated using a slight generalization of Eq. (\ref{ssho_explicit}), 
in which the limits of integration in the ordered integral are 
interpreted in this manner, and in which the integrand is generalized
to allow for the existence of different statistical segment lengths in 
different blocks.

We also define functions
\begin{eqnarray}
   \sshid_{i_1,\ldots,i_{n}}^{(n)}(\kv_{1},\cdots,\kv_{n}) & \equiv &
   N^{-n} \sshi_{i_1,\ldots,i_{n}}^{(n)}(\kv_{1},\cdots,\kv_{n})
   \\ 
   \sshod_{i_1,\ldots,i_{n}}^{(n)}(\kv_{1},\cdots,\kv_{n}) 
   & \equiv &
   N^{-n} \ssho_{i_1,\ldots,i_{n}}^{(n)}(\kv_{1},\cdots,\kv_{n}) 
   \quad.
\end{eqnarray}
Explicit expressions for these quantities may be written as functions 
of the 2n dimensionless wavevectors $\Kv_{ai} = \kv_{a}b_{i}\sqrt{N/6}$,
and of $f_{1}$ and $f_{2}$. Because the existence of two statistical 
segment lengths $b_{1}$ and $b_{2}$ provides two ways to 
non-dimensionalize each wavevector, we will write
$\sshid_{i_1,\ldots,i_{n}}^{(n)}$ and $\sshod_{i_1,\ldots,i_{n}}^{(n)}$
for diblock copolymers as functions of $\kv_{1},\ldots,\kv_{n}$, rather 
than as functions of dimensionless wavevectors.

\subsection{Three Point Correlations}

\subsubsection{Homopolymer}
The dimensionless ordered integral $\sshod^{(3)}$ for a homopolymer 
is given exactly, for arbitrary wavevector arguments, by
\begin{eqnarray}
   \sshod^{(3)}(\Kv_{1},\Kv_{2},\Kv_{3}) 
   & = & 
   \frac{1}{2}\frac{g(K_{1}^2)-g(K_{3}^2)}{K_{3}^{2}-K_{1}^{2}}
\end{eqnarray}
Each of the ordered integrals required in (\ref{ssh3_homo_sum}) 
may be evaluated using this general result. To obtain an asymptotic 
expansion of each of the resulting integrals, we drop all terms that 
are proportional to $e^{-Q^{2}}$ (which are not analytic functions 
of $1/Q$ in the limit $1/Q \rightarrow 0$), and expand the remaining 
terms in powers of $1/Q$. To ${\cal O}(Q^{-4})$, this yields
\begin{eqnarray}
   \sshod(\Kv,\Qv_{-},-\Qv_{+}) 
   & \simeq & 
   \gf(K^{2}) \left [ \frac{1}{2Q^{2}} - \frac{\alpha K}{2Q^{3}} \right ]
   \nonumber \\
   & + & 
   \left [ \frac{1}{2}K^2 (3+4\alpha^2) g(K^{2}) - 4 \right ]
   \frac{1}{4 Q^{4} } 
   \nonumber \\
   \sshod(\Qv_{-},\Kv,-\Qv_{+}) & \simeq & \frac{1}{Q^{4}}
\end{eqnarray}
An expression for $\sshod(\Kv,-\Qv_{+},\Qv_{-})$ may be obtained 
by taking $\Qv \rightarrow -\Qv$, and $\alpha \rightarrow -\alpha$,
in the expression for $\sshod(\Kv,\Qv,_{-},-\Qv_{+})$. Terms 
linear in $\alpha$ cancel upon adding permutations, giving Eqs. 
(\ref{ssh3_0_homo}) and (\ref{ssh3_1_homo}).

\subsubsection{Diblock Copolymers}
For a diblock copolymer, we need the four functions 
$\sshi^{(3)}_{iii}(\kv,\qv_{-},-\qv_{+})$,
$\sshi^{(3)}_{ijj}(\kv,\qv_{-},-\qv_{+})$, 
$\sshi^{(3)}_{iij}(\kv,\qv_{-},-\qv_{+})$, and
$\sshi^{(3)}_{iji}(\kv,\qv_{-},-\qv_{+})$ for $i \neq j$. 
The function $\sshi^{(3)}_{iji}(\kv,\qv_{-},-\qv_{+})$ may be 
obtained by taking $\qv \rightarrow -\qv$ in our result for
$\sshi^{(3)}_{iij}(\kv,\qv_{-},-\qv_{+})$. The function 
$\sshi^{(3)}_{iii}$ for a diblock copolymer with an $i$ block 
of length $f_{i}N$ is simply equal to the corresponding 
function for a homopolymer of type $i$ and length $f_{i}N$. 

To calculate the remaining two functions, in which one of the 
species indices is different from the other two, we may start 
from the general result
\begin{equation}
   \sshod^{(3)}_{ijj}(\kv_{a},\kv_{b},\kv_{c})
   = \ef_{i}(K_{ai}^{2})
   \frac{\ef_{j}(K_{aj}^2) - \ef_{j}(K_{cj}^2)}{K_{cj}^{2}-K_{aj}^{2}}
\end{equation}
where 
$K_{ai}^{2} \equiv |\kv_{a}|^{2}b_{i}^{2}N/6$ and and  
$K_{ci}^{2} \equiv |\kv_{c}|^{2}b_{i}^{2}N/6$. 
To calculate $\sshod^{(3)}_{jji}$, we note the symmetry
$\sshod^{(3)}_{jji}(\kv_{c},\kv_{b},\kv_{a}) =
 \sshod^{(3)}_{ijj}(\kv_{a},\kv_{b},\kv_{c})$. Evaluating 
and expanding the required integrals yields
\begin{eqnarray}
   \sshod^{(3)}_{ijj}(\kv,\qv_{-},-\qv_{+})
   & \simeq & 
   \ef_{i}(K_{i}^{2})\ef_{j}(K_{j}^{2})
   \left [ \frac{1}{Q_{j}^{2}} - \frac{\alpha K}{Q_{j}^{3}} \right ]
   \nonumber \\ 
   & + &
   \left [ 
   \ef_{j}(K_{j}^{2}) K_{j}^2 (3+4\alpha^2) - 4 \right ]
   \frac{ \ef_{i}(K_{i}^{2}) }{4 Q_{j}^{4} } 
   \nonumber \\
   \sshod_{iij}(\kv,\qv_{-},-\qv_{+}) & \simeq & 
   \ef_{i}(K_{i}^{2}) \frac{1} {Q_{j}^{4}}
\end{eqnarray}
Expressions for $\sshod^{(3)}_{ijj}(\kv,-\qv_{+},\qv_{-})$ 
and $\sshod^{(3)}_{iij}(\kv,-\qv_{+},\qv_{-})$ may be obtained 
by taking $\qv \rightarrow -\qv$ in the above to integrals. 
The quantities $\sshod_{iij}(\qv_{-},\kv,-\qv_{+})$ and
$\sshod_{iij}(-\qv_{+},\kv,\qv_{-})$ are ${\cal O}(Q^{-6})$,
and so may be neglected.

\subsection{Four Point Correlations}

\subsubsection{Homopolymers}
The ordered integrals required to calculate the four point function 
$\psid^{(4)}(\Kv,-\Kv,\Qv,-\Qv)$ may be expressed as
\begin{eqnarray}
   \psiod^{(4)}(\Kv,\pm\Qv,\mp\Qv,-\Kv) & = &
   \int\limits_{0}^{1}d^{4}\underline{\sd} \;
   e^{-K^{2}\sd_{43}}
   H_{\pm}(\sd_{32})
   e^{-K^{2}\sd_{21}} 
   \nonumber \\
   \psiod^{(4)}(\Kv,\pm\Qv,-\Kv,\mp\Qv) & = &
   \int\limits_{0}^{1}d^{4}\underline{\sd} \;
   e^{-K^{2}\sd_{43}}
   H_{\pm}(\sd_{32})
   e^{-Q^{2}\sd_{21}} 
   \nonumber \\
   \psiod^{(4)}(\pm\Qv,\Kv,-\Kv,\mp\Qv) & = &
   \int\limits_{0}^{1}d^{4}\underline{\sd} \;
   e^{-Q^{2}\sd_{43}}
   H_{\pm}(\sd_{32})
   e^{-Q^{2}\sd_{21}} 
   \nonumber
\end{eqnarray}
where 
\begin{equation}
   H_{\pm}(\sd_{32}) \equiv
   e^{-Q_{\pm}^{2}\sd_{32}} - 
   e^{-(Q^{2}+K^{2})\sd_{32}} 
\end{equation}
and $\Qv_{\pm} \equiv \Qv \pm \Kv$.
The dominant behavior of each of these intgrals can be obtained by noting 
that $\sd_{32}$ is confined to very small values by the exponential factors 
in $H(\sd_{32})$, and that we may thus approximate the integral with 
respect to $\sd_{3}$ over the domain $\sd_{2} < \sd_{3} < \sd_{4}$ to 
a first approximation by an integral over $\sd_{2} < \sd_{3} < \infty$.
This approximation yields a common factor
\begin{eqnarray}
   \int\limits_{0}^{\infty} \! d\sd_{32} \; H_{\pm}(s_{32}) 
   & = &
   \frac{1}{Q_{\pm}^{2}} - \frac{1}{Q^{2} + K^{2}} 
   \nonumber 
   \\
   & \simeq &
   \frac{\mp 2K\alpha}{Q^{3}} + \frac{4(K\alpha)^{2}}{Q^{4}} + \cdots
   \quad \label{Hint}
\end{eqnarray}
in the integrals
\begin{eqnarray}
   \psiod^{(4)}(\Kv,\pm\Qv,\mp\Qv,-\Kv) & \simeq & 
   -\frac{1}{2}
   \frac{\partial g(K^{2})}{\partial(K^{2})} \;
   \int_{0}^{\infty} \! ds \; H_{\pm}(s)
   \nonumber \\
   \psio^{(4)}(\Kv,\pm\Qv,-\Kv,\mp\Qv) & \simeq &
   \frac{1}{2 Q^{2}}
   g(K^{2}) \;
   \int_{0}^{\infty} \! ds \; H_{\pm}(s)
   \nonumber \\
   \psio^{(4)}(\pm\Qv,\Kv,-\Kv,\mp\Qv) & \simeq &
   \frac{1}{Q^{4}} \;
   \int_{0}^{\infty} \! ds \; H_{\pm}(s)
   \quad.
\end{eqnarray}
Upon adding the four permutations in which $\pm\Kv$ are the first
and last arguments, and using the expansion given in Eq. (\ref{Hint}) 
for the remaining integral, terms that are proportional to $\alpha$ 
and ${\cal O}(Q^{-3})$ cancel. This leaves a leading order contribution 
of ${\cal O}(Q^{-4})$. All other permutations lead to contributions 
of ${\cal O}(Q^{-6})$ or higher, which do not lead to UV divergent
contributions to $\delta\ssh^{(2)}$ in the one-loop approximation.

\subsubsection{Diblock Copolymers}
Consider a diblock in which block $1$ extends from $0 < s < f_{1}N$ 
and block $2$ is $f_{1}N < s < N$, and let 
$\psid_{ijkl}^{(4)}(\kv,-\kv,\qv,-\qv) \equiv 
N^{-4} \psi_{ijkl}^{(4)}(\kv,-\kv,\qv,-\qv)$. 
As for homopolymers, we may express $\psid_{ijkl}^{(4)}(\kv,-\kv,\qv,-\qv)$ 
as a sum of ordered integrals. To calculate an ordered integral
$\psiod_{i_1 \cdots i_4}^{(4)}(\kv_{1},\kv_{2},\kv_{3},\kv_{4})$ for a 
diblock, we require that the integrals over each monomer index $s_{i}$ 
be constrained to the block specified by the corresponding monomer type 
index, as well as $s_{1} < s_{2} < s_{3} < s_{4}$, and thus that 
$i_1 \leq i_2 \leq i_3 \leq i_4$. As for homopolymers, we find that 
$\psiod_{ijkl}^{(4)}(\kv,-\kv,\qv,-\qv) = 0$ for ordered integrals
in which $\pm\kv$ are the first two or the last two wavevector 
arguments.  

As found for homopolymers, we find that the the dominant contributions 
to $\psid_{ijkl}^{(4)}$ are ${\cal O}(Q^{-4})$, and arise from ordered 
integrals in which $\pm\kv$ are the first and last arguments of 
$\psiod_{ijkl}^{(4)}$.  For diblock copolymers, the ${\cal O}(Q^{-4})$ 
contributions are obtained only from ordered integrals of the form
$\psiod_{ikkj}(\pm\kv,\pm\qv,\mp\qv,\pm\kv)$, where $\pm\qv$ are 
associated with monomers in the same block. The sum of the four ordered 
integrals of the form $\psio_{1111}^{(4)}(\pm\kv,\pm\qv,\mp\qv,\pm\kv)$ 
yield a contribution to $\psi_{1111}^{(4)}(\kv,\qv,-\qv,-\kv)$ equal to 
that obtained from for homopolymer of length $f_{1}N$. The resulting 
contribution to $\psid_{1111}^{(4)}(\kv,\qv,-\qv,-\kv)$ is
\begin{eqnarray}
    \psid^{(4)}_{1111} & \simeq  &
    \frac{16 K^{2}_{1} \alpha^{2}}{Q^{4}_{1}}
    \int\limits_{0}^{f_{1}}d\sd_{2}\int\limits_{0}^{\sd_{2}}d\sd_{1}\;
    \sd_{21} \; e^{-K^{2}_{1}\sd_{21}}
    \nonumber \\
    & \simeq &   -
    \frac{8 K^{2} \alpha^{2}}{Q^{4}}
    \frac{\partial \gf_{11} }{\partial (K^{2})}
    \quad.
\end{eqnarray}
The dominant contributions to $\psid_{1112}(\kv,\qv,-\qv,-\kv)$ arise 
from the ordered integrals $\psiod_{1112}(\pm\kv,\pm\qv,\mp\qv,-\kv)$, 
which yield
\begin{eqnarray}
    \psid^{(4)}_{1112} & \simeq  &
    \frac{16 K^{2}_{1} \alpha^{2}}{Q^{4}_{1}}
    \int\limits_{0}^{f_{2}} d\sd_{2}' \; e^{-K^{2}_{2}\sd_{2}'}
    \int\limits_{0}^{f_{1}} d\sd_{1}' \; \sd_{1}'
    \; e^{-K^{2}_{1}\sd_{1}'} 
    \nonumber \\
    & \simeq &   -
    \frac{8 K^{2}_{1} \alpha^{2}}{Q^{4}_{1}}
    \frac{\partial \gf_{12} }{\partial (K^{2}_{1})}
    \quad,
\end{eqnarray}
where $\sd_{2}' = \sd_{2} - f_{1}$ and $\sd_{1}' = f_{1} - \sd_{2}$.
Corresponding approximations for 
$\psid_{2222}^{(4)}(\pm\kv,\pm\qv,\mp\qv,-\kv)$
$\psid_{1222}^{(4)}(\pm\kv,\pm\qv,\mp\qv,-\kv)$
can be obtained by analogy, by switching the labelling of blocks $1$ 
and $2$. The dominant contributions are thus all of the form
\begin{equation}
    \psid^{(4)}_{ikkj} \simeq 
    \frac{8 K^{2}_{k} \alpha^{2}}{Q^{4}_{k}}
    \frac{\partial \gf_{ij} }{\partial (K^{2}_{k})}
\end{equation}

\bibliography{correlations}

\end{document}

%% file: definitions.tex
\def\eqsp{\;\;}

\def\la{\langle}
\def\ra{\rangle}

\def\rv{{\bf r}}
\def\kv{{\bf k}}
\def\qv{{\bf q}}
\def\ev{{\bf e}}

\def\xv{{\bf x}}

\def\Rv{{\bf R}}
\def\Kv{{\bf K}}
\def\Qv{{\bf Q}}
\def\Iv{{\bf I}}
\def\qplv{{\qv_{+}}}
\def\qmv{{\qv_{-}}}

\def\cmon{c}

\def\cpol{\rho}
\def\cpoli{\tilde{\cpol}}

\def\Ur{U}
\def\Uh{U}
\def\Uhb{{\bf \Uh}}

\def\barBZero{B_{0}}
\def\barchiZero{\chi_{0}}
\def\barchi{\chi}
\def\barchieff{\chi_{\rm eff}}
\def\barchia{\chi_{a}}

\def\Scc{S}
\def\Sccb{{\bf \Scc}}

\def\Scci{\tilde{S}}

\def\Ss{\Omega}
\def\Ssh{\Ss}

\def\Sshi{\tilde{\Ss}}

\def\Sshib{\tilde{\bf \Ss}}

\def\ssr{\omega}
\def\ssh{\ssr}
\def\sshi{\tilde{\ssh}}

\def\ssho{\underline{\ssh}}

\def\sd{\hat{s}}
\def\sshid{\hat{\ssh}}
\def\sshod{\underline{\hat{\ssh}}}
\def\psid{\hat{\psi}}
\def\psiod{\underline{\hat{\psi}}}

\def\ssch{\psi}
\def\sschi{\tilde{\ssch}}

\def\psio{\underline{\psi}}

\def\Ch{C}

\def\hr{h}
\def\hh{h}
\def\hir{\tilde{h}}

\def\hSCF{\tilde{h}}

\def\Vh{V}
\def\Rhi{\tilde{R}}
\def\Ehi{\tilde{E}}

\def\Jr{J}
\def\Jh{\hat{J}}

\def\CGint{N}

\def\saddle{s}
\def\Jrs{J^{\saddle}}

\def\Zp{Z}
\def\Zid{\tilde{\Zp}}

\def\Zc{Z}

\def\Gm{F}

\def\gm{f}

\def\Gr{G}

\def\Gh{\Gr}
\def\Ghi{\tilde{\Gh}}
\def\Ghb{{\bf \Gh}}
\def\Ghib{\tilde{\Ghb}}

\def\Lf{L}

\def\Lm{\Lambda}

\def\Sgh{\Sigma}

\def\Dh{D}

\def\gf{g}
\def\ef{e}
 
